\documentclass[11pt,a4paper]{article}
\usepackage{authblk}
\usepackage{lmodern}
\usepackage{jheppub}
\usepackage{amsmath}
\usepackage{mathtools}
\usepackage{tikz}
\usepackage{empheq}
\usepackage{enumitem}
\usepackage{graphics}
\usepackage{wrapfig}
\usepackage{caption}
\usepackage{natbib}
\usepackage{xcolor}
\usepackage{framed}
\usepackage{array}
\usepackage{dsfont}
\usepackage{comment}
\definecolor{shadecolor}{gray}{0.925}

\usepackage[cbgreek]{textgreek}

\def\sideremark#1{\ifvmode\leavevmode\fi\vadjust{\vbox to0pt{\vss
 \hbox to 0pt{\hskip\hsize\hskip1em
 \vbox{\hsize3cm\tiny\raggedright\pretolerance10000
 \noindent #1\hfill}\hss}\vbox to8pt{\vfil}\vss}}}%

\newcommand{\bi}{\begin{itemize}}
\newcommand{\ei}{\end{itemize}}
\newcommand{\bea}{\begin{align}}
\newcommand{\eea}{\end{align}}
\newcommand{\be}{\begin{equation}}
\newcommand{\ee}{\end{equation}}

\makeatletter
\renewcommand*\env@matrix[1][\arraystretch]{%
  \edef\arraystretch{#1}%
  \hskip -\arraycolsep
  \let\@ifnextchar\new@ifnextchar
  \array{*\c@MaxMatrixCols c}}
\makeatother
\author{S\'ebastien MALHERBE}
\author{\, Paolo PERGOLA}
\author{\, Charlotte SLEIGHT}
\author{\, Massimo TARONNA}

\affiliation{Dipartimento di Fisica ``Ettore Pancini'', Universit\`a degli Studi di Napoli Federico II, \\Monte S. Angelo, Via Cintia, 80126 Napoli, Italy}

\affiliation{INFN, Sezione di Napoli, Monte S. Angelo, Via Cintia, 80126 Napoli, Italy}

\emailAdd{sebastien.malherbe@unina.it, paolo.pergola@unina.it, charlotte.sleight@na.infn.it, massimo.taronna@unina.it}

\title{\centering \huge Celestial Mellin Amplitudes II: \\ Spin-1 and Spin-2 Exchange, and Beyond}

\abstract{We further develop celestial Mellin amplitudes \cite{Pacifico:2024dyo} to capture the exchange of particles with spin. For massless external scalar fields, we derive closed-form celestial Mellin amplitudes for a spin-1 and spin-2 exchange—both massless and massive—and extract the associated OPE data in the direct-channel. For gauge-boson and graviton exchanges, we identify in the celestial OPE the spin-1 current and the spin-2 stress tensor, together with the boundary gauge boson and graviton. Perturbative time-ordered correlators involving spinning fields can be obtained from those of scalars through the action of differential operators on the external points. In Mellin space these actions become finite-difference (shift) operators in the Mellin variables, which we use to extend our results to arbitrary integer spin.}

\begin{document}

\begin{flushright}    
\texttt{}
\end{flushright}

\maketitle

\newpage

\section{Introduction}

Celestial holography \cite{Strominger:2017zoo,Raclariu:2021zjz,Pasterski:2021rjz,Pasterski:2021raf,McLoughlin:2022ljp,Donnay:2023mrd} aims to realise the holographic principle in asymptotically flat spacetime by taking the holographic screen to be the codimension-two celestial sphere at null infinity. The Lorentz group of Minkowski space acts on the celestial sphere as the global conformal group, and the resulting celestial observables take the form of correlation functions in a conformal field theory (CFT). A key question, then, is how a celestial CFT encodes the physics of Minkowski space. 

\vskip 4pt
The fundamental observable in asymptotically flat space is the S-matrix. \emph{Celestial amplitudes} are scattering amplitudes expressed in a conformal primary basis \cite{He:2015zea,Cheung:2016iub,Kapec:2016jld,Pasterski:2016qvg,Pasterski:2017kqt,Pasterski:2017ylz}, where momentum labels are replaced by conformal weights and insertion points on the celestial sphere. Much insight into their structure has come from translating insights from asymptotic symmetries and S-matrix theory into the conformal basis see e.g. \cite{Lam:2017ofc,Stieberger:2018edy,Donnay:2018neh,Stieberger:2018onx,Pate:2019mfs,Adamo:2019ipt,Puhm:2019zbl,Guevara:2019ypd,Fotopoulos:2019vac,Casali:2020vuy,Arkani-Hamed:2020gyp,Strominger:2021mtt,Adamo:2021lrv,Costello:2022wso,Adamo:2024mqn}. Further progress also comes from taking the flat-space limit of AdS/CFT \cite{deGioia:2022fcn,deGioia:2023cbd,deGioia:2024yne} and from exploiting a hyperbolic foliation of Minkowski space \cite{deBoer:2003vf,Cheung:2016iub,Ball:2019atb,Casali:2022fro,Iacobacci:2022yjo,Sleight:2023ojm,Bu:2023cef,Hao:2023wln,Melton:2023hiq,Melton:2023bjw,Iacobacci:2024nhw,Melton:2024jyq,Melton:2024akx,Donnay:2025yoy,Pacifico:2025emk}. 

\vskip 4pt
In anti-de Sitter (AdS), however, there is no standard notion of an S-matrix and holographic correlators in AdS/CFT can be defined as the off-shell extrapolation of time-ordered correlation functions to the boundary. Inspired by this perspective, \emph{celestial correlators} were introduced in \cite{Sleight:2023ojm} as the off-shell extrapolation of time-ordered bulk correlators to the celestial sphere. The extrapolation prescription naturally arises from the hyperbolic foliation $X = R {\hat X}$ with radial coordinate $R>0$ (defined in section \ref{subsec::NaC}), where celestial correlators are defined as the extrapolation of Mellin transformed, time-ordered correlators of fields $\phi_i$ in the radial variables:\footnote{Celestial amplitudes can also be defined using the Mellin transform with respect to the radial direction \cite{Sleight:2023ojm}, where the latter implements the conformal change of basis for both massive and massless particles.}
\begin{align}\label{ccdefn}
    \left\langle O_{1}(Q_1)\ldots O_{n}(Q_n)\right\rangle =\prod_i \lim_{{\hat X}_i\to Q_i}\,\int^\infty_0 \frac{{\rm d}R_i}{R_i}\,R_i^{\Delta_i}\left\langle\phi_1(R_1\hat{X}_1)\ldots \phi_n(R_n\hat{X}_n)\right\rangle.
\end{align}  
The Feynman rules for celestial correlators are inherited from those of bulk time-ordered correlators and were given for scalar field theories in \cite{Iacobacci:2024nhw}. Moreover, they share structural features with AdS holography: at least perturbatively, celestial correlators for a $(d+2)$-dimensional Minkowski space $\mathbb{M}^{d+2}$ admit a reformulation \cite{Iacobacci:2024nhw} in terms of boundary correlators in $(d+1)$-dimensional Euclidean AdS space (EAdS$_{d+1}$), which allows to import techniques and results from AdS/CFT—which so far has been explored in \cite{Pacifico:2024dyo,Pacifico:2025emk}. 

\vskip 4pt 
The Mellin-amplitude representation \cite{Mack:2009mi,Mack:2009gy} of conformal correlators has been instrumental in the study of holographic correlators in AdS and in the conformal bootstrap, starting from the work \cite{Penedones:2010ue} and subsequent developments e.g. \cite{Fitzpatrick:2011ia,Paulos:2011ie,Fitzpatrick:2012cg,Costa:2012cb,Gopakumar:2016wkt,Gopakumar:2016cpb,Aharony:2016dwx,Rastelli:2017udc,Sleight:2018ryu,Gopakumar:2018xqi,Sleight:2019ive}. Motivated by this and by structural parallels with AdS holography, the recent work \cite{Pacifico:2024dyo} introduced a Mellin-amplitude representation for celestial correlators \eqref{ccdefn}—the \emph{celestial Mellin amplitude}. Perturbative celestial Mellin amplitudes mirror key features of AdS Mellin amplitudes \cite{Pacifico:2024dyo}: they are meromorphic in the Mellin variables; contact interactions give polynomials; and particle exchange appears as a specific family of poles encoding the exchanged single particle state on the celestial sphere.

\vskip 4pt
In this work we further develop the Mellin-amplitude representation of celestial correlators \cite{Pacifico:2024dyo} and use it to capture the exchange of particles with spin in Minkowski space. Perturbative, position-space, correlators involving spinning fields can be obtained from those of scalars through the action of differential operators on the external points. In Mellin space, these actions become finite-difference (shift) operations in the Mellin variables. Focusing for simplicity on massless external scalars,\footnote{The extension to massive external scalars is briefly discussed in appendix \ref{app::extmass}.} we derive closed-form celestial Mellin amplitudes for spin-1 and spin-2 exchange—both massless and massive—and show how to extract the associated OPE data in the direct-channel. Using the difference operations we extend the construction to any integer spin $J$: for the massless exchange we obtain a closed-form result together with the complete set of OPE data, while for the massive exchange we compute a subset of OPE coefficients and characterise the remaining structures. These results extend to general exchanged integer spin-$J$ those obtained for scalar field theories ($J=0$) in \cite{Pacifico:2024dyo}.

\vskip 4pt
For external massless scalars with scaling dimensions $\Delta_i$, we find that exchange of a massive spin-$J$ field of mass $m$ appears, in the direct-channel OPE on the celestial sphere, as infinite families of primary operators with all spins $k=0,\ldots,J$ and scaling dimensions:
\begin{subequations}
 \label{twrmJ}
\begin{align}
    \Delta^{(n,k)}_{12} = \Delta_1+\Delta_2 + (2-d-k)+2n, \qquad n = 0, \ldots, \infty,\\
    \Delta^{(n,k)}_{34} = \Delta_3+\Delta_4 + (2-d-k)+2n, \qquad n = 0, \ldots, \infty.
\end{align}   
\end{subequations}
The corresponding celestial Mellin amplitude $M_{\Delta_1 \Delta_2 \Delta_3 \Delta_4}\left(s_{12}, s_{13}\right)$ in the Mellin variables $s_{12}, s_{13}$ has a residue expansion of the form:\footnote{The $\ldots$ denote descendant contributions and ${\cal Q}_{\tau^{(n,k)}_{12}, J,0}\left(s_{13}\right)$ are kinematic polynomials of degree $J$ in $s_{13}$. Mellin amplitudes and the relevant notation are introduced in section \ref{sec::CMA}.}
\begin{multline}
    M_{\Delta_1 \Delta_2 \Delta_3 \Delta_4}\left(s_{12}, s_{13}\right) = \sum^{J}_{k=0}\sum^\infty_{n=0} a^{(m)}_{\tau^{(n,k)}_{12},\,k} \left[ \frac{{\cal Q}_{\tau^{(n,k)}_{12}, k,0}\left(s_{13}\right)}{s_{12}-\Delta^{(n,k)}_{12}+k}+\ldots\right]\\+ \Delta_{1}, \Delta_2 \rightarrow \Delta_{3}, \Delta_4,
\end{multline}
with twists $\tau^{(n,k)}_{ij}=\Delta^{(n,k)}_{ij}-k$. For exchanged spin $J=1, 2$, we determine the OPE coefficients $a^{(m)}_{\tau^{(0,k)}_{ij},\,k}$ of the leading twist operators $(n=0)$ for each $k$ and, when $J > 2$, for the highest spin family $k=J$. 
 
 \vskip 4pt
 When the exchanged particle is also massless, the towers \eqref{twrmJ} collapse to a single shadow pair of spin-$J$ primaries with dimensions:
\begin{equation}\label{exchopmlJINT}
   \Delta^{(0,J)}_{12} = \Delta_1+\Delta_2 + (2-d-J), \qquad \Delta^{(0,J)}_{34}= \Delta_3+\Delta_4 + (2-d-J),
\end{equation}
which are shadow of one another by virtue of the following constraint on the scaling dimensions $\Delta_i$ in the massless case:
\begin{equation}
    -3d +4-2J + \sum_i \Delta_i = 0.
\end{equation}
That there are no lower spin contributions can be understood from the fact that massless particles in $\mathbb{M}^{d+2}$ have $SO(d)$ spin, as do irreducible representations of the conformal group on the celestial sphere. The celestial correlator is given (up to contact terms) by a single conformal partial wave $\mathcal{F}_{\Delta,\, J}$ \eqref{CPW}:
\begin{multline}
    \left\langle O_{1}(Q_1)O_{2}(Q_2)O_{3}(Q_3) O_{4}(Q_4)\right\rangle_{m^2=0,\,J} \\ = 2\pi i\,\delta\left(\frac{-3d +4-2J + \sum_i \Delta_i}{2} \right)a^{\prime\,(m=0)}_{\tau^{(0)}_{12}, J} \mathcal{F}_{\Delta^{(0,J)}_{12},\, J}\left(Q_1,Q_2,Q_3,Q_4\right),
\end{multline}
where the coefficient $a^{\prime\,(m=0)}_{\tau^{(0)}_{12}, J}$ is given in \eqref{mass_lim_exch_res_sum_spin_J_2}. For massless external scalars, the boundary two-point function only has support for $\Delta_i = \frac{d}{2}$ \cite{Pacifico:2025emk}. For $\Delta_1=\Delta_2=\frac{d}{2}$, the exchanged operators \eqref{exchopmlJINT} correspond precisely to the boundary gauge boson and boundary current of spin-$J$:
\begin{equation}
    \Delta^{(0,J)}_{12} = 2-J, \qquad \Delta^{(0,J)}_{34} = J+d-2,
\end{equation}
and likewise for $\Delta_3=\Delta_4=\frac{d}{2}$. For spins $J=1$ and $J=2$ in particular, this would correspond to the boundary gauge boson and graviton respectively, together with the boundary currents.

\vskip 4pt
Along the way we develop practical tools to compute perturbative time-ordered correlators with spinning fields in Minkowski space. Any Feynman diagram with spin or derivative couplings can be obtained by acting with differential operators on the external points of the corresponding scalar diagram. In the Mellin-amplitude representation of Minkowski time-ordered correlators \cite{Mack:2009mi}, these differential actions become finite-difference operations in the Mellin variables. Celestial Mellin amplitudes then follow by taking the boundary (extrapolate) limit \eqref{ccdefn}, which in Mellin space is implemented by constraints on the Mellin variables encoding the conformal kinematics on the celestial sphere.

\vskip 4pt
The paper is organised as follows. Section \ref{sec::CMA} reviews the Mellin representation of conformal correlators \cite{Mack:2009mi,Mack:2009gy}. Section \ref{sec::reviewscalarMA} summarises the computation of Mellin amplitudes for perturbative celestial correlators in scalar field theory \cite{Pacifico:2024dyo}. Section \ref{subsec::extractingOPE} explains how to extract OPE data for particle exchange from the celestial Mellin-amplitude representation, and Section \ref{subsec::OPEinCCFT} discusses the resulting OPE structure in celestial CFT. Section \ref{sec::spin1} treats spin-1 exchange: massive exchange in Section \ref{subsec::massivespin1}, gauge-boson exchange in Section \ref{subsec::gaugeboson}, and the associated OPE data in Section \ref{subsec::opedataspin1}. Section \ref{sec::spin2} treats spin-2 exchange: massive exchange in Section \ref{subsec::massivespin2}, graviton exchange in Section \ref{subsec::graviton}, and the corresponding OPE data in Section \ref{subsec::OPEdataspin2}. In section \ref{sec::higherspin} we extend these results to arbitrary exchanged spin $J$. In appendix \ref{app::extmass} we discuss the extension to massive external scalars. Various technical details are relegated to the remaining appendices.

\newpage
\subsection{Notation and conventions.}
\label{subsec::NaC}

We work in $(d+2)$-dimensional Minkowski space $\mathbb{M}^{d+2}$ with coordinates $X^\mu$ where $\mu = 0, 1, \ldots, d+1$, and ``mostly plus" metric $\eta = \left(- + \ldots +\right)$. The extrapolation to the celestial sphere is defined via the hyperbolic foliation:
\begin{equation}
    X = R {\hat X}, \qquad X^2 = \sigma R^2,
\end{equation}
where $R>0$ is a radial coordinate and $\hat X^2=\sigma$, $\sigma\in\{-1,+1\}$. The celestial sphere is the projective cone of light rays:
\begin{equation}\label{nullcone}
    Q^2=0, \qquad Q \equiv \lambda Q, \qquad \lambda \in \mathbb{R}^+,
\end{equation}
which describe $d$-dimensional spheres $S^-_d$ ($Q^0<0$) and $S^+_d$ ($Q^0>0$) in the infinite past and infinite future respectively.\footnote{This can be made manifest by introducing projective coordinates:
\begin{align}
    \xi_1 = Q^1/Q^0, \quad \xi_2 = Q^2/Q^0, \quad \ldots \quad, \quad \xi_{d+1} = Q^{d+1}/Q^0,
\end{align}
which satisfy
\begin{equation}
    \xi^2_1+\ldots+\xi^2_{d+1}-1=0.
\end{equation}}
The physical $d$-dimensional space can be obtained by considering the Euclidean section: 
\begin{equation}\label{QPS}
    Q = 
    \left(\frac{1+\vec{y}^2}{2},\vec{y},\frac{1-\vec{y}^2}{2}\right), \qquad \vec{y} \in \mathbb{R}^d. 
\end{equation}
 For example, the separation of points $\vec{y}_1$, $\vec{y_2}$ on the sphere is encoded by:
\begin{equation}\label{sep}
    Q_1 \cdot Q_2 = -\frac{1}{2}\left(\vec{y}_1 - \vec{y}_2\right)^2.
\end{equation}
More details on the formulation of conformal kinematics in $\mathbb{R}^d$ on the projective null cone of $\mathbb{M}^{d+2}$ can be found e.g. in \cite{Rychkov:2016iqz}.

\newpage

\section{Celestial Mellin amplitudes}
\label{sec::CMA}

Mellin amplitudes for conformal correlation functions were introduced by Mack in \cite{Mack:2009mi,Mack:2009gy}. For a conformal correlation function of operators $O_i$ with scaling dimension $\Delta_i$, its Mellin amplitude $M_{\Delta_1 \ldots \Delta_n}\left(\delta_{ij}\right)$ is defined by:\footnote{In this work conformal kinematics in $\mathbb{R}^d$ is formulated on the projective cone \eqref{nullcone} of light rays $Q_i$ in $\mathbb{M}^{d+2}$. To project to the physical space one uses \eqref{QPS} and \eqref{sep}.}
\begin{equation}\label{MR}
\left\langle O_{1}(Q_1)\ldots O_{n}(Q_n)\right\rangle = \int^{c+i\infty}_{c-i\infty}  \frac{{\rm d}\delta_{ij}}{2\pi i} M_{\Delta_1 \ldots \Delta_n}\left(\delta_{ij}\right)\prod_{i<j} \Gamma\left(\delta_{ij}\right)\left(-2\, Q_i \cdot Q_j\right)^{-\delta_{ij}},
\end{equation}
where the integration contour runs parallel to the imaginary axis. Conformal covariance requires that the Mellin integration variables satisfy the following constraints:
\begin{equation}\label{MC}
    \sum^n_{j=1} \delta_{ij}=0, \qquad  \delta_{ij} =  \delta_{ji},  \qquad \delta_{ii}=-\Delta_i.
\end{equation}
The operator product expansion (OPE) is encoded into simple analytic properties of the Mellin integrand. Consider the OPE
\begin{equation}\label{ope}
    O_{i}O_j \sim \sum_{\Delta,\,J} c_{ij\Delta,\,J} \left(O_{\Delta,\,J} + \text{descendants}\right).
\end{equation}
The contributions from the operator $O_{\Delta,\,J}$ of scaling dimension $\Delta$ and spin-$J$ are encoded in the Mellin integrand by the following simple poles:
\begin{equation}
    \delta_{ij} = \frac{\Delta_i+\Delta_j-\Delta-J}{2}-m, \qquad m = 0, 1, 2, \ldots\,, 
\end{equation}
where the satellite poles with $m>0$ are the contributions from the descendant operators. The poles from the $\Gamma$-functions in the Mellin integrand \eqref{MR} generate contributions from  towers of composite ``double-trace" operators of the schematic form
\begin{equation}\label{DT}
    \left[O_i O_j\right]_{n,\ell} \sim O_i \partial_{i_1} \ldots  \partial_{i_\ell}  (\partial^ 2)^ n O_j+\ldots\,, \qquad n = 0, 1, 2, \ldots\,, 
\end{equation}
which have scaling dimensions
\begin{equation}
    \Delta_i+ \Delta_j+2n+\ell\,.
\end{equation}
This feature makes Mellin amplitudes particularly suitable for large $N$ theories and the corresponding perturbative calculations of holographic correlation functions in AdS space, where the double-trace operators \eqref{DT} appear in the OPE \eqref{ope}. For a large class of diagrams, the $\Gamma$-function poles automatically account for the double-trace contributions \cite{Penedones:2010ue}.

\vskip 4pt 
Mellin amplitudes enjoy many structural similarities to flat space scattering amplitudes. The constraints \eqref{MC} on the Mellin variables $\delta_{ij}$ can be solved by introducing fictitious momenta $p_i$,
\begin{equation}
    \delta_{ij} = p_i \cdot p_j,
\end{equation}
which obey ``momentum conservation" and the ``on-shell" condition:
\begin{equation}
   \sum\limits^n_{i=1} p_i = 0, \qquad  p^2_i=-\Delta_i.
\end{equation}
Focusing on four-point functions ($n=4$), there are two independent ``Mandelstam" variables \cite{Costa:2012cb}:\footnote{The other Mellin variables can be expressed in terms of $\delta_{12}$ and $\delta_{23}$ via:
\begin{subequations}
    \begin{align}
    \delta_{14}&= \Delta_1 - \delta_{12}-\delta_{13}, \hspace*{2.35cm} 
     \delta_{23}=  - \delta_{12}-\delta_{13} + \frac{\Delta_1+\Delta_2+\Delta_3-\Delta_4}{2},\\
     \delta_{24}&=  \delta_{13} + \frac{\Delta_2+\Delta_4-\Delta_1-\Delta_3}{2}, \qquad 
     \delta_{34}=  \delta_{12} + \frac{\Delta_3+\Delta_4-\Delta_1-\Delta_2}{2}.
\end{align}
\end{subequations}}
\begin{subequations}\label{MBmandel}
 \begin{align}
     s_{12} &= -\left(p_1+p_2\right)^2= \Delta_1+\Delta_2-2\delta_{12},\\
     s_{13} &= -\left(p_1 + p_3\right)^2 - \Delta_1-\Delta_4 =\Delta_3-\Delta_4-2\delta_{13}.
\end{align}   
\end{subequations}
Taking $s_{12}$ and $s_{13}$ as independent Mellin variables, conformal four-point functions have the following Mellin representation:
\begin{multline}\label{MBst}
\left\langle O_{1}(Q_1) O_{2}(Q_2) O_{3}(Q_3) O_{4}(Q_4)\right\rangle \\ = \frac{1}{\left(-2 Q_1 \cdot Q_2\right)^{\frac{\Delta_1+\Delta_2}{2}}\left(-2 Q_3 \cdot Q_4\right)^{\frac{\Delta_3+\Delta_4}{2}}}\left(\frac{Q_2 \cdot Q_4}{Q_1 \cdot Q_4}\right)^{\frac{\Delta_1-\Delta_2}{2}}\left(\frac{Q_1 \cdot Q_4}{Q_1 \cdot Q_3}\right)^{\frac{\Delta_3-\Delta_4}{2}}\\ \times \int^{+i\infty}_{-i\infty}\frac{{\rm d}s_{12}{\rm d}s_{13}}{\left(4\pi i\right)^2}\, u^{\frac{s_{12}}{2}}v^{-\left(\frac{s_{12}+s_{13}}{2}\right)}\,\rho\left(s_{12},s_{13}\right)\,M_{\Delta_1 \Delta_2 \Delta_3 \Delta_4}\left(s_{12},s_{13}\right),
\end{multline}
with conformal invariant cross ratios
\begin{equation}\label{crossratios}
    u = \frac{Q_1 \cdot Q_2\,Q_3 \cdot Q_4}{Q_1 \cdot Q_3 Q_2 \cdot Q_4}, \qquad v = \frac{Q_1 \cdot Q_4\,Q_2 \cdot Q_3}{Q_1 \cdot Q_3 Q_2 \cdot Q_4},
\end{equation}
and the $\Gamma$-functions are contained in the measure
\begin{multline}\label{MMeasure}
\rho\left(s_{12},s_{13}\right)=\Gamma\left(\frac{\Delta_1+\Delta_2-s_{12}}{2}\right)\Gamma\left(\frac{\Delta_3+\Delta_4-s_{12}}{2}\right)\\ \times \Gamma\left(\frac{\Delta_{34}-s_{13}}{2}\right)\Gamma\left(\frac{-\Delta_{12}-s_{13}}{2}\right)\Gamma\left(\frac{s_{12}+s_{13}}{2}\right)\Gamma\left(\frac{s_{12}+s_{13}+\Delta_{12}-\Delta_{34}}{2}\right),
\end{multline}
where $\Delta_{ij}=\Delta_i-\Delta_j$. The conformal block expansion of conformal four-point functions in the s-channel takes the form \cite{Costa:2012cb} 
\begin{equation}\label{MAcbe}
\rho\left(s_{12},s_{13}\right)M_{\Delta_1 \Delta_2 \Delta_3 \Delta_4}\left(s_{12}, s_{13}\right) = {\tilde \rho}\left(s_{12},s_{13}\right) \sum_{\Delta,J} a_{\tau,J}\left[\sum^\infty_{m=0}\frac{{\cal Q}_{\tau, J,m}\left(s_{13}\right)}{s_{12}-\tau-2m}+\ldots\right],
\end{equation}
where the $\ldots$ denote the entire function piece of the conformal block (see e.g. \cite{Fitzpatrick:2012cg}) and we introduced the reduced Mellin measure\footnote{This is obtained from the measure \eqref{MMeasure} by stripping off the $\Gamma$ functions with poles encoding contributions from double-trace operators \eqref{DT}.} 
\begin{multline}\label{rrho}
{\tilde \rho}\left(s_{12},s_{13}\right)=\Gamma\left(\frac{\Delta_{34}-s_{13}}{2}\right)\Gamma\left(\frac{-\Delta_{12}-s_{13}}{2}\right)\\ \times \Gamma\left(\frac{s_{12}+s_{13}}{2}\right)\Gamma\left(\frac{s_{12}+s_{13}+\Delta_{12}-\Delta_{34}}{2}\right).
\end{multline}
The contribution from a primary operator $O_{\Delta,J}$ of scaling dimension $\Delta$ and spin-$J$, and its descendants ($m>0$), are encoded in the poles
\begin{equation}
    s_{12} = \tau+2m, \qquad m=0,1,2,\ldots,
\end{equation}
where $\tau = \Delta-J$ is the twist. Analogous to the partial wave expansion of the flat-space S-matrix, the residues of these poles are kinematic polynomials ${\cal Q}_{\tau, J,m}\left(s_{13}\right)$ of degree $J$ in the Mellin variable $s_{13}$. Their explicit form is given in appendix \ref{a::mack_pol}. 
For the OPE \eqref{ope} the conformal block coefficients $a_{\tau,J}$ factorise as:
\begin{equation}\label{discbope}
    a_{\tau,J}=c_{12\Delta,J}\,c_{34\Delta,J}.
\end{equation}

\vskip 4pt
Mellin amplitudes for holographic correlators in AdS make transparent the analogy with flat space scattering amplitudes \cite{Penedones:2010ue,Fitzpatrick:2011ia,Paulos:2011ie}: Mellin amplitudes for contact Witten diagrams are polynomial in the Mellin variables and for a tree level exchange Witten diagram the corresponding Mellin amplitude only has poles corresponding to the ``single-trace" operator dual to the field being exchanged in AdS. For example, the Mellin amplitude for an s-channel exchange Witten
diagram with an exchanged field of conformal dimension $\Delta$ and spin $J$ has the following simple analytic structure \cite{Costa:2012cb}: 
\begin{equation}\label{MAAdSexch}
    M_{\Delta_1 \Delta_2 \Delta_3 \Delta_4}\left(s_{12}, s_{13}\right) = c_{12\Delta,J}c_{34\Delta,J} \sum^\infty_{m=0} \frac{{\cal Q}_{\tau, J,m}\left(s_{13}\right)}{s_{12}-\tau-2m} + P_{J-1}\left(s_{12}, s_{13}\right),
\end{equation}
 where $P_{J-1}\left(s_{12}, s_{13}\right)$ is degree $J-1$ polynomial in $s_{12}$ and $s_{13}$.

\vskip 4pt
 Inspired by the utility of Mellin amplitudes for holographic correlators in AdS and CFT, the recent work \cite{Pacifico:2024dyo} proposed to consider the Mellin representation \eqref{MR} for celestial correlators \eqref{ccdefn}. This follows naturally from the Mellin representation of time-ordered correlation functions in the bulk Minkowski space, which takes the form \cite{Mack:2009mi}
\begin{equation}\label{tcMA}
    \left\langle\phi_1(X_1)\ldots \phi_n(X_n)\right\rangle = \int^{i\infty}_{-i\infty}   \frac{{\rm d}\delta_{ij}}{2\pi i}\,  M\left(\delta_{ij}\right) \prod_{i<j}\Gamma\left(\delta_{ij}\right)\left[\left(Y_i-Y_j\right)^2+i \epsilon\right]^{-\delta_{ij}},
\end{equation}
with Mellin amplitude $M\left(\delta_{ij}\right)$, where for bulk correlation functions the Mellin variables are subject only to $\delta_{ij}=\delta_{ji}$. To obtain the corresponding celestial correlator, the external points are extrapolated to the celestial sphere according to the prescription \cite{Sleight:2023ojm}, 
\begin{align}\label{ccdefn2}
    \left\langle O_{1}(Q_1)\ldots O_{n}(Q_n)\right\rangle =\prod_i \lim_{{\hat X}_i\to Q_i}\,\int^\infty_0 \frac{{\rm d}R_i}{R_i}\,R_i^{\Delta_i}\left\langle\phi_1(R_1\hat{X}_1)\ldots \phi_n(R_n\hat{X}_n)\right\rangle.
\end{align} 
The Mellin transform with respect to the radial coordinates $R_i$ introduces the constraints \eqref{MC} for conformal covariance:
\begin{equation}\label{confcon}
    \int_0^\infty\frac{{\rm d}R_i}{R_i} R^{\Delta_i}_iR_i^{-\sum_{j\neq i}\delta_{ij}}=(2\pi i)\delta\left(\Delta_i-\sum_{j\neq i}\delta_{ij}\right),
\end{equation}
which also implies that
\begin{equation}\label{constconf}
  \sum_{i<j}\delta_{ij} = \frac{1}{2}\sum_{i}\Delta_i.
\end{equation}
As expected by conformal symmetry, the Mellin representation of celestial correlators \eqref{ccdefn2} therefore takes the same form \eqref{MR} but with an $i \epsilon$ prescription originating from the time ordering of the underlying bulk correlation functions \cite{Pacifico:2024dyo}:
\begin{equation}\label{MAcc1}
\left\langle O_{1}(Q_1)\ldots O_{n}(Q_n)\right\rangle = \int^{c+i\infty}_{c-i\infty}  \frac{{\rm d}\delta_{ij}}{2\pi i} M_{\Delta_1 \ldots \Delta_n}\left(\delta_{ij}\right)\prod_{i<j} \Gamma\left(\delta_{ij}\right)\left(-2\, Q_i \cdot Q_j+i\epsilon\right)^{-\delta_{ij}}.
\end{equation}
The \emph{celestial Mellin amplitudes} $M_{\Delta_1 \ldots \Delta_n}\left(\delta_{ij}\right)$ therefore follow from the Mellin representation $M\left(\delta_{ij}\right)$ of the corresponding time-ordered correlation functions \eqref{tcMA} in Minkowski space by imposing the conformal constraints \eqref{constconf} on the Mellin variables $\delta_{ij}$.

\vskip 4pt
Like Mellin amplitudes for AdS Witten diagrams, celestial Mellin amplitudes \eqref{MAcc1} are meromorphic functions of the Mellin variables $\delta_{ij}$, where contact interactions are polynomial in the Mellin variables and particle exchanges are encoded by specific sets of poles. In the next section we review these properties and the techniques to compute perturbative celestial Mellin amplitudes in scalar field theories \cite{Pacifico:2024dyo}, in the view to extend them to fields with spin in later sections.

\subsection{Review: Scalar field theories}
\label{sec::reviewscalarMA}

In this section we review the approach to determine perturbative celestial Mellin amplitudes in scalar field theories presented in \cite{Pacifico:2024dyo}. This will serve as a basis for the extension to fields with spin in later sections through the action of differential operators on the external points (in position space) and finite difference operations (in Mellin space).

\vskip 4pt
A useful tool to determine the Mellin representation \eqref{tcMA} of time-ordered correlators in Minkowski space is the Schwinger parameterisation of the Feynman propagator in position space. For a scalar field of mass $m$ in $\mathbb{M}^{d+2}$, this is:
\begin{subequations}\label{SPm}
\begin{align}
    G^{(m)}_{T}\left(X,Y\right)&=\frac{1}{2 \pi^{\frac{d+2}{2}}}\left(\frac{m}{2}\right)^{\frac{d}{2}} \frac{1}{\left[\left(X-Y\right)^2+i\epsilon\right]^{\frac{d}{4}}}K_{\frac{d}{2}}\left(m\left[\left(X-Y\right)^2+i\epsilon\right]^{\frac{1}{2}}\right)\\&= \int^{+i\infty}_{-i\infty}\frac{{\rm d}s}{2\pi i}\frac{1}{4\pi^{\frac{d+2}{2}}}\Gamma\left(s-\tfrac{d}{4}\right)\left(\frac{m}{2}\right)^{-2s+\frac{d}{2}}i^{-(s+\tfrac{d}{4})} \\
& \hspace*{5cm}\times \int^{\infty}_0 \frac{{\rm d}t}{t} t^{s+\tfrac{d}{4}}\, \exp \left[i t \left(X-Y\right)^2\right],  \nonumber
\end{align}    
\end{subequations}
which, for mass $m=0$, reduces to:
\begin{subequations}\label{MLschw}
 \begin{align}
    G^{(0)}_{T}\left(X,Y\right)&=\frac{\Gamma\left(\frac{d}{2}\right)}{4 \pi^{\frac{d+2}{2}}} \frac{1}{\left[\left(X-Y\right)^2+i\epsilon\right]^{\frac{d}{2}}}\\&=\frac{i^{-\frac{d}{2}}}{4 \pi^{\frac{d+2}{2}}} \int^{\infty}_0 \frac{{\rm d}t}{t} t^{\frac{d}{2}} \exp \left[i t \left(X-Y\right)^2\right].
\end{align}   
\end{subequations}
Using Schwinger parameterisation, integrals over internal points in any given Feynman diagram are given by (nested) Gaussian integrals of the form:
\begin{align}\label{gaussb}
    E_{t_1,\,t_2,\, \ldots ,\,t_{n}}\left(Y_1,Y_2,\ldots,Y_n\right) &=\int {\rm d}^{d+2}X\, \exp\left[i\sum\limits^n_{i=1}t_i\left(X-Y_i\right)^2\right],\\ \nonumber 
    &= i^{\frac{d}{2}} \pi^{\frac{d+2}{2}} \left(t_1+ \ldots + t_n\right)^{-\frac{d+2}{2}} \exp\left[\frac{i}{t_1+ \ldots + t_n}\sum\limits^n_{i<j}t_it_j\left(Y_i-Y_j\right)^2\right].
\end{align}
The Mellin amplitude \eqref{tcMA} for the corresponding contribution to time-ordered correlators follows from the Mellin transform of the exponential function:
\begin{equation}\label{MTexp}
     \exp\left[it\left(Y_i-Y_j\right)^2\right] = \int^{+i \infty}_{-i\infty} \frac{{\rm d} \delta}{2\pi i} \Gamma\left(\delta\right) \left[-it \left(Y_i-Y_j\right)^2+\epsilon\right]^{-\delta}.
\end{equation}
 The corresponding celestial Mellin amplitude \eqref{MAcc1} then follows by imposing the conformal constraints \eqref{confcon}. In the following we give some examples.

\paragraph{Contact diagrams.} Consider the $n$-point contact diagram generated by the following non-derivative vertex of scalar fields,
\begin{equation}\label{nptcontnond}
     {\cal V}_{12 \ldots n-1 m} = g_{12 \ldots n-1 \varphi}\, \phi_1 \phi_2 \ldots \phi_{n-1} \varphi.
\end{equation}
For the purposes of this work it will be sufficient to take the scalars $\phi_i$ to be massless and the scalar $\varphi$ to have generic mass $m$. The corresponding contact diagram reads
\begin{multline}\label{contnpt}
  {\cal A}^{{\cal V}_{12 \ldots n-1 m}}\left(Y_1,Y_2,\ldots,Y_n\right) \\ = -i g_{12 \ldots n-1 \varphi} \int {\rm d}^{d+2}X\, G^{(0)}_{T}\left(X,Y_1\right) \ldots G^{(0)}_{T}\left(X,Y_{n-1}\right) G^{(m)}_{T}\left(X,Y_{n}\right), 
\end{multline}
which, employing Schwinger parameterisation \eqref{SPm}, can be expressed in the form
\begin{multline}
    {\cal A}^{{\cal V}_{12 \ldots n-1 m}}\left(Y_1,Y_2,\ldots,Y_n\right) =-ig_{12 \ldots n-1 \varphi} \left(\frac{1}{4\pi^{\frac{d+2}{2}}}i^{-\frac{d}{2}}\right)^{n-1}\, \left(\frac{1}{4\pi^{\frac{d+2}{2}}}\right)  \\ \times \int^{+i\infty}_{-i\infty}\frac{{\rm d}s_n}{2\pi i}\Gamma\left(s_n-\tfrac{d}{4}\right)\left(\frac{m}{2}\right)^{-2s_n+\frac{d}{2}} i^{-(s_n+\frac{d}{4})} \\ \times \int^{\infty}_0 \frac{{\rm d}t_n}{t_n} t^{s_n+\frac{d}{4}}_n \prod^{n-1}_{i=1} \frac{{\rm d}t_i}{t_i} t^{\frac{d}{2}}_i\, E_{t_1,\,t_2,\, \ldots ,\,t_{n}}\left(Y_1,Y_2,\ldots,Y_n\right),
\end{multline}
in terms of the Gaussian integral \eqref{gaussb}. The Mellin representation is obtained from the Mellin transform of the exponential function as in \eqref{MTexp}, so that the Gaussian integral \eqref{gaussb} takes the form
\begin{multline}
    E_{t_1,\,t_2,\, \ldots ,\,t_{n}}\left(Y_1,Y_2,\ldots,Y_n\right) = i^{\frac{d}{2}} \pi^{\frac{d+2}{2}} \int^{i\infty}_{-i\infty} \frac{{\rm d}\delta_{ij}}{2\pi i}\,\prod_{i<j}\Gamma\left(\delta_{ij}\right)\left[-i t_i t_j\left(Y_i-Y_j\right)^2+\epsilon\right]^{-\delta_{ij}} \\ 
    \times \left(t_1+ \ldots + t_n\right)^{-\frac{d+2}{2}+\sum_{i<j}\delta_{ij}
    }.
\end{multline}
The integrals over the Schwinger parameters can be evaluated using the approach given in \cite{Pacifico:2024dyo}, which we review in appendix \ref{app::schwint}, giving the contact diagram in the form \eqref{tcMA}:
\begin{multline}\label{MAbulkcont}
  {\cal A}^{{\cal V}_{12 \ldots n-1 m}}\left(Y_1,Y_2,\ldots,Y_n\right)
  \\ = \int^{i\infty}_{-i\infty}   \frac{{\rm d}\delta_{ij}}{2\pi i}\,  M^{{\cal V}_{12 \ldots n-1 m}}\left(\delta_{ij}\right) \prod_{i<j}\Gamma\left(\delta_{ij}\right)\left[\left(Y_i-Y_j\right)^2+i \epsilon\right]^{-\delta_{ij}},
\end{multline}
with Mellin amplitude
\begin{multline}\label{MAbulkcontexpl}
    M^{{\cal V}_{12 \ldots n-1 m}}\left(\delta_{ij}\right)=g_{12 \ldots n-1 \varphi}i^{\frac{d}{2}-1} \pi^{\frac{d+2}{2}} \left(\frac{1}{4\pi^{\frac{d+2}{2}}}i^{-\frac{d}{2}}\right)^{n-1}\, \left(\frac{1}{4\pi^{\frac{d+2}{2}}}\right)  \\ \times \left(\frac{m}{2}\right)^{\frac{2d(n-1)-4}{2}-2\sum_{i<j}\delta_{ij}} i^{\frac{d(n-2)}{2}-1} \Gamma\left(-\tfrac{d(n-1)}{2}+1+\sum_{i<j}\delta_{ij}\right) \\ \times  \frac{\Gamma\left(-\tfrac{d(n-2)}{2}+1+\sum_{i<j}\delta_{ij}-\sum_{i\neq n}\delta_{in}\right)\prod\limits^{n-1}_{i=1}\Gamma\left(\frac{d}{2}-\sum_{j\neq i}\delta_{ij}\right)}{\Gamma\left(\frac{d+2}{2}-\sum_{i<j}\delta_{ij}
    \right)}.
\end{multline}

\vskip 4pt
As explained at the end of the last section, the corresponding celestial Mellin amplitude \eqref{MAcc1} then follows by imposing the conformal constraints \eqref{confcon} on the Mellin variables:
\begin{equation} \nonumber
   {\cal A}^{{\cal V}_{1 \ldots n-1 m}}_{\Delta_1  \ldots \Delta_{n-1} \Delta_n}\left(Q_1,\ldots,Q_n\right)
   = \int^{c+i\infty}_{c-i\infty}  \frac{{\rm d}\delta_{ij}}{2\pi i} M^{{\cal V}_{1 \ldots n-1 m}}_{\Delta_1 \ldots \Delta_n}\left(\delta_{ij}\right)\prod_{i<j} \Gamma\left(\delta_{ij}\right)\left(-2\, Q_i \cdot Q_j+i\epsilon\right)^{-\delta_{ij}},
\end{equation}
with celestial Mellin amplitude \cite{Pacifico:2024dyo}:
\begin{multline}
    M^{{\cal V}_{1 \ldots n-1 m}}_{\Delta_1 \ldots \Delta_{n-1} \Delta_n}\left(\delta_{ij}\right)=g_{12 \ldots n}i^{\frac{d}{2}-1} \pi^{\frac{d+2}{2}} \left(\frac{1}{4\pi^{\frac{d+2}{2}}}i^{-\frac{d}{2}}\right)^{n-1}\, \left(\frac{1}{4\pi^{\frac{d+2}{2}}}\right)  \\ \times i^{\frac{d(n-2)}{2}-1}\left(\frac{m}{2}\right)^{\frac{2d(n-1)-4}{2}-\sum_{i}\Delta_{i}}  \Gamma\left(-\tfrac{d(n-1)}{2}+1+\tfrac{1}{2}\sum_{i}\Delta_{i}\right) \\ \times  \frac{\Gamma\left(-\tfrac{d(n-2)}{2}+1+\tfrac{1}{2}\sum_{i}\Delta_{i}-\Delta_{n}\right)\prod\limits^{n-1}_{i=1}\Gamma\left(\frac{d}{2}-\Delta_i\right)}{\Gamma\left(\frac{d+2}{2}-\tfrac{1}{2}\sum_{i}\Delta_{i}
    \right)}.
\end{multline}

\vskip 4pt
For three-point diagrams ($n=3$) the constraints \eqref{confcon} on the Mellin variables $\delta_{ij}$ can be solved completely,\footnote{Namely, 
\begin{equation}
    \delta_{12} = \frac{\Delta_1+\Delta_2-\Delta_3}{2}, \qquad \delta_{13} = \frac{\Delta_1+\Delta_3-\Delta_2}{2}, \qquad \delta_{23} = \frac{\Delta_2+\Delta_3-\Delta_1}{2}. 
\end{equation}}
so that the three-point contact diagram can be brought into the standard form
\begin{multline}
    {\cal A}^{{\cal V}_{1 2 \varphi}}_{\Delta_1  \Delta_2 \Delta_3}\left(Q_1,Q_2,Q_3\right) \\ =  \frac{C^{(m)}_{\Delta_1  \Delta_2 \Delta_3}}{\left(-2 Q_1 \cdot Q_2+i\epsilon\right)^{\frac{\Delta_1+\Delta_2-\Delta_3}{2}}\left(-2 Q_1 \cdot Q_3+i\epsilon\right)^{\frac{\Delta_3+\Delta_1-\Delta_2}{2}}\left(-2 Q_2 \cdot Q_3+i\epsilon\right)^{\frac{\Delta_2+\Delta_3-\Delta_1}{2}}},
\end{multline}
with coefficient: 
\begin{multline}\label{opemml0}
 C^{(m)}_{\Delta_1  \Delta_2 \Delta_3} = -\frac{1}{2} g_{123}\,\pi^{\frac{d+2}{2}} \left(\frac{1}{4\pi^{\frac{d+2}{2}}}\right)^{3} \left(\frac{m}{2}\right)^{2d-2-\sum_{i}\Delta_{i}}  \Gamma\left(\frac{2-2d+\sum_{i}\Delta_{i}}{2}\right) \\ \times  \frac{\Gamma\left(\frac{2-d+\Delta_1+\Delta_2-\Delta_3}{2}\right)\Gamma\left(\frac{d}{2}-\Delta_1\right)\Gamma\left(\frac{d}{2}-\Delta_2\right)}{\Gamma\left(\frac{d+2-\Delta_1-\Delta_2-\Delta_3}{2}
    \right)}\\
 \times\Gamma \left(\frac{\Delta_1 + \Delta_2 - \Delta_3}{2} \right)
  \Gamma \left(\frac{\Delta_1 - \Delta_2 + \Delta_3}{2}  \right)
  \Gamma \left(\frac{-\Delta_1 + \Delta_2 + \Delta_3}{2} \right),
\end{multline}
which was first given in \cite{Pacifico:2025emk}.

\paragraph{Tree-level exchange diagrams.}
Consider the tree-level exchange of a scalar field $\varphi$ of mass $m$ between massless scalars $\phi_i$ mediated by non-derivative cubic vertices:
\begin{equation}\label{cubicv}
    {\cal V}_{12 \varphi} = g_{12}\, \phi_1 \phi_2 \varphi, \qquad {\cal V}_{34 \varphi} = g_{34}\, \phi_3 \phi_4 \varphi.
\end{equation}
This reads 
\begin{multline}\label{exchdef}
   {\cal A}^{{\cal V}_{12 m}{\cal V}_{34 m}}\left(Y_1,Y_2,Y_3,Y_4\right) = (-i g_{12})(-i g_{34}) \int {\rm d}^{d+2}X_1 {\rm d}^{d+2}X_2\, G^{(0)}_{T}\left(X_1,Y_1\right)G^{(0)}_{T}\left(X_1,Y_2\right) \\ \times G^{(m)}_{T}\left(X_1,X_2\right)G^{(0)}_{T}\left(X_2,Y_3\right)G^{(0)}_{T}\left(X_2,Y_4\right).
\end{multline}
Employing Schwinger parameterisation \eqref{SPm} and \eqref{MLschw} of the Feynman propagators, this is given by
\begin{align}\label{exchmsbar}
   & {\cal A}^{{\cal V}_{12 m}{\cal V}_{34 m}}\left(Y_1,Y_2,Y_3,Y_4\right)
    \\  &\hspace*{1.75cm}= - g_{12}g_{34} \left(\frac{1}{4\pi^{\frac{d+2}{2}}}i^{-\frac{d}{2}}\right)^4\, \left(\frac{1}{4\pi^{\frac{d+2}{2}}}\left(\frac{m}{2}\right)^{\frac{d}{2}}\right)  \int^{+i\infty}_{-i\infty}\frac{{\rm d}{\bar s}}{2\pi i}\Gamma\left({\bar s}-\tfrac{d}{4}\right)\left(\frac{m}{2}\right)^{-2{\bar s}} i^{-({\bar s}+\frac{d}{4})} \nonumber \\ &\hspace*{6.5cm} \times \int^{\infty}_0 \frac{{\rm d}{\bar t}}{{\bar t}} {\bar t}^{{\bar s}+\frac{d}{4}} \prod^4_{i=1} \frac{{\rm d}t_i}{t_i} t^{\frac{d}{2}}_i\, E_{t_1,\,t_2\,|\,{\bar t}\,|\,t_3,\,t_4}\left(Y_1,Y_2,Y_3,Y_4\right), \nonumber
\end{align}
where ${\bar t}$ is the Schwinger parameter for the internal line and $t_i$ for the external lines. This reduces the integrals over the two internal points to two nested Gaussian integrals:
\begin{align} \nonumber
\hspace*{-0.25cm}E_{t_1,\,t_2\,|\,{\bar t}\,|\,t_3,\,t_4}\left(Y_1,Y_2,Y_3,Y_4\right) &= \int {\rm d}^{d+2}X_1 {\rm d}^{d+2}X_2\, \exp\left[it_1\left(X_1-Y_1\right)^2+i t_2\left(X_1-Y_2\right)^2\right.\\ & \nonumber \left.\hspace*{2.3cm}+i {\bar t}\left(X_1-X_2\right)^2+i t_3\left(X_2-Y_3\right)^2+i t_4\left(X_2-Y_4\right)^2\right]\\
  &= i^{d} \pi^{d+2} \left[\left(t_1+t_2\right)\left(t_3+t_4\right)+{\bar t}
\left(t_1+t_2+t_3+t_4\right)\right]^{-\frac{d+2}{2}} \nonumber \\ & \hspace*{-2.5cm}\times \exp\left[i\frac{{\bar t} t_1 t_3 \left(Y_1-Y_3\right)^2+{\bar t} t_1 t_4 \left(Y_1-Y_4\right)^2+{\bar t} t_2 t_3 \left(Y_2-Y_3\right)^2+{\bar t} t_2 t_4 \left(Y_2-Y_4\right)^2}{\left(t_1+t_2\right)\left(t_3+t_4\right)+{\bar t}
\left(t_1+t_2+t_3+t_4\right)}\right. \nonumber \\ & \hspace*{-1cm} \left. +i\frac{t_1t_2\left({\bar t}+t_1+t_2\right)\left(Y_1-Y_2\right)^2+t_3t_4\left({\bar t}+t_3+t_4\right)\left(Y_3-Y_4\right)^2}{\left(t_1+t_2\right)\left(t_3+t_4\right)+{\bar t}
\left(t_1+t_2+t_3+t_4\right)}\right]. \label{gaussexchint}
\end{align}
As for the contact diagram example above, the Mellin representation \eqref{tcMA} for the exchange diagram is obtained from the Mellin transform \eqref{MTexp} of the exponential function, so that the Gaussian integrals \eqref{gaussexchint} take the form
\begin{multline}\label{schwintexch}
    E_{t_1,\,t_2\,|\,{\bar t}\,|\,t_3,\,t_4}\left(Y_1,Y_2,Y_3,Y_4\right)=i^{d} \pi^{d+2}  \int^{i\infty}_{-i\infty}   \frac{{\rm d}\delta_{ij}}{2\pi i}\, \prod_{i<j} \left( t_i t_j\right)^{-\delta_{ij}}\Gamma\left(\delta_{ij}\right)\left[-i\left(Y_i-Y_j\right)^2+\epsilon\right]^{-\delta_{ij}} \\ \hspace*{-0.25cm} \times {\bar t}^{-\left(\delta_{13}+\delta_{14}+\delta_{23}+\delta_{24}\right)}\left({\bar t}+t_1+t_2\right)^{-\delta_{12}}\left({\bar t}+t_3+t_4\right)^{-\delta_{34}} \\ \times \left[\left(t_1+t_2\right)\left(t_3+t_4\right)+{\bar t}
\left(t_1+t_2+t_3+t_4\right)\right]^{-\frac{d+2}{2}+\sum_{i<j}\delta_{ij}}.
\end{multline}
The integrals over the Schwinger parameters can be evaluated using the approach given \cite{Pacifico:2024dyo}, which we review in appendix \ref{app::schwint}. This generates a generalised ${}_3F_2$ hypergeometric function, and a Dirac delta function that eliminates the integral over ${\bar s}$ coming from the massive Feynman propagator in \eqref{exchmsbar}. This gives the exchange diagram \eqref{exchdef} in the form:
\begin{align}\label{MAbulkc}
 \hspace*{-0.5cm}  {\cal A}^{{\cal V}_{12m}{\cal V}_{34 m}}\left(Y_1,Y_2,Y_3,Y_4\right)
   = \int^{i\infty}_{-i\infty}   \frac{{\rm d}\delta_{ij}}{2\pi i}\,  M^{{\cal V}_{12 m}{\cal V}_{34 m}}\left(\delta_{ij}\right)\prod_{i<j}\Gamma\left(\delta_{ij}\right)\left[\left(Y_i-Y_j\right)^2+i \epsilon\right]^{-\delta_{ij}},
\end{align}
with Mellin amplitude
\begin{multline}\label{Mbulkexch}
    M^{{\cal V}_{12 m}{\cal V}_{34 m}}\left(\delta_{ij}\right) = \frac{1}{4} g_{12}g_{34} \pi^{\frac{d+2}{2}} \prod^4_{i=1}\frac{\Gamma\left(\frac{d}{2}-\sum\limits_{j\ne i}\delta_{ij}\right)}{4\pi^{\frac{d+2}{2}}}\\ \times \,  \Gamma\left(\frac{4-3d+2\sum_{i<j}\delta_{ij}}{2}\right)\left(\frac{m}{2}\right)^{3d-4-2\sum_{i<j}\delta_{ij}} 
    \\ \times  \frac{\Gamma\left(\delta_{12}+\frac{2-d}{2}\right)\Gamma\left(\delta_{34}+\frac{2-d}{2}\right)}{\Gamma\left(\delta_{12}+\frac{d+2}{2}-\sum_{i<j}\delta_{ij}\right)\Gamma\left(\delta_{34}+\frac{d+2}{2}-\sum_{i<j}\delta_{ij}\right)}\\ \times  {}_3F_2\left(\begin{matrix}\delta_{12}+\frac{2-d}{2},\delta_{34}+\frac{2-d}{2},\frac{d+2}{2}-\sum_{i<j}\delta_{ij}\\\delta_{12}+\frac{d+2}{2}-\sum_{i<j}\delta_{ij},\delta_{34}+\frac{d+2}{2}-\sum_{i<j}\delta_{ij} \end{matrix};1\right).
\end{multline}

\vskip 4pt
Like for the contact diagram above, the Mellin amplitude \eqref{MR} for the corresponding celestial correlator follows naturally from the Mellin amplitude \eqref{Mbulkexch} for the corresponding bulk correlator upon extrapolating the external points to the celestial sphere, which imposes the conformal constraints \eqref{confcon} on the Mellin variables. This gives \cite{Pacifico:2024dyo}:
 \begin{multline}\label{MAexchdij}
    M^{{\cal V}_{12 m}{\cal V}_{34 m}}_{\Delta_1 \Delta_2\Delta_3\Delta_4}\left(\delta_{ij}\right)=\frac{1}{4}g_{12}g_{34}\,\pi^{\frac{d+2}{2}}\,\prod^4_{i=1}\left(\frac{\Gamma\left(\tfrac{d}{2}-\Delta_i\right)}{4\pi^{\frac{d+2}{2}}}\right)\left(\frac{m}{2}\right)^{3d-4-\sum_i\Delta_i}\Gamma\left(\frac{-3d+4+\sum_i\Delta_i}{2}\right)\\ \hspace*{-1.25cm} \times \frac{\Gamma\left(\delta_{12}+\frac{2-d}{2}\right)\Gamma\left(\delta_{34}+\frac{2-d}{2}\right)}{\Gamma\left(\delta_{12}-\frac{\beta}{2}\right)\Gamma\left(\delta_{34}-\frac{\beta}{2}\right)}
    {}_3F_2\left(\begin{matrix}\delta_{12}+\frac{2-d}{2},\delta_{34}+\frac{2-d}{2},-\frac{\beta}{2}\\\delta_{12}-\frac{\beta}{2},\delta_{34}-\frac{\beta}{2} \end{matrix};1\right),
\end{multline}  
where we defined $\beta=-(d+2)+\sum_i \Delta_i$. In terms of the Mellin ``Mandelstam variables" \eqref{MBmandel} this reads,
\begin{multline}\label{scTexch}
    M^{{\cal V}_{12 m}{\cal V}_{34 m}}_{\Delta_1 \Delta_2\Delta_3\Delta_4}\left(s_{12}, s_{13}\right)=\frac{1}{8}g_{12}g_{34}\,\pi^{\frac{d+2}{2}}\,\prod^4_{i=1}\left(\frac{\Gamma\left(\tfrac{d}{2}-\Delta_i\right)}{4\pi^{\frac{d+2}{2}}}\right)\\ \; \times \left(\frac{m}{2}\right)^{3d-4-\sum_i\Delta_i}\Gamma\left(\frac{-3d+4+\sum_i\Delta_i}{2}\right) T_{e}\left(d-\Delta_1-\Delta_2,d-\Delta_3-\Delta_4,\frac{\beta}{2},0,0\right),
\end{multline}
where for later purposes we introduced the compact notation:
\begin{multline}\label{Tscexch}
    T_e\left(a_1,a_2,b,b_{12},b_{34}\right) =  \frac{\Gamma\left(\tfrac{\Delta_3+\Delta_4-s_{12}}{2}-b-b_{12}-a_1\right)\Gamma\left(\tfrac{\Delta_1+\Delta_2-s_{12}}{2}-b-b_{34}-a_2\right)}{\Gamma\left(\tfrac{\Delta_3+\Delta_4-s_{12}}{2}-b-b_{12}\right)\Gamma\left(\tfrac{\Delta_1+\Delta_2-s_{12}}{2}-b-b_{34}\right)}\\ \times  {}_3F_2\left(\begin{matrix}\tfrac{\Delta_1+\Delta_2-s_{12}}{2}-b-b_{34}-a_2,\tfrac{\Delta_3+\Delta_4-s_{12}}{2}-b-b_{12}-a_1,-b\\\tfrac{\Delta_1+\Delta_2-s_{12}}{2}-b-b_{34},\tfrac{\Delta_3+\Delta_4-s_{12}}{2}-b-b_{12} \end{matrix};1\right).
\end{multline}

\vskip 4pt
\paragraph{Massless exchange.} When the exchanged particle is also massless there are further simplifications owing to the power law nature of the  corresponding Feynman propagator. At the level of the Mellin representation \eqref{MAbulkc} this results in an additional constraint among the Mellin variables, which can be seen from the massless limit:
\begin{equation}\label{mlconstrd}
    \lim_{m\to0}\Gamma\left(\frac{4-3d+2\sum_{i<j}\delta_{ij}}{2}\right)\left(\frac{m}{2}\right)^{3d-4-2\sum_{i<j}\delta_{ij}} = 2\pi i\, \delta \left(\frac{4-3d+2\sum_{i<j}\delta_{ij}}{2}\right),
\end{equation}
so that 
\begin{multline}\label{Mlbulkexch}
    M^{{\cal V}_{12 m=0}{\cal V}_{34 m=0}}\left(\delta_{ij}\right) = \frac{1}{4} g_{12}g_{34} \pi^{\frac{d+2}{2}} \prod^4_{i=1}\left(\frac{\Gamma\left(\frac{d}{2}-\sum\limits_{j\ne i}\delta_{ij}\right)}{4\pi^{\frac{d+2}{2}}} \right)2\pi i\, \delta \left(\frac{4-3d+2\sum_{i<j}\delta_{ij}}{2}\right)
    \\ \times  \frac{\Gamma\left(\delta_{12}+\frac{2-d}{2}\right)\Gamma\left(\delta_{34}+\frac{2-d}{2}\right)}{\Gamma\left(\delta_{12}+\frac{d+2}{2}-\sum_{i<j}\delta_{ij}\right)\Gamma\left(\delta_{34}+\frac{d+2}{2}-\sum_{i<j}\delta_{ij}\right)}\\ \times  {}_3F_2\left(\begin{matrix}\delta_{12}+\frac{2-d}{2},\delta_{34}+\frac{2-d}{2},\frac{d+2}{2}-\sum_{i<j}\delta_{ij}\\\delta_{12}+\frac{d+2}{2}-\sum_{i<j}\delta_{ij},\delta_{34}+\frac{d+2}{2}-\sum_{i<j}\delta_{ij} \end{matrix};1\right).
\end{multline}
The massless constraint \eqref{mlconstrd} on the Mellin variables simplifies the generalised hypergeometric function in the Mellin amplitude \eqref{Mbulkexch}, where for integer $d \ge 3$ we can use Saalsch\"utz theorem:
\begin{equation}\label{saal}
    { }_3 F_2\left(\begin{array}{c}
a, b,3-d \\
c, 1+a+b-c+3-d
\end{array} ; 1\right)=\frac{(c-a)_{d-3}(c-b)_{d-3}}{(c)_{d-3}(c-a-b)_{d-3}}. 
\end{equation}
This gives (for integer $d \ge 3$):
\begin{multline}\label{Mlbulkexchsimp}
    M^{{\cal V}_{12 m=0}{\cal V}_{34 m=0}}\left(\delta_{ij}\right) =\frac{1}{4}  g_{12}g_{34} \pi^{\frac{d+2}{2}} \prod^4_{i=1}\left(\frac{\Gamma\left(\frac{d}{2}-\sum\limits_{j\ne i}\delta_{ij}\right)}{4\pi^{\frac{d+2}{2}}}\right)  2\pi i\, \delta \left(\frac{4-3d+2\sum_{i<j}\delta_{ij}}{2}\right)
   \\ \times  \frac{\Gamma \left(\frac{d}{2}-1\right) \Gamma (1-\delta_{34}) \Gamma \left(-\frac{d}{2}+\delta_{12}+1\right) \Gamma \left(-\frac{d}{2}+\delta_{34}+1\right) \Gamma \left(\frac{d}{2}+\delta_{12}-\delta_{34}-1\right)}{\Gamma \left(2-\frac{d}{2}\right) \Gamma (\delta_{12}) \Gamma (d-\delta_{34}-2) \Gamma (-d+\delta_{34}+3) \Gamma \left(-\frac{d}{2}+\delta_{12}-\delta_{34}+2\right)}.
\end{multline}

Taking the boundary limit \eqref{ccdefn2} to obtain the corresponding celestial correlator, the constraint \eqref{constconf} from conformal covariance in turn leads to a constraint on the external scaling dimensions in the corresponding celestial correlator:
\begin{equation}\label{cnstrDeltascexch}
    \sum_{i}\Delta_i = 3d-4.
\end{equation}
Using Saalsch\"utz theorem \eqref{saal}, the celestial Mellin amplitude \eqref{MAexchdij} (for integer $d \ge 3$) reads \cite{Pacifico:2024dyo}:
\begin{multline}\label{mldge3}
    M^{{\cal V}_{12 m=0}{\cal V}_{34 m=0}}_{\Delta_1 \Delta_2\Delta_3\Delta_4}\left(s_{12},s_{13}\right)=\frac{1}{8}g_{12}g_{34}\,\pi^{\frac{d+2}{2}}\,\prod^4_{i=1}\frac{\Gamma\left(\tfrac{d}{2}-\Delta_i\right)}{4\pi^{\frac{d+2}{2}}}\,\\ \times (-1)^{3-d} \frac{\Gamma\left(\frac{d}{2}-1\right)}{\Gamma\left(2-\frac{d}{2}\right)} \frac{\Gamma\left(2d-3-\Delta_1-\Delta_2\right)}{\Gamma\left(d-\Delta_1-\Delta_2\right)} \\ \times 2\pi i\, \delta\left(\frac{-3d+4+\sum_i\Delta_i}{2}\right) \frac{\Gamma \left(\frac{2-d+\Delta_3+\Delta_4-s_{12}}{2}\right) \Gamma \left(\frac{2-d+\Delta_1+\Delta_2-s_{12}}{2}\right)}{\Gamma\left(\frac{\Delta_3+\Delta_4-s_{12}}{2}\right)\Gamma\left(\frac{\Delta_1+\Delta_2-s_{12}}{2}\right)},
\end{multline}
which is vanishing for even values of $d >3$. This is proportional to the Mellin amplitude for a conformal partial wave corresponding to the exchange of two (shadow) scalar operators with scaling dimensions:
\begin{align}
    \Delta^{(0)}_{12}=\Delta_1+\Delta_2+(2-d), \qquad 
   \Delta^{(0)}_{34}=\Delta_3+\Delta_4+(2-d),
\end{align}
which are shadow of one another by virtue of the Dirac delta constraint \eqref{cnstrDeltascexch} on the scaling dimensions $\Delta_i$. More explicitly we have:
 \begin{multline}\label{spin0MACPW}
    M^{{\cal V}_{12 m=0}{\cal V}_{34 m=0}}_{\Delta_1 \Delta_2\Delta_3\Delta_4}\left(s_{12},s_{13}\right) = 
     2\pi i\, \delta\left(\frac{-3d + 4 + \sum_i \Delta_i}{2} \right)a^{\prime\,(m=0)}_{\Delta^{(0)}_{12}, 0} \\ \times \mathcal{F}_{\Delta_1 + \Delta_2 + (2-d), 0}(s_{12}, s_{13}),
\end{multline}
with
\begin{multline}
    a^{\prime\,(m=0)}_{\Delta^{(0)}_{12}, 0} =  -\frac{1}{8}g_{12}g_{34} \pi^{\frac{d+2}{2}}  \prod_i \frac{\Gamma\left(\frac{d}{2}- \Delta_i\right)}{4\pi^{\frac{d+2}{2}}}\\ \nonumber 
        \times\frac{\Gamma \left(\frac{d}{2}-1\right) \Gamma \left(\frac{-\Delta_1-\Delta_2+\Delta_3+\Delta_4}{2}\right)\Gamma \left(-\frac{d}{2}+\Delta_1+1\right) \Gamma \left(-\frac{d}{2}+\Delta_2+1\right)}{\Gamma\left(2-\frac{d}{2}\right)\Gamma (d-\Delta_1-\Delta_2) \Gamma (2-d+\Delta_1+\Delta_2)}\\ \nonumber 
        \times\Gamma \left(\frac{2-d+\Delta_1+\Delta_2+\Delta_3-\Delta_4}{2} \right) \Gamma \left(\frac{2-d+\Delta_1+\Delta_2-\Delta_3+\Delta_4}{2} \right)\\
        \times\Gamma \left(\frac{d-2 -\Delta_1-\Delta_2+\Delta_3+\Delta_4}{2}\right). \nonumber 
\end{multline}
The Mellin amplitude representation for conformal partial waves ${\cal F}_{\Delta,0}$ is reviewed in appendix \ref{app::CPW}.

\vskip 4pt
 In later sections we confirm that this feature of massless exchanges extends to massless particles with arbitrary integer spin.

\subsection{Extracting the OPE data} 
\label{subsec::extractingOPE}

\begin{figure}[t]
    \centering
    \includegraphics[width=0.9\textwidth]{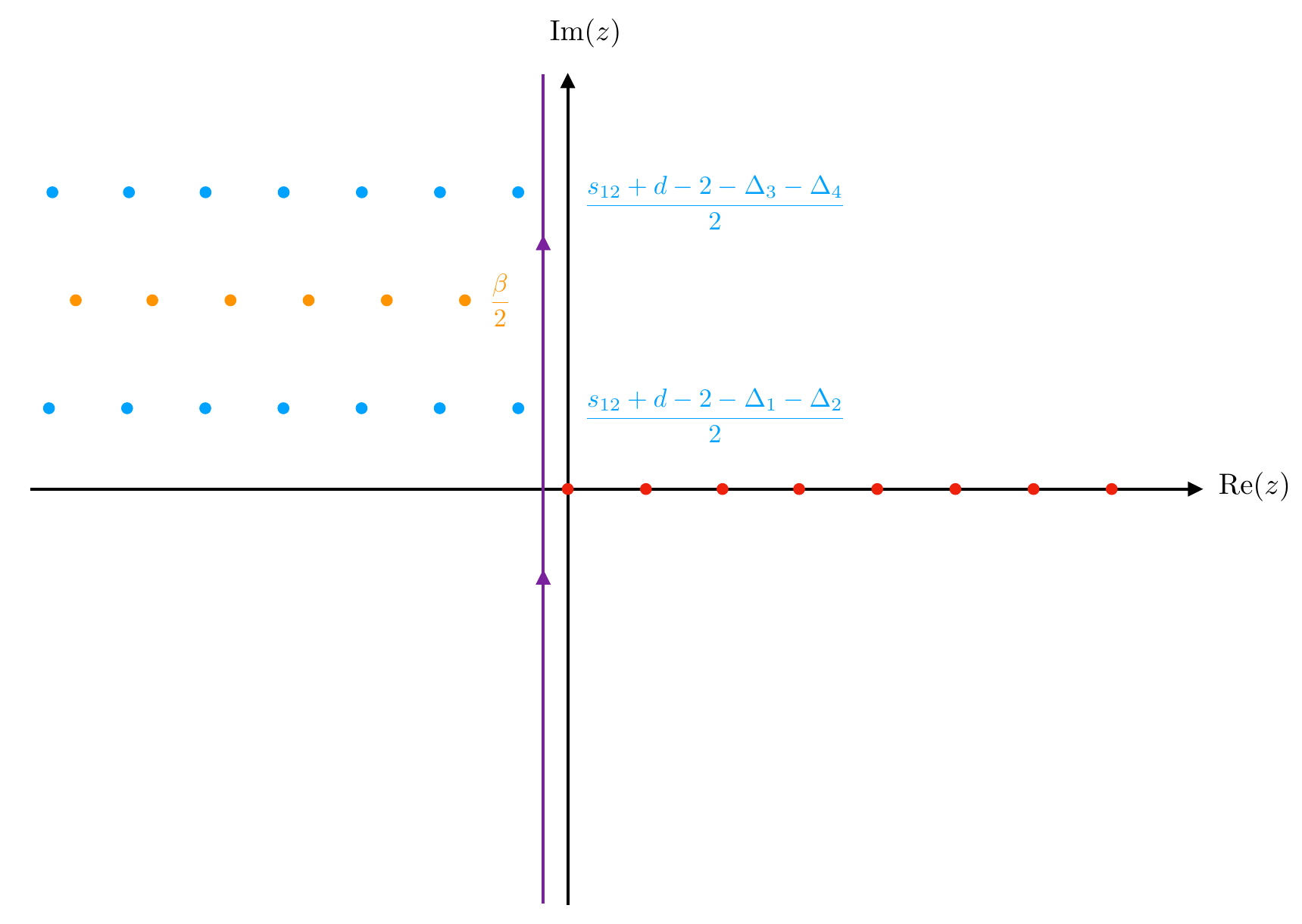}
    \caption{Poles in $z$ complex plane and integration contour (purple line) used in the Mellin-Barnes integral \eqref{Mexchs12s13_scalar}. The poles \eqref{s_12_poles_scalar1} in $s_{12}$ arise from the pinching of the integration contour between red and blue poles of $z$ as shown in the figure: as $s_{12}$ varies, the blue poles move and collide with red poles. For simplicity of the presentation, in the figure we took the scaling dimensions to lie on the Principal Series $\Delta_i = \frac{d}{2}+i\nu_i$, $\nu_i \in \mathbb{R}$.}
    \label{fig::pole_pinching_scalar}
\end{figure}

In this section we illustrate how to extract OPE data from the Mellin amplitude \eqref{MAexchdij} for exchange diagram contributions to celestial correlators \eqref{ccdefn}. As reviewed at the beginning of section \ref{sec::CMA}, the contributions to the conformal block expansion in the s-channel are encoded by the poles in the Mellin variable $s_{12}$. To determine the latter it is useful to employ the Mellin-Barnes representation of the generalised hypergeometric function:
\begin{multline}\label{MB3F2}
    {}_3F_2\left(\begin{matrix}a,b,c\\e,f \end{matrix};1\right) = \frac{\Gamma\left(e\right)\Gamma\left(f\right)}{\Gamma\left(a\right)\Gamma\left(b\right)\Gamma\left(c\right)} \int^{+i\infty}_{-i\infty} \frac{{\rm d}z}{2\pi i}\frac{\Gamma\left(z+a\right)\Gamma\left(z+b\right)\Gamma\left(z+c\right)}{\Gamma\left(z+e\right)\Gamma\left(z+f\right)}\\ \times \Gamma\left(-z\right)\left(-1\right)^z.
\end{multline}
This Mellin representation in fact naturally arises from the integration over the Schwinger parameters appearing in the exchange diagram \eqref{schwintexch}, which we review in appendix \ref{app::schwint}. As a Mellin-Barnes integral, the exchange Mellin amplitude \eqref{MAexchdij} reads \cite{Pacifico:2024dyo}:
\begin{multline}\label{Mexchs12s13_scalar}
    M^{{\cal V}_{12 m}{\cal V}_{34 m}}_{\Delta_1 \Delta_2\Delta_3\Delta_4}\left(s_{12},s_{13}\right)=\frac{1}{8}g_{12}g_{34}\,\pi^{\frac{d+2}{2}}\,\prod^4_{i=1}\frac{\Gamma\left(\tfrac{d}{2}-\Delta_i\right)}{4\pi^{\frac{d+2}{2}}}\,\\ \times \left(\frac{m}{2}\right)^{3d-4-\sum_i\Delta_i}\Gamma\left(\frac{-3d+4+\sum_i\Delta_i}{2}\right)\\ \times \frac{1}{\Gamma\left(-\frac{\beta}{2}\right)}\int^{+i\infty}_{-i\infty}\frac{{\rm d}z}{2\pi i}\frac{\Gamma\left(z+\tfrac{1}{2}(2-d+\Delta_1+\Delta_2-s_{12})\right)\Gamma\left(z+\tfrac{1}{2}(2-d+\Delta_3+\Delta_4-s_{12})\right)}{\Gamma\left(z+\tfrac{1}{2}(d+2-\Delta_3-\Delta_4-s_{12})\right)\Gamma\left(z+\tfrac{1}{2}(d+2-\Delta_1-\Delta_2-s_{12})\right)}\\ \times \Gamma\left(z-\tfrac{\beta}{2}\right)\Gamma\left(-z\right)\left(-1\right)^z.
\end{multline}
The poles in $z$ are located at ($p, q \in \mathbb{N}$):
\begin{subequations}\label{0z_poles}
    \begin{alignat}{1}
        &z=p, \label{0z_n_poles}\\
        &z=\frac{d-2+s_{12}-\Delta_1-\Delta_2}{2}-q,\label{0z_Delta_12_poles}\\
        &z=\frac{d-2+s_{12}-\Delta_3-\Delta_4}{2}-q,\label{0z_Delta_34_poles}\\
        &z=\frac{\beta}{2}-q.\label{0z_betapoles}
    \end{alignat}
\end{subequations}
The poles in $s_{12}$ arise from the pinching of the poles \eqref{0z_Delta_12_poles} and \eqref{0z_Delta_34_poles} with the poles \eqref{0z_n_poles}. This occurs for (see figure \ref{fig::pole_pinching_scalar}):\footnote{Note that the poles in $\beta$ arising from the pinching of \eqref{0z_betapoles} and \eqref{0z_n_poles} are canceled by the $1/\Gamma\left(-\frac{\beta}{2}\right)$ factor in \eqref{Mexchs12s13_scalar}.}
\begin{equation}\label{s_12_poles_scalar1}
    \qquad s_{12} = \Delta_1 + \Delta_2 +2-d+2(p+q), \qquad s_{12} = \Delta_3 + \Delta_4 +2- d + 2(p+q).
\end{equation}
Comparing with the form of the s-channel conformal block expansion \eqref{MAcbe} in Mellin space, these correspond to two infinite families of scalar operators with scaling dimensions \cite{Pacifico:2024dyo}: 
\begin{subequations}\label{scfam}
 \begin{align}
    \Delta^{(n)}_{12}&=\Delta_1+\Delta_2+(2-d)+2n_{12}, \qquad n_{12}=0,1,2,\ldots,\\
   \Delta^{(n)}_{34}&=\Delta_3+\Delta_4+(2-d)+2n_{34}, \qquad n_{34}=0,1,2,\ldots\,,
\end{align}   
\end{subequations}
which encode the exchange of a massive scalar field in $\mathbb{M}^{d+2}$. Note that operators with $n_{12}, n_{34}>0$ mix with the descendants of operators with lower values of $n_{12}, n_{34}$.\footnote{Note that this is the same type of mixing as the one among the double-trace operators \eqref{DT}.}  

\vskip 4pt
To evaluate the residues of the poles \eqref{s_12_poles_scalar1} in $s_{12}$  we close the $z$-integration contour to the right, which encloses the poles at $z\in\mathbb{N}$ along the positive real axis. This gives: 
{\allowdisplaybreaks\begin{multline}
    M^{{\cal V}_{12 m}{\cal V}_{34 m}}_{\Delta_1 \Delta_2\Delta_3\Delta_4}\left(s_{12},s_{13}\right)=\frac{1}{4}g_{12}g_{34}\,\pi^{\frac{d+2}{2}}\,\prod^4_{i=1}\frac{\Gamma\left(\tfrac{d}{2}-\Delta_i\right)}{4\pi^{\frac{d+2}{2}}}\\
\times\left(\frac{m}{2}\right)^{3d-4-\sum_i\Delta_i}\Gamma\left(\frac{-3d+4+\sum_i\Delta_i}{2}\right)\\
\times \left[\sum_{p=0}^{+\infty}\frac{(-1)^p}{p!}\left(-\frac{\beta}{2}\right)_p\sum_{q=0}^{+\infty}\frac{(-1)^q}{q!}\frac{\Gamma\left(-q-\frac{\Delta_1+\Delta_2-\Delta_3-\Delta_4}{2}\right)}{\Gamma\left(d-q-\Delta_1-\Delta_2\right)\Gamma\left(d-q-\frac{\Delta_1+\Delta_2+\Delta_3+\Delta_4}{2}\right)}\right. \\  \left.\times \frac{1}{s_{12}-(\Delta_1+\Delta_2-d+2+2(p+q))} + \left(\Delta_{1,2} \leftrightarrow \Delta_{3,4} \right)\right].
\end{multline}}Since there is no mixing for the leading two operators $(n_{12}=n_{34}=0)$ in the family \eqref{scfam}, their conformal block coefficients \eqref{MAcbe} can be immediately read off from the residues of the poles for $p=q=0$:
\begin{multline}\label{scalar_OPE_coeff}
 a^{(m)}_{\Delta^{(0)}_{12},0}  = -\frac{1}{8}g_{12}g_{34} \pi^{\frac{d+2}{2}} \prod_i \frac{\Gamma\left(\frac{d}{2}- \Delta_i\right)}{4\pi^{\frac{d+2}{2}}}\\ \times  \left(\frac{m}{2}\right)^{3d-4-\sum_i\Delta_i}\Gamma\left(\frac{-3d+4+\sum_i\Delta_i}{2}\right)\\
        \times\frac{\Gamma \left(\frac{d}{2}-1\right) \Gamma \left(\frac{-\Delta_1-\Delta_2+\Delta_3+\Delta_4}{2} \right)\Gamma \left(-\frac{d}{2}+\Delta_1+1\right) \Gamma \left(-\frac{d}{2}+\Delta_2+1\right)}{\Gamma (d-\Delta_1-\Delta_2) \Gamma (2-d+\Delta_1+\Delta_2) \Gamma \left(\frac{2 d-\Delta_1-\Delta_2-\Delta_3-\Delta_4}{2}\right)}\\
        \times\Gamma \left(\frac{2-d+\Delta_1+\Delta_2+\Delta_3-\Delta_4}{2} \right) \Gamma \left(\frac{2-d+\Delta_1+\Delta_2-\Delta_3+\Delta_4}{2} \right)\\
        \times\Gamma \left(\frac{d-2 -\Delta_1-\Delta_2+\Delta_3+\Delta_4}{2}\right),
\end{multline}
and similar for $a^{(m)}_{\Delta^{(0)}_{34},0}$. In the massless limit we have \eqref{mlconstrd}:
\begin{subequations}\label{mlcbcoeff}
 \begin{align}
     a^{(0)}_{\Delta^{(0)}_{12},0} & = \lim_{m \to 0}  a^{(m)}_{\Delta^{(0)}_{12},0}  \\
     &= 2\pi i \,\delta\left(\frac{-3d+4+\sum_i\Delta_i}{2}\right) a^{\prime\,(m=0)}_{\Delta^{(0)}_{12}, 0},
\end{align}   
\end{subequations}
where $a^{\prime\,(m=0)}_{\Delta^{(0)}_{12}, 0}$ is the coefficient of the CPW \eqref{spin0MACPW} in the massless exchange, as expected.

\vskip 4pt
To extract the OPE coefficients of operators \eqref{scfam} with $n_{12}, n_{34} >0 $, one proceeds iteratively due to the mixing with the descendants of operators with lower values of $n_{12}, n_{34}$. In other words, the residue of the poles \eqref{s_12_poles_scalar1} for $p+q \ne 0$ contains contributions from descendants of operators appearing at lower values of $p+q$, which should be subtracted. This procedure can be implemented efficiently through the use of Casimir operators to project away contributions from operators below a given twist \cite{Sleight:2018epi}. 

\vskip 4pt
Note that the conformal block expansion of the celestial correlator will also contain contributions from double-trace operators \eqref{DT}, as for Witten diagrams in AdS. In the Mellin representation \eqref{MBst}, these are encoded by poles in the measure \eqref{MMeasure} at
\begin{equation}\label{DTpoles}
    s_{12} = \Delta_1+\Delta_2+2n, \qquad s_{12} = \Delta_3+\Delta_4+2n, \qquad n\in \mathbb{N}.
\end{equation}
While in this work we focus on poles of the type \eqref{s_12_poles_scalar1} encoding the exchanged single-particle state in Minkowski space, the contributions of double-trace operators can be extracted in the same way.

\subsection{OPE in Celestial CFT} 
\label{subsec::OPEinCCFT}

In celestial CFT one expects the OPE expansion of scalar operators $O_i$ and $O_j$ takes the form 
\begin{equation}
    O_{i}O_j \sim \int^{c+i\infty}_{c-i\infty} \frac{{\rm d}\Delta}{2\pi i}\, c_{ij\Delta} \left(O_\Delta + \text{descendants}\right),
\end{equation}
decomposing into unitary representations of the Lorentz group, which includes the Principal series representations $\Delta = \frac{d}{2}+i\nu$, $\nu \in \mathbb{R}$.\footnote{In the context of celestial amplitudes, see \cite{Pasterski:2017kqt, Guevara:2021tvr}.}${}^{,}$\footnote{The choice of $c \in \mathbb{R}$ accommodates, via analytic continuation, possible contributions from discrete UIRs of $SO(d+1,1)$ \cite{Dobrev:1977qv}.} This in turn gives rise to a conformal block expansion of celestial four-point functions of the form:
\begin{equation}
    \langle O_{1}\left(Q_1\right)O_{2}\left(Q_2\right)O_{3}\left(Q_3\right)O_{4}\left(Q_4\right) \rangle = \int^{c+i\infty}_{c-i\infty} \frac{{\rm d}\Delta}{2\pi i} c_{12\Delta}c_{34\Delta}\, {\cal G}^{12,34}_{\Delta}\left(Q_1,Q_2,Q_3,Q_4\right),
\end{equation}
where ${\cal G}^{12,34}_{\Delta}$ is the direct channel conformal block for an operator of scaling dimension $\Delta$. The OPE coefficients in (perturbative) celestial CFT are meromorphic functions of $\Delta$ (see \cite{Pacifico:2025emk}) and an expansion as a discrete sum of conformal blocks can be obtained by closing the contour to the right, where the conformal blocks are exponentially suppressed as $\Delta \to \infty $.\footnote{Note that these same steps give the conformal block expansion of a conformal four-point function from its conformal partial wave expansion.} 

\vskip 4pt
We can test this expectation by comparing the conformal block coefficient \eqref{scalar_OPE_coeff} extracted from the four-point exchange \eqref{exchdef} above with what one would obtain from the OPE coefficients induced by the corresponding three-point contact subdiagram. The latter was given in equation \eqref{opemml0}, which for convenience we repeat below:
\begin{multline}\label{opemml}
 C^{(m)}_{\Delta_1  \Delta_2 \Delta} = - \frac{1}{2}g_{123}\, \pi^{\frac{d+2}{2}} \left(\frac{1}{4\pi^{\frac{d+2}{2}}}\right)^{3} \left(\frac{m}{2}\right)^{2d-2-\Delta_1-\Delta_2-\Delta}  \Gamma\left(\frac{2-2d+\Delta_1+\Delta_2+\Delta}{2}\right) \\ \times  \frac{\Gamma\left(\frac{2-d+\Delta_1+\Delta_2-\Delta}{2}\right)\Gamma\left(\frac{d}{2}-\Delta_1\right)\Gamma\left(\frac{d}{2}-\Delta_2\right)}{\Gamma\left(\frac{d+2-\Delta_1-\Delta_2-\Delta}{2}
    \right)}\\
 \times\Gamma \left(\frac{\Delta_1 + \Delta_2 - \Delta}{2} \right)
  \Gamma \left(\frac{\Delta_1 - \Delta_2 + \Delta}{2} \right)
  \Gamma \left(\frac{-\Delta_1 + \Delta_2 + \Delta}{2} \right),
\end{multline}
and the free theory two-point function reads \cite{Sleight:2023ojm}:
\begin{subequations}
 \begin{align}
\langle O_{\Delta_1}\left(Q_1\right)O_{\Delta_2}\left(Q_2\right) \rangle &=\frac{C^{(m)}_{\Delta_1}}{\left(-2 Q_1 \cdot Q_2+i\epsilon\right)^{\Delta_1}}  \,2\pi i \,\delta\left(\Delta_1-\Delta_2\right),\\
   C^{(m)}_{\Delta}&= \frac{1}{4\pi^{\frac{d+2}{2}}} \left(\frac{m}{2}\right)^{d-2\Delta} \Gamma\left(\Delta\right)\Gamma\left(\Delta - \frac{d}{2}\right).
\end{align}   
\end{subequations}
The three-point coefficients \eqref{opemml} have poles to the right of the contour at ($n=0,1,2,\ldots$):
\begin{subequations}\label{polesC3sc}
 \begin{align}
    \Delta &= \Delta_1+\Delta_2+2n, \hspace*{1.75cm} \quad \Delta = \Delta_3+\Delta_4+2n, \label{DTpolesC3sc}\\
    \Delta &= \Delta_1+\Delta_2+2-d+2n, \quad \quad \Delta = \Delta_3+\Delta_4+2-d+2n,
\end{align}   
\end{subequations}
where the first line correspond to the double-trace operators \eqref{DT} and the second the operators \eqref{scfam} encoding the single particle exchange of a massive scalar in Minkowski space. The conformal block coefficient \eqref{scalar_OPE_coeff} is indeed recovered from the following residue at $\Delta=\Delta^{(0)}_{12}=\Delta_1+\Delta_2+2-d$:
\begin{equation}
     a^{(m)}_{\Delta^{(0)}_{12},0} = \text{Res}_{\Delta=\Delta^{(0)}_{12}}\left[\frac{C^{(m)}_{\Delta_1\Delta_2\Delta}C^{(m)}_{\Delta\Delta_3\Delta_4}}{C^{(m)}_{\Delta}}\right],
\end{equation}
and likewise for the residue of the pole at $\Delta=\Delta^{(0)}_{34}=\Delta_3+\Delta_4+2-d$, providing an independent check of the OPE data extracted directly from the celestial Mellin amplitude \eqref{Mexchs12s13_scalar} for the exchange diagram. This is to be contrasted with the relation \eqref{discbope} between OPE and conformal block coefficients when the OPE expansion takes the standard discrete form \eqref{ope}---as in AdS/CFT.

\vskip 4pt
It is instructive to consider the case that the exchanged scalar is also massless $(m=0)$, where the scaling dimensions are subject to constraints. From the three- and two-point coefficients we have \cite{Pacifico:2025emk}
\begin{align} \nonumber
 C^{(m = 0)}_{\Delta_i  \Delta_j \Delta}:& \quad \lim_{m \to 0} \left(\frac{m}{2}\right)^{2d-2-\Delta_i-\Delta_j-\Delta}  \Gamma\left(\tfrac{2-2d+\Delta_i+\Delta_j+\Delta}{2}\right) =   2 \pi i\, \delta\left(\tfrac{2-2d+\Delta_i+\Delta_j+\Delta}{2}\right),  \\ \label{ml2pt} C^{(m = 0)}_{\Delta}:& \hspace*{3.4cm} \lim_{m \to 0} \left(\frac{m}{2}\right)^{d-2\Delta} \Gamma\left(\Delta - \tfrac{d}{2}\right) =  2\pi i\, \delta\left(\Delta-\tfrac{d}{2}\right).
\end{align}
The latter in particular implies that for massless scalars in Minkowski space the free theory two-point function of scalar operators on the celestial sphere only has support for operators with scaling dimension $\Delta=\frac{d}{2}$. In this case the massless conformal block coefficient \eqref{mlcbcoeff} is recovered via 
\begin{equation}
    a^{\prime\,(m=0)}_{\Delta^{(0)}_{12},0} = \text{Res}_{\Delta=\Delta^{(0)}_{12}}\left[\frac{C^{\,\prime\,(m=0)}_{\Delta_1\Delta_2\Delta}C^{\,\prime\,(m=0)}_{\Delta\Delta_3\Delta_4}}{C^{\,\prime\,(m=0)}_{\Delta}}\right],
\end{equation}
where the ${}^\prime$ denotes coefficients with the delta function constraint stripped off:
\begin{subequations}
 \begin{align}\label{4ptcbconstr}
   a^{(m=0)}_{\Delta^{(0)}_{12},0} &= 2\pi i \,\delta\left(\frac{-3d+4+\sum_i\Delta_i}{2}\right)\, a^{\prime\,(m=0)}_{\Delta^{(0)}_{12},0},\\
   C^{(m=0)}_{\Delta_i\Delta_j\Delta}&= 2 \pi i\, \delta\left(\tfrac{2-2d+\Delta_i+\Delta_j+\Delta}{2}\right) C^{\,\prime\,(m=0)}_{\Delta_i\Delta_j\Delta},\\ 
   C^{(m=0)}_{\Delta}&= 2\pi i\, \delta\left(\Delta-\tfrac{d}{2}\right) C^{\,\prime\,(m=0)}_{\Delta}.
\end{align}   
\end{subequations}

By virtue of the four-point constraint \eqref{4ptcbconstr} on the scaling dimensions, we have that for integer $d>2$ the residues of the double-trace poles \eqref{DTpolesC3sc} are vanishing:
\begin{subequations}
 \begin{align}
    a^{(m=0)}_{\Delta_1+\Delta_2+2n,0} &= 2\pi i \,\delta\left(\tfrac{-3d+4+\sum_i\Delta_i}{2}\right)\, \text{Res}_{\Delta=\Delta_1+\Delta_2+2n,0}\left[\frac{C^{\,\prime\,(m=0)}_{\Delta_1\Delta_2\Delta}C^{\,\prime\,(m=0)}_{\Delta\Delta_3\Delta_4}}{C^{\,\prime\,(m=0)}_{\Delta}}\right] = 0,\\
    a^{(m=0)}_{\Delta_3+\Delta_4+2n,0} &= 2\pi i \,\delta\left(\tfrac{-3d+4+\sum_i\Delta_i}{2}\right)\, \text{Res}_{\Delta=\Delta_3+\Delta_4+2n,0}\left[\frac{C^{\,\prime\,(m=0)}_{\Delta_1\Delta_2\Delta}C^{\,\prime\,(m=0)}_{\Delta\Delta_3\Delta_4}}{C^{\,\prime\,(m=0)}_{\Delta}}\right] = 0.
\end{align}   
\end{subequations}
This is in accordance with the fact that the massless exchange is given by a conformal partial wave \eqref{mldge3} for integer $d>2$ and therefore does not contain double-trace contributions. In particular, one can write:
\begin{multline}
    M^{{\cal V}_{12 \varphi}{\cal V}_{34 \varphi}}_{\Delta_1 \Delta_2\Delta_3\Delta_4}\left(s_{12},s_{13}\right)=  2\pi i \,\delta\left(\frac{-3d+4+\sum_i\Delta_i}{2}\right)\,\\ \times \text{Res}_{\Delta=\Delta^{(0)}_{12}}\left[\frac{C^{\,\prime\,(m=0)}_{\Delta_1\Delta_2\Delta}C^{\,\prime\,(m=0)}_{\Delta\Delta_3\Delta_4}}{C^{\,\prime\,(m=0)}_{\Delta}}\right]  {\cal F}_{\Delta^{(0)}_{12},0}\left(s_{12},s_{13}\right),
\end{multline}
where we recall that $\Delta^{(0)}_{12}$ and $\Delta^{(0)}_{34}$ are shadow due to the four-point constraint \eqref{4ptcbconstr}, and where ${\cal F}_{\Delta^{(0)}_{12},0}$ refers to the Mellin representation of a scalar Conformal Partial Wave, whose properties and  expression are recalled in appendix \ref{app::CPW}. 

\vskip 4pt
It would be interesting to compare the structure of the celestial OPE from celestial correlators with analogous results in the context of celestial amplitudes \cite{Lam:2017ofc,Nandan:2019jas,Fan:2021isc,Atanasov:2021cje,Guevara:2021tvr,Melton:2021kkz,Fan:2022vbz,Garcia-Sepulveda:2022lga,Chang:2023ttm,Liu:2024vmx}.

\section{Spin-1 exchange}
\label{sec::spin1}

 In this section we consider the celestial Mellin amplitude for the four-point exchange of massive (section \ref{subsec::massivespin1}) and massless (section \ref{subsec::gaugeboson}) spin-1 fields between massless scalars in $\mathbb{M}^{d+2}$, extracting the direct channel OPE data on the celestial sphere in section \ref{subsec::opedataspin1}.

 \begin{figure}[t]
    \centering
    \includegraphics[width=0.4\textwidth]{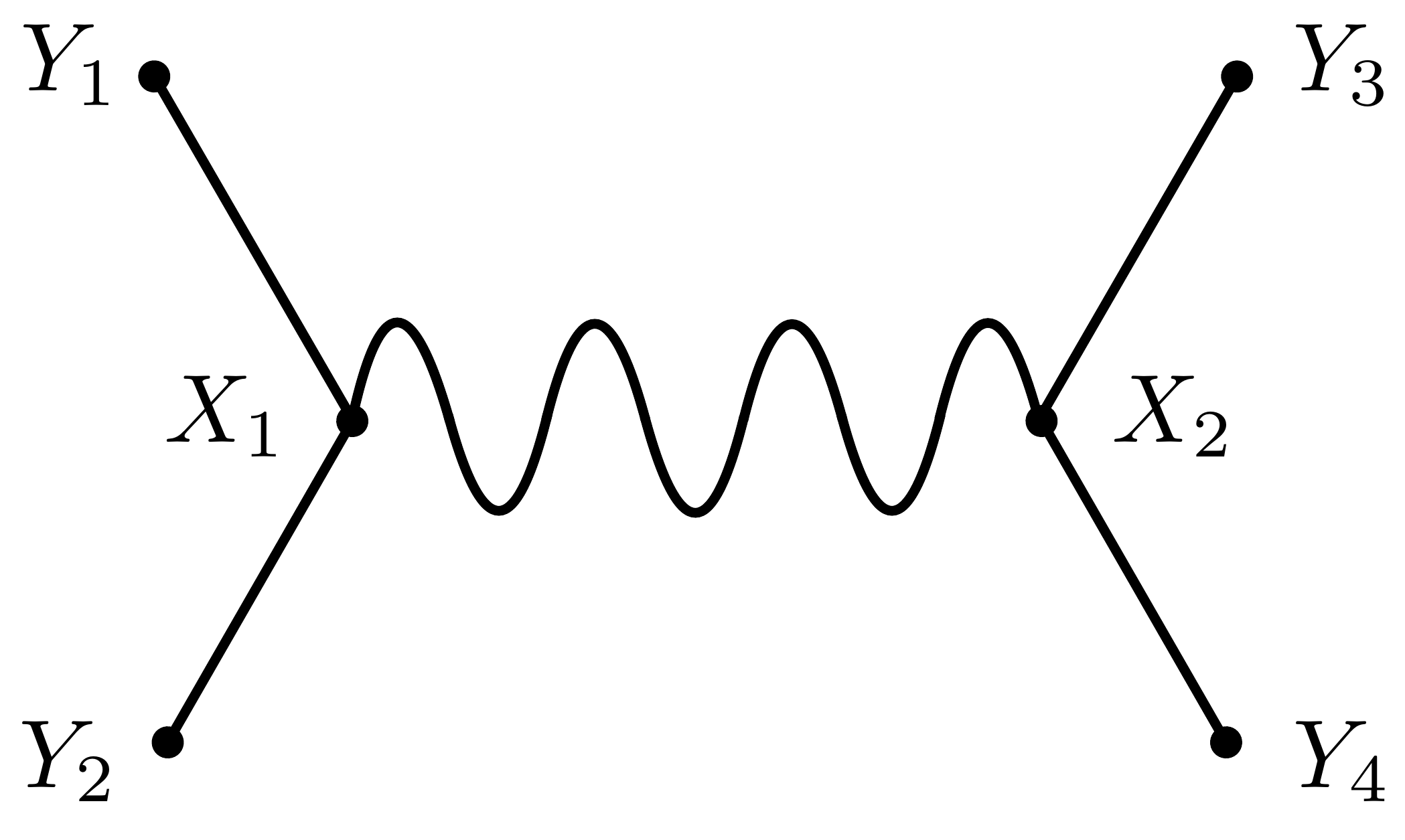}
    \caption{In this work we consider the four-point exchange of fields with spin-1 and 2 (wavy line) between external massless scalars (solid lines) in ($d$+2)-dimensional Minkowski space. The external points $Y_i$ are extrapolated to the celestial sphere according to the prescription \eqref{ccdefn2}.}
    \label{fig::exch_spin}
\end{figure}

\subsection{Massive spin-1}
\label{subsec::massivespin1}

The equation for the propagator of a massive spin-1 field in Minkowski space is
\begin{equation}
    \left(\partial^2-m^2\right)G^{(m)\,\mu \nu}\left(X,Y\right) = - \mathcal{P}_{m^2}^{\mu \nu}\, \delta^{d+2}\left(X-Y\right),
\end{equation}
with projector
\begin{equation}\label{spin_1_projector}
    \mathcal{P}_{m^2}^{\mu \nu}=\eta^{\mu \nu}+\frac{1}{m^2}\frac{\partial}{\partial X_{\mu}}\frac{\partial}{\partial Y_{\nu}}.
\end{equation}
Since the projector commutes with the Laplacian, the massive spin-1 Feynman propagator can be expressed in terms of its spin-0 counterpart via:
\begin{equation}\label{spin_1_propagator}
    G^{(m)\,\mu \nu}_T\left(X,Y\right)=\mathcal{P}_{m^2}^{\mu \nu}G^{(m)}_T\left(X,Y\right).
\end{equation}
Using this representation, perturbative contributions to correlation functions involving massive spin-1 fields can be expressed as differential operators in the external points acting on their scalar counterparts. In Mellin space, the action of derivatives translates into a difference operation in the Mellin variables. In the following this will be demonstrated explicitly for the four-point tree-level exchange of a massive spin-1 field.

 \vskip 4pt
 Consider the four-point exchange of a massive spin-1 field $A_\mu$ between massless scalars $\phi_i$ built from vertices:\footnote{We employ the shorthand notation:
\begin{equation}
    f \overleftrightarrow{\partial_\mu} g := \left(\partial_\mu f\right)g -  f\left(\partial_\mu g\right).
\end{equation}}${}^{,}$\footnote{\label{foo::cubicimpr}Note that cubic vertices involving two scalars and a spinning field are unique on-shell \cite{Metsaev:2005ar}, so that diagrams computed by vertices obtained from the above by the addition of terms proportional to the free equations of motion differ from diagrams generated from the above by contact terms. For this reason, in the main text we will often neglect contributions from contact terms, which are discussed in more detail in appendix \ref{app::contactterms}. The choice \eqref{spin-1v} facilitates comparison with the gauge boson exchange in section \ref{subsec::gaugeboson}. \label{fnt:onshell3pt}}                      
\begin{equation}\label{spin-1v}
    {\cal V}_{12 A} = g_{12}\, A^\mu \phi_1 \overleftrightarrow{\partial_\mu} \phi_2, \qquad {\cal V}_{34 A} = g_{34}\, A^\mu\phi_3\overleftrightarrow{\partial_\mu} \phi_4.
\end{equation}
The corresponding contribution to the time ordered correlator of the $\phi_i$ reads
\begin{multline}\label{spin-1exchdef}
   {\cal A}^{{\cal V}_{12 A}{\cal V}_{34 A}}\left(Y_1,Y_2,Y_3,Y_4\right) = (-i g_{12})(-i g_{34}) \int {\rm d}^{d+2}X_1 {\rm d}^{d+2}X_2\,\left(G^{(0)}_T(Y_1,X_1)\overleftrightarrow{\partial_\mu}G^{(0)}_T(X_1,Y_2) \right) \\
    \times G^{(m)\,\mu\nu}_T(X_1,X_2)\left(G^{(0)}_T(Y_3,X_2)\overleftrightarrow{\partial_\nu}G^{(0)}_T(X_2,Y_4) \right).
\end{multline}
Using invariance under translations, all derivatives in the exchange integrand can be chosen to act on the external points $Y_i$. This implies that the exchange diagram \eqref{spin-1exchdef} can be expressed as a differential operator in the external points acting on the scalar exchange diagram \eqref{exchdef}:
\begin{align}\nonumber
    {\cal A}^{{\cal V}_{12 A}{\cal V}_{34 A}}\left(Y_1,Y_2,Y_3,Y_4\right)&=\biggl(\left(\partial_{Y_1}-\partial_{Y_2}\right)\cdot\left(\partial_{Y_3}-\partial_{Y_4}\right)+\\
    & \hspace*{1.35cm}+\frac{1}{m^2}\left(\partial^2_{Y_1}-\partial^2_{Y_2}\right)\left(\partial^2_{Y_3}-\partial^2_{Y_4}\right)\biggr){\cal A}^{{\cal V}_{12 m}{\cal V}_{34 m}}\left(Y_1,Y_2,Y_3,Y_4\right), \nonumber\\
    &\approx \left(\partial_{Y_1}-\partial_{Y_2}\right)\cdot\left(\partial_{Y_3}-\partial_{Y_4}\right){\cal A}^{{\cal V}_{12 m}{\cal V}_{34 m}}\left(Y_1,Y_2,Y_3,Y_4\right),\label{spin-1exchexpansion}
\end{align}
where in the second equality we neglected terms proportional to the free equations of motion, which throughout we denote by the symbol $\approx$. The latter give contact term contributions to the exchange by virtue of the propagator equation for the external massless scalar fields:
\begin{equation}
    \partial^2_{Y_i}G^{(0)}_T\left(X,Y_i\right)=-\delta^{d+2}\left(X-Y_i\right).
\end{equation}
Such contributions are discussed in more detail in appendix \ref{app::contactterms}, while in the main text we focus on contributions encoding the exchanged single particle state.

\vskip 4pt
Following the scalar exchange reviewed in section \ref{sec::reviewscalarMA}, we consider the Mellin representation \eqref{tcMA} of the bulk time-ordered correlation function:
\begin{multline}\label{MAspin1}
  {\cal A}^{{\cal V}_{12 A}{\cal V}_{34 A}}\left(Y_1,Y_2,Y_3,Y_4\right)
  \\ = \int^{i\infty}_{-i\infty}   \frac{{\rm d}\delta_{ij}}{2\pi i}\,  M^{{\cal V}_{12 A}{\cal V}_{34 A}}\left(\delta_{ij}\right)\prod_{i<j}\Gamma\left(\delta_{ij}\right)\left[\left(Y_i-Y_j\right)^2+i \epsilon\right]^{-\delta_{ij}}.
\end{multline}
In Mellin space the action of derivatives in $Y_i$ translates into shifts in the Mellin variables $\delta_{ij}$. Defining
\begin{equation}\label{D}
    {\cal D} := \left(\partial_{Y_1}-\partial_{Y_2}\right)\cdot\left(\partial_{Y_3}-\partial_{Y_4}\right),
\end{equation}
it is straightforward to verify that the action of ${\cal D}$ on a Poincar\'e invariant function $f\left(Y_i\right)$ in $\mathbb{M}^{d+2}$ translates into the following finite difference operation in Mellin space, where ${\tilde f}\left(\delta_{ij}\right)$ is the Mellin amplitude of $f\left(Y_i\right)$:
{\allowdisplaybreaks\begin{multline}\label{recursion}
  \Delta_{{\cal D}}\left[{\tilde f}\left(\delta_{ij}\right)\right]=  -2 (2 (\delta_{12}+\delta_{13}+\delta_{34})-d-2) {\tilde f}(\delta_{12},\delta_{13}-1,\delta_{14},\delta_{23},\delta_{24},\delta_{34})\\+2  (2 (\delta_{12}+\delta_{14}+\delta_{34})-d-2) {\tilde f}(\delta_{12},\delta_{13},\delta_{14}-1,\delta_{23},\delta_{24},\delta_{34})\\+2  (2 (\delta_{12}+\delta_{23}+\delta_{34})-d-2) {\tilde f}(\delta_{12},\delta_{13},\delta_{14},\delta_{23}-1,\delta_{24},\delta_{34})\\-2 (2 (\delta_{12}+\delta_{24}+\delta_{34})-d-2) {\tilde f}(\delta_{12},\delta_{13},\delta_{14},\delta_{23},\delta_{24}-1,\delta_{34})\\-4  (\delta_{13}-\delta_{14}-\delta_{23}+\delta_{24}) \left( {\tilde f}(\delta_{12},\delta_{13},\delta_{14},\delta_{23},\delta_{24},\delta_{34}-1)+{\tilde f}(\delta_{12}-1,\delta_{13},\delta_{14},\delta_{23},\delta_{24},\delta_{34})\right)\\+4 \delta_{12} {\tilde f}(\delta_{12}+1,\delta_{13},\delta_{14}-1,\delta_{23}-1,\delta_{24},\delta_{34})
  -4 \delta_{24} {\tilde f}(\delta_{12}-1,\delta_{13},\delta_{14}-1,\delta_{23},\delta_{24}+1,\delta_{34})\\+4 \delta_{23} {\tilde f}(\delta_{12}-1,\delta_{13}-1,\delta_{14},\delta_{23}+1,\delta_{24},\delta_{34})+4 \delta_{14}  {\tilde f}(\delta_{12}-1,\delta_{13},\delta_{14}+1,\delta_{23},\delta_{24}-1,\delta_{34})\\-4 \delta_{13}  {\tilde f}(\delta_{12}-1,\delta_{13}+1,\delta_{14},\delta_{23}-1,\delta_{24},\delta_{34})-4 \delta_{34}  {\tilde f}(\delta_{12},\delta_{13}-1,\delta_{14},\delta_{23},\delta_{24}-1,\delta_{34}+1)\\+4 \delta_{23}  {\tilde f}(\delta_{12},\delta_{13}-1,\delta_{14},\delta_{23}+1,\delta_{24}-1,\delta_{34})+4 \delta_{14}  {\tilde f}(\delta_{12},\delta_{13}-1,\delta_{14}+1,\delta_{23},\delta_{24}-1,\delta_{34})\\+4 \delta_{14} {\tilde f}(\delta_{12},\delta_{13}-1,\delta_{14}+1,\delta_{23},\delta_{24},\delta_{34}-1)+4 \delta_{34} {\tilde f}(\delta_{12},\delta_{13},\delta_{14}-1,\delta_{23}-1,\delta_{24},\delta_{34}+1)\\-4 \delta_{24}  {\tilde f}(\delta_{12},\delta_{13},\delta_{14}-1,\delta_{23}-1,\delta_{24}+1,\delta_{34})-4 \delta_{24} {\tilde f}(\delta_{12},\delta_{13},\delta_{14},\delta_{23}-1,\delta_{24}+1,\delta_{34}-1)\\+4 \delta_{23} {\tilde f}(\delta_{12},\delta_{13},\delta_{14},\delta_{23}+1,\delta_{24}-1,\delta_{34}-1)-4 \delta_{13}  {\tilde f}(\delta_{12},\delta_{13}+1,\delta_{14}-1,\delta_{23}-1,\delta_{24},\delta_{34})\\-4 \delta_{13} {\tilde f}(\delta_{12},\delta_{13}+1,\delta_{14}-1,\delta_{23},\delta_{24},\delta_{34}-1)-4 \delta_{12} {\tilde f}(\delta_{12}+1,\delta_{13}-1,\delta_{14},\delta_{23},\delta_{24}-1,\delta_{34})\\+8 \delta_{23} {\tilde f}(\delta_{12}-1,\delta_{13},\delta_{14},\delta_{23}+1,\delta_{24},\delta_{34}-1)-8 \delta_{24} {\tilde f}(\delta_{12}-1,\delta_{13},\delta_{14},\delta_{23},\delta_{24}+1,\delta_{34}-1)\\+8 \delta_{14} {\tilde f}(\delta_{12}-1,\delta_{13},\delta_{14}+1,\delta_{23},\delta_{24},\delta_{34}-1)-8 \delta_{13} {\tilde f}(\delta_{12}-1,\delta_{13}+1,\delta_{14},\delta_{23},\delta_{24},\delta_{34}-1).
\end{multline}}In Mellin space, the differential relationship \eqref{spin-1exchexpansion} between the spin-1 and spin-0 exchange then becomes the finite difference operation on the spin-0 exchange Mellin amplitude:
\begin{equation}
    M^{{\cal V}_{12 A}{\cal V}_{34 A}}\left(\delta_{ij}\right) \approx \Delta_{{\cal D}}\left[M^{{\cal V}_{12 m}{\cal V}_{34 m}}\left(\delta_{ij}\right)\right].
\end{equation}

\vskip 4pt
As detailed in section \ref{sec::CMA}, the Mellin amplitude for the corresponding celestial correlator is obtained from the Mellin amplitude \eqref{MAspin1} of the bulk correlation function simply by imposing the conformal constraints \eqref{confcon} on the Mellin variables. This gives:
{\allowdisplaybreaks
\begin{flalign}\label{mass_spin_1_exch_dij}
&M^{{\cal V}_{12 A}{\cal V}_{34 A}}_{\Delta_1 \Delta_2\Delta_3\Delta_4}\left(\delta_{ij}\right)=-\frac{1}{2}g_{12}g_{34}\,\pi^{\frac{d+2}{2}}\,\prod^4_{i=1}\frac{\Gamma\left(\tfrac{d}{2}-\Delta_i\right)}{4\pi^{\frac{d+2}{2}}}\\
&\nonumber\times \left(\frac{m}{2}\right)^{-2+3d-\sum_i\Delta_i}\Gamma\left(\frac{2-3d+\sum_i\Delta_i}{2}\right)\left(\Delta_1-\Delta_2\right)\left(\Delta_3-\Delta_4\right)\\
    &\nonumber\times\left[\,\frac{\left(\delta_{13}-\delta_{14}-\delta_{23}+\delta_{24}\right)\Gamma \left(\tfrac{-d+2\delta_{12}-\Delta_1-\Delta_2+\Delta_3+\Delta_4}{2} \right)\Gamma \left(\tfrac{-d+2\delta_{34}+\Delta_1+\Delta_2-\Delta_3-\Delta_4}{2} \right) }{\left(\Delta_1-\Delta_2\right)(\Delta_3-\Delta_4)\Gamma \left(\tfrac{d+2\delta_{12}-\Delta_1-\Delta_2-\Delta_3-\Delta_4}{2} \right)\Gamma \left(\tfrac{d+2\delta_{34}-\Delta_1-\Delta_2-\Delta_3-\Delta_4}{2} \right)} \right.\\
    &\nonumber\phantom{xxxxxxxxxxxxxxxxx}
   \times {}_3F_2\left(\begin{matrix}\tfrac{-d+2\delta_{12}-\Delta_1-\Delta_2+\Delta_3+\Delta_4}{2},\tfrac{-d+2\delta_{34}+\Delta_1+\Delta_2-\Delta_3-\Delta_4}{2},-\frac{\beta}{2}-1\\\tfrac{d+2\delta_{12}-\Delta_1-\Delta_2-\Delta_3-\Delta_4}{2},\tfrac{d+2\delta_{34}-\Delta_1-\Delta_2-\Delta_3-\Delta_4}{2}\end{matrix};1\right)
    \\
    &\nonumber \phantom{xxx}+\frac{\left(\delta_{13}-\delta_{14}+\delta_{23}-\delta_{24}\right)\Gamma \left(\tfrac{-d+2\delta_{12}-\Delta_1-\Delta_2+\Delta_3+\Delta_4}{2} \right)\Gamma \left(\tfrac{-d+2\delta_{34}+\Delta_1+\Delta_2-\Delta_3-\Delta_4}{2} \right) }{(\Delta_3-\Delta_4)\Gamma \left(\tfrac{d+2\delta_{12}-\Delta_1-\Delta_2-\Delta_3-\Delta_4}{2} \right)\Gamma \left(\tfrac{d+2\delta_{34}-\Delta_1-\Delta_2-\Delta_3-\Delta_4+2}{2} \right)}\\
    &\nonumber\phantom{xxxxxxxxxxxxxxxxxxxx}\, \times {}_3F_2\left(\begin{matrix}\tfrac{-d+2\delta_{12}-\Delta_1-\Delta_2+\Delta_3+\Delta_4}{2},\tfrac{-d+2\delta_{34}+\Delta_1+\Delta_2-\Delta_3-\Delta_4}{2},-\frac{\beta}{2}\\\tfrac{d+2\delta_{12}-\Delta_1-\Delta_2-\Delta_3-\Delta_4}{2},\tfrac{d+2\delta_{34}-\Delta_1-\Delta_2-\Delta_3-\Delta_4+2}{2}\end{matrix};1\right)\\
    &\nonumber \phantom{xxx} + \, \frac{\left(\delta_{13}+\delta_{14}-\delta_{23}-\delta_{24}\right)\Gamma \left(\tfrac{-d+2\delta_{12}-\Delta_1-\Delta_2+\Delta_3+\Delta_4}{2} \right)\Gamma \left(\tfrac{-d+2\delta_{34}+\Delta_1+\Delta_2-\Delta_3-\Delta_4}{2} \right) }{(\Delta_1-\Delta_2)\Gamma \left(\tfrac{d+2\delta_{12}-\Delta_1-\Delta_2-\Delta_3-\Delta_4+2}{2} \right)\Gamma \left(\tfrac{d+2\delta_{34}-\Delta_1-\Delta_2-\Delta_3-\Delta_4}{2} \right)}\\
    &\nonumber\phantom{xxxxxxxxxxxxxxxxxxxx} \times {}_3F_2\left(\begin{matrix}\tfrac{-d+2\delta_{12}-\Delta_1-\Delta_2+\Delta_3+\Delta_4}{2},\tfrac{-d+2\delta_{34}+\Delta_1+\Delta_2-\Delta_3-\Delta_4}{2},-\frac{\beta}{2}\\\tfrac{d+2\delta_{12}-\Delta_1-\Delta_2-\Delta_3-\Delta_4+2}{2},\tfrac{d+2\delta_{34}-\Delta_1-\Delta_2-\Delta_3-\Delta_4}{2}\end{matrix};1\right)\\
    &\nonumber \phantom{xxx} +\,\frac{\left(\delta_{13}+\delta_{14}+\delta_{23}+\delta_{24}\right)\Gamma \left(\tfrac{-d+2\delta_{12}-\Delta_1-\Delta_2+\Delta_3+\Delta_4}{2} \right)\Gamma \left(\tfrac{-d+2\delta_{34}+\Delta_1+\Delta_2-\Delta_3-\Delta_4}{2} \right)}{\Gamma \left(\tfrac{d+2\delta_{12}-\Delta_1-\Delta_2-\Delta_3-\Delta_4+2}{2} \right)\Gamma \left(\tfrac{d+2\delta_{34}-\Delta_1-\Delta_2-\Delta_3-\Delta_4+2}{2} \right)}  \\
    &\nonumber\phantom{xxxxxxxxxxxxxxxx} \times {}_3F_2\left(\begin{matrix}\tfrac{-d+2\delta_{12}-\Delta_1-\Delta_2+\Delta_3+\Delta_4}{2},\tfrac{-d+2\delta_{34}+\Delta_1+\Delta_2-\Delta_3-\Delta_4}{2},-\frac{\beta}{2}+1\\\tfrac{d+2\delta_{12}-\Delta_1-\Delta_2-\Delta_3-\Delta_4+2}{2},\tfrac{d+2\delta_{34}-\Delta_1-\Delta_2-\Delta_3-\Delta_4+2}{2}\end{matrix};1\right)\\
    &\nonumber \phantom{xxx} -\, \frac{\left(-\beta+\delta_{13}+\delta_{14}+\delta_{23}+\delta_{24}\right)\Gamma \left(\tfrac{-d+2\delta_{12}-\Delta_1-\Delta_2+\Delta_3+\Delta_4+2}{2} \right)\Gamma \left(\tfrac{-d+2\delta_{34}+\Delta_1+\Delta_2-\Delta_3-\Delta_4+2}{2} \right) }{\Gamma \left(\tfrac{d+2\delta_{12}-\Delta_1-\Delta_2-\Delta_3-\Delta_4+4}{2} \right)\Gamma \left(\tfrac{d+2\delta_{34}-\Delta_1-\Delta_2-\Delta_3-\Delta_4+4}{2} \right)} \\
    &\left.\nonumber\phantom{xxxxxxxxxxxxxxxx}
  \times  {}_3F_2\left(\begin{matrix}\tfrac{-d+2\delta_{12}-\Delta_1-\Delta_2+\Delta_3+\Delta_4+2}{2},\tfrac{-d+2\delta_{34}+\Delta_1+\Delta_2-\Delta_3-\Delta_4+2}{2},-\frac{\beta}{2}+1\\\tfrac{d+2\delta_{12}-\Delta_1-\Delta_2-\Delta_3-\Delta_4+4}{2},\tfrac{d+2\delta_{34}-\Delta_1-\Delta_2-\Delta_3-\Delta_4+4}{2}\end{matrix};1\right)
    \right].
\end{flalign}
}To extract the operator content in the s-channel we work in terms of the Mellin ``Mandelstam variables" \eqref{MBmandel}. Using the compact notation \eqref{Tscexch}, the Mellin amplitude reads:
{\allowdisplaybreaks
\begin{flalign}\label{mass_spin_1_exch_s12s13}
&M^{{\cal V}_{12 A}{\cal V}_{34 A}}_{\Delta_1 \Delta_2\Delta_3\Delta_4}\left(s_{12},s_{13}\right)=-\frac{1}{4}g_{12}g_{34}\,\pi^{\frac{d+2}{2}}\,\prod^4_{i=1}\frac{\Gamma\left(\tfrac{d}{2}-\Delta_i\right)}{4\pi^{\frac{d+2}{2}}}\\
&\nonumber\times\left(\frac{m}{2}\right)^{-2+3d-\sum_i\Delta_i}\Gamma\left(\frac{2-3d+\sum_i\Delta_i}{2}\right)\left(\Delta_1-\Delta_2\right)\left(\Delta_3-\Delta_4\right)\\
    &\nonumber\times\left[\,\frac{\left(-s_{12}-2s_{13}-\Delta_1+\Delta_2+\Delta_3-\Delta_4\right) }{\left(\Delta_1-\Delta_2\right)(\Delta_3-\Delta_4)}T_{e}\left(d-\Delta_1-\Delta_2,d-\Delta_3-\Delta_4,\frac{\beta}{2}+1,0,0\right)\right.\\
    &\nonumber\phantom{xxxxxx}+T_{e}\left(d-\Delta_1-\Delta_2,d-\Delta_3-\Delta_4+1,\frac{\beta}{2},1,0\right)\\
    &\nonumber\phantom{xxxxxx}+T_{e}\left(d-\Delta_1-\Delta_2+1,d-\Delta_3-\Delta_4,\frac{\beta}{2},0,1\right)\\
    &\nonumber\phantom{xxxxxx}+s_{12}\,T_{e}\left(d-\Delta_1-\Delta_2+1,d-\Delta_3-\Delta_4+1,\frac{\beta}{2}-1,1,1\right)\\
    &\left.\nonumber\phantom{xxxxxx}-\left(2+d+s_{12}-\sum_i\Delta_i\right)\,T_{e}\left(d-\Delta_1-\Delta_2+1,d-\Delta_3-\Delta_4+1,\frac{\beta}{2}-1,0,0\right)\right].
\end{flalign}
}This is a degree one polynomial in $s_{13}$ and the poles in $s_{12}$ are at
\begin{equation}\label{spin1mpoless12}
  s_{12} = \Delta_1 + \Delta_2 -d+2n, \qquad s_{12} = \Delta_3 + \Delta_4 - d + 2n, \qquad n \in \mathbb{N},
\end{equation}
which, owing to the finite difference operation \eqref{recursion}, are shifted by -2 with respect to those \eqref{s_12_poles_scalar1} for the massive scalar exchange. These poles correspond to operators on the co-dimension 2 celestial sphere of spins 1 and 0, and twists:
\begin{subequations}
 \begin{align}
    \tau^{(n)}_{12} &= \Delta_1 + \Delta_2-d+2n_{12}, \qquad n_{12}=0,1,2,\ldots,\\ \tau^{(n)}_{34} &= \Delta_3 + \Delta_4 -d+2n_{34}, \qquad n_{34}=0,1,2,\ldots\,,
\end{align}   
\end{subequations}
encoding the bulk massive spin-1 exchanged single particle state. In section \ref{subsec::opedataspin1} we extract the corresponding OPE coefficients.

\subsection{Gauge Boson}
\label{subsec::gaugeboson}

One proceeds in a similar way for a gauge boson exchange. In the $R_\xi$ gauge, the position space Feynman propagator is given by:
\begin{multline}\label{gbprop}
    \left(G_T\right)_{\mu \nu}(X,Y) = \frac{\Gamma\left(\frac{d}{2}\right)}{4 \pi^{\frac{d+2}{2}}}\left[\frac{\eta_{\mu \nu}}{\left[(X-Y)^2 + i \epsilon \right]^{\frac{d}{2}}} \right. \\ \left. + \frac{1 - \xi}{2(d - 2)}\partial^Y_\mu \partial^Y_\nu \left( \frac{1}{\left[(X-Y)^2 + i \epsilon \right]^{\frac{d}{2}-1}} \right) \right].
\end{multline}

We consider the four-point exchange of a gauge boson $A^\mu$ between minimally coupled massless scalar fields $\phi_i$, which is mediated by the cubic vertices
\begin{equation}
    {\cal V}_{12 A} = g_{12}\, A^\mu \phi_1 \overleftrightarrow{\partial_\mu} \phi_2, \qquad {\cal V}_{34 A} = g_{34}\, A^\mu\phi_3\overleftrightarrow{\partial_\mu} \phi_4,
\end{equation}
which reads
\begin{multline}\label{gbexchfrsc}
    \mathcal{A}^{\mathcal{V}_{12A}\mathcal{V}_{34A}}(Y_1,Y_2,Y_3,Y_4) = -g_{12}g_{34} \int {\rm d}^{d+2}X_1 {\rm d}^{d+2}X_2 \left(G^{(0)}_T(Y_1,X_1)\overleftrightarrow{\partial_\mu}G^{(0)}_T(X_1,Y_2) \right) \\
    \times G^{\mu\nu}_T(X_1,X_2)\left(G^{(0)}_T(Y_3,X_2)\overleftrightarrow{\partial_\nu}G^{(0)}_T(X_2,Y_4) \right).
\end{multline}

\vskip 4pt
The gauge dependent terms in the gauge boson propagator on the second line of \eqref{gbprop} contribute only contact terms to the exchange process, which are discussed in more detail in appendix \ref{app::contactterms}. We therefore take $\xi = 1$, in which case the propagator \eqref{gbprop} is proportional to that \eqref{MLschw} of a massless scalar field. 

\vskip 4pt
Following the same steps as for the massive spin-1 exchange, the gauge boson exchange can be expressed up to contact terms as the differential operator \eqref{D} in the external points acting on the exchange diagram \eqref{Mlbulkexch} for a massless scalar:
\begin{equation}\label{eq: exchange diagram of gauge field mellin amplitude}
    \mathcal{A}^{\mathcal{V}_{12A}\mathcal{V}_{34A}}(Y_1,Y_2,Y_3,Y_4) \approx {\cal D}\mathcal{A}^{\mathcal{V}_{12m=0}\mathcal{V}_{34m=0}}(Y_1,Y_2,Y_3,Y_4),
\end{equation}
Note that this is the same differential relationship as the one \eqref{spin-1exchexpansion} between massive spin-1 and spin-0 exchange diagrams. The corresponding Mellin amplitude for the gauge boson exchange is therefore given by the same finite difference operation \eqref{recursion}, which acts on the Mellin amplitude \eqref{Mlbulkexch} for the massless scalar exchange. The corresponding celestial Mellin amplitude is therefore simply
\begin{multline}
M^{{\cal V}_{12 A}{\cal V}_{34 A}}_{\Delta_1 \Delta_2\Delta_3\Delta_4}\left(s_{12},s_{13}\right)=-\frac{1}{4}g_{12}g_{34}\,\pi^{\frac{d+2}{2}}\,\prod^4_{i=1}\frac{\Gamma\left(\tfrac{d}{2}-\Delta_i\right)}{4\pi^{\frac{d+2}{2}}}\\
\nonumber\times (2\pi i)\delta\left(\frac{-3d + 2 + \sum_i \Delta_i}{2} \right)\left(\Delta_1-\Delta_2\right)\left(\Delta_3-\Delta_4\right)\\
    \nonumber\times\left[\,\frac{\left(3d-2-s_{12}-2(s_{13}+\Delta_1+\Delta_4)\right) }{\left(\Delta_1-\Delta_2\right)(\Delta_3-\Delta_4)}T_{e}\left(d-\Delta_1-\Delta_2,d-\Delta_3-\Delta_4,d-1,0,0\right)\right.\\
    \nonumber+T_{e}\left(d-\Delta_1-\Delta_2,d-\Delta_3-\Delta_4+1,d-2,1,0\right)\\
    \nonumber+T_{e}\left(d-\Delta_1-\Delta_2+1,d-\Delta_3-\Delta_4,d-2,0,1\right)\\
    \nonumber+s_{12}\,T_{e}\left(d-\Delta_1-\Delta_2+1,d-\Delta_3-\Delta_4+1,d-3,1,1\right)\\
    \left.\nonumber-\left(2d-4-s_{12}\right)\,T_{e}\left(d-\Delta_1-\Delta_2+1,d-\Delta_3-\Delta_4+1,d-3,0,0\right)\right],
\end{multline}
where the difference operation results in a shift in the constraint on the external scaling dimensions $\Delta_i$ with respect to that \eqref{mldge3} for the massless scalar exchange.\footnote{This can also be traced back to the powers of $1/R_i$ arising from the differential operators in $Y_i$, which in turn shift the argument of the Dirac delta function \eqref{confcon} in the extrapolation step.} This expression can also be obtained by taking the massless limit of the Mellin amplitude \eqref{mass_spin_1_exch_s12s13} for the massive spin-1 exchange, using that:
\begin{equation}\label{spin1mlconstr}
   \lim_{m\rightarrow0}\left(\frac{m}{2}\right)^{-2+3d-\sum_i\Delta_i}\Gamma\left(\frac{2-3d+\sum_i\Delta_i}{2}\right) = 2\pi i\, \delta\left(\frac{2-3d+\sum_i\Delta_i}{2}\right). 
\end{equation}

\vskip 4pt
As for the exchange of a massless scalar reviewed in section \ref{sec::reviewscalarMA}, the constraint \eqref{spin1mlconstr} on the external scaling dimensions leads to simplification of the generalised hypergeometric functions \eqref{Tscexch}. For integer $d \geq 3$ one can apply Saalsch\"utz theorem \eqref{saal}, giving:\footnote{This simplification for massless exchanges can already be understood from the relation \eqref{eq: exchange diagram of gauge field mellin amplitude} between the gauge boson and massless scalar exchange in the bulk, where the Mellin amplitude \eqref{Mlbulkexch} of the latter simplifies \eqref{Mlbulkexchsimp} to a ratio of $\Gamma$-functions for integer $d \geq 3$ using Saalsch\"utz theorem.}
 \begin{multline}\label{mlspin1MA}
    M^{\mathcal{V}_{12A^\mu}\mathcal{V}_{34A_\mu}}_{\Delta_1 \Delta_2 \Delta_3 \Delta_4}(s_{12},s_{13}) = 
     2\pi i\, \delta\left(\frac{-3d + 2 + \sum_i \Delta_i}{2} \right)a^{\prime\,(m=0)}_{\Delta^{(0,1)}_{12}, J=1} \\ \times \mathcal{F}_{\Delta_1 + \Delta_2 + (1-d), 1}(s_{12}, s_{13}),
\end{multline}
with coefficient\footnote{Note that the Mellin amplitude is vanishing for even $d\geq 2$, implying the vanishing of the corresponding celestial correlator for non-null separated points $Q_i \cdot Q_j \ne 0$.}
\begin{multline}\label{spin1mlOPE}
    a^{\prime\,(m=0)}_{\Delta^{(0,1)}_{12}, 1}=-g_{12}g_{34}\,\pi^{\frac{d+2}{2}} \prod^4_{i=1}\left(\frac{\Gamma\left(\tfrac{d}{2}-\Delta_i\right)}{4\pi^{\frac{d+2}{2}}}\right)\frac{2  \Gamma \left(\frac{d}{2}\right)   \Gamma\left(1-\frac{d}{2}+\Delta_1\right)\Gamma\left(1-\frac{d}{2}+\Delta_2\right) }{\Gamma (-d+\Delta_1+\Delta_2+2)}\\
    \times(-1)^{-d}\frac{ \Gamma (d-\Delta_3) \Gamma (d-\Delta_4) \Gamma (-d+\Delta_1+\Delta_2)\Gamma \left(\frac{3 d}{2}-\Delta_1-\Delta_2\right)}{\Gamma \left(1-\frac{d}{2}\right)  \Gamma (-2 d+\Delta_1+\Delta_2+2) \left(-\frac{3 d}{2}+\Delta_1+\Delta_2+1\right)}.  
\end{multline}
which is proportional to the Mellin amplitude for a spin-1 conformal partial wave (defined in appendix \ref{app::CPW}) encoding the exchange of two (shadow) spin-1 operators with scaling dimensions:
\begin{align}
    \Delta^{(0,1)}_{12}=\Delta_1+\Delta_2+(1-d), \qquad 
   \Delta^{(0,1)}_{34}=\Delta_3+\Delta_4+(1-d).
\end{align}
These are shadow of one another by virtue of the constraint \eqref{spin1mlconstr} on the external scaling dimensions.

\subsection{OPE data}
\label{subsec::opedataspin1}

In the previous sections we determined the celestial Mellin amplitude for a four-point (massive and massless) spin-1 exchange between massless scalars. In this section we determine the direct channel OPE data.

\paragraph{Massive spin-1.}  As outlined in section \ref{subsec::extractingOPE}, the OPE data can be extracted from the residues of the poles in $s_{12}$. These are at:
\begin{equation}\label{spin1exchs12poles}
   s_{12} = \Delta_1 + \Delta_2 -d+2(p+q), \qquad s_{12} = \Delta_3 + \Delta_4 - d + 2(p+q), \qquad p, q \in \mathbb{N},
\end{equation}
corresponding to spin-1 and spin-0 operators of twists:
\begin{equation}\label{twistfamspin1}
    \tau^{(n)}_{12}= \Delta_1+\Delta_2-d+2n, \qquad \tau^{(n)}_{34}= \Delta_3+\Delta_4-d+2n, \qquad n \in \mathbb{N}.
\end{equation}
The conformal block expansion \eqref{MAcbe} of the Mellin amplitude then takes the form:\footnote{Note that we leave implicit the contributions from poles \eqref{DTpoles} encoding double-trace operators \eqref{DT}. In this work we focus on the contributions from the exchanged single particle state in Minkowski space.} 
\begin{multline}\label{cbespin1}
\rho\left(s_{12},s_{13}\right)M^{{\cal V}_{12 A}{\cal V}_{34 A}}_{\Delta_1 \Delta_2\Delta_3\Delta_4}\left(s_{12}, s_{13}\right) = {\tilde \rho}\left(s_{12},s_{13}\right) \left[ \sum^\infty_{n=0} a^{(m)}_{\tau^{(n)}_{12},1}\sum^\infty_{m=0}\frac{{\cal Q}_{\tau^{(n)}_{12},1,m}\left(s_{13}\right)}{s_{12}-\tau^{(n)}_{12}-2m} \right. \\
\left. +\sum^\infty_{n=0} a^{(m)}_{\tau^{(n)}_{12},0}\sum^\infty_{m=0}\frac{{\cal Q}_{\tau^{(n)}_{12},0,m}\left(s_{13}\right)}{s_{12}-\tau^{(n)}_{12}-2m} + \left(\Delta_{1,2} \leftrightarrow \Delta_{3,4}\right) \right].
\end{multline}
Note that there is now also mixing among spin-1 and spin-0 operators (and their descendants), in addition to the mixing among operators in the same spin family discussed at the end of section \ref{subsec::extractingOPE}. This can be dealt with systematically, as illustrated in the following.

\vskip 4pt
To extract the residues of the poles \eqref{spin1exchs12poles}, we proceed as for the scalar exchange in section \ref{subsec::extractingOPE}. Employing the Mellin-Barnes representation \eqref{MB3F2} for the generalised hypergeometric functions, the spin-1 exchange Mellin amplitude \eqref{mass_spin_1_exch_s12s13} reads:
{\allowdisplaybreaks\begin{align} \nonumber
& M^{{\cal V}_{12 A}{\cal V}_{34 A}}_{\Delta_1 \Delta_2\Delta_3\Delta_4}\left(s_{12},s_{13}\right)=-\frac{1}{4}g_{12}g_{34}\,\pi^{\frac{d+2}{2}}\,\left(\Delta_1-\Delta_2\right)\left(\Delta_3-\Delta_4\right)\prod^4_{i=1}\frac{\Gamma\left(\tfrac{d}{2}-\Delta_i\right)}{4\pi^{\frac{d+2}{2}}}\\ \nonumber
&\times \left(\frac{m}{2}\right)^{-2+3d-\sum_i\Delta_i}\Gamma\left(\frac{2-3d+\sum_i\Delta_i}{2}\right)\int^{+i\infty}_{-i\infty}\frac{{\rm d}z}{2\pi i}\Gamma\left(-z\right)\left(-1\right)^z\\ \nonumber
    &\times\left[\frac{\left(-s_{12}-2s_{13}-\Delta_1+\Delta_2+\Delta_3-\Delta_4\right) }{\left(\Delta_1-\Delta_2\right)\left(\Delta_3-\Delta_4\right)\Gamma \left(-\tfrac{\beta}{2}-1 \right)}\right.\\ \nonumber
    & \phantom{xxxxxxxxxxxx} \times \frac{\Gamma\left(z+\tfrac{-d-s_{12}+\Delta_1+\Delta_2}{2}\right)\Gamma\left(z+\tfrac{-d-s_{12}+\Delta_3+\Delta_4}{2}\right)\Gamma\left(z-\tfrac{\beta}{2}-1\right)}{\Gamma\left(z+\tfrac{d-s_{12}-\Delta_1-\Delta_2}{2}\right)\Gamma\left(z+\tfrac{d-s_{12}-\Delta_3-\Delta_4}{2}\right)}\\ \nonumber
    &\phantom{xxx}+\frac{\Gamma\left(z+\tfrac{-d-s_{12}+\Delta_1+\Delta_2}{2}\right)\Gamma\left(z+\tfrac{-d-s_{12}+\Delta_3+\Delta_4}{2}\right)\Gamma\left(z-\tfrac{\beta}{2}\right)}{\Gamma\left(-\tfrac{\beta}{2}\right)\Gamma\left(z+\tfrac{d-s_{12}-\Delta_1-\Delta_2}{2}\right)\Gamma\left(z+\tfrac{d-s_{12}-\Delta_3-\Delta_4+2}{2}\right)}\\ \nonumber
    &\phantom{xxx}+\frac{\Gamma\left(z+\tfrac{-d-s_{12}+\Delta_1+\Delta_2}{2}\right)\Gamma\left(z+\tfrac{-d-s_{12}+\Delta_3+\Delta_4}{2}\right)\Gamma\left(z-\tfrac{\beta}{2}\right)}{\Gamma\left(-\tfrac{\beta}{2}\right)\Gamma\left(z+\tfrac{d-s_{12}-\Delta_1-\Delta_2+2}{2}\right)\Gamma\left(z+\tfrac{d-s_{12}-\Delta_3-\Delta_4}{2}\right)}\\ \nonumber
    &\phantom{xxx}+\frac{s_{12}}{\Gamma\left(-\tfrac{\beta}{2}+1\right)}\frac{\Gamma\left(z+\tfrac{-d-s_{12}+\Delta_1+\Delta_2}{2}\right)\Gamma\left(z+\tfrac{-d-s_{12}+\Delta_3+\Delta_4}{2}\right)\Gamma\left(z-\tfrac{\beta}{2}+1\right)}{\Gamma\left(z+\tfrac{d-s_{12}-\Delta_1-\Delta_2+2}{2}\right)\Gamma\left(z+\tfrac{d-s_{12}-\Delta_3-\Delta_4+2}{2}\right)}\\ \nonumber
    &\phantom{xxx}-\frac{\left(2+d+s_{12}-\sum_i\Delta_i\right) }{\Gamma\left(-\tfrac{\beta}{2}+1\right)}\\
    & \phantom{xxxxxxxx} \left.\times\frac{\Gamma\left(z+\tfrac{-d-s_{12}+\Delta_1+\Delta_2+2}{2}\right)\Gamma\left(z+\tfrac{-d-s_{12}+\Delta_3+\Delta_4+2}{2}\right)\Gamma\left(z-\tfrac{\beta}{2}+1\right)}{\Gamma\left(z+\tfrac{d-s_{12}-\Delta_1-\Delta_2+4}{2}\right)\Gamma\left(z+\tfrac{d-s_{12}-\Delta_3-\Delta_4+4}{2}\right)}\right]. \label{mass_spin_1_exch_integr}
\end{align}}This has the following poles in $z$:
\begin{subequations}\label{z_poles}
    \begin{alignat}{1}
        &z=p,\label{z_n_poles}\\
        &z=\frac{d+s_{12}-\Delta_1-\Delta_2}{2}-q,\label{z_Delta_12_poles}\\
        &z=\frac{d+s_{12}-\Delta_3-\Delta_4}{2}-q,\label{z_Delta_34_poles}\\
        &z=\frac{\beta}{2}+1-q,
    \end{alignat}
\end{subequations}
with $p, q=0,1,2,\dots\,$ . The poles \eqref{spin1exchs12poles} in $s_{12}$ arise from the pinching of the poles \eqref{z_Delta_12_poles} and \eqref{z_Delta_34_poles} with the poles \eqref{z_n_poles}. To evaluate their residues, we write:
{\allowdisplaybreaks\begin{multline}
    M^{{\cal V}_{12 A}{\cal V}_{34 A}}_{\Delta_1 \Delta_2\Delta_3\Delta_4}\left(s_{12},s_{13}\right)=-\frac{1}{4}g_{12}g_{34}\,\pi^{\frac{d+2}{2}}\,\left(\Delta_1-\Delta_2\right)\left(\Delta_3-\Delta_4\right)\prod^4_{i=1}\frac{\Gamma\left(\tfrac{d}{2}-\Delta_i\right)}{4\pi^{\frac{d+2}{2}}}\\
\times\left(\frac{m}{2}\right)^{-2+3d-\sum_i\Delta_i}\Gamma\left(\frac{2-3d+\sum_i\Delta_i}{2}\right)\int^{+i\infty}_{-i\infty}\frac{{\rm d}z}{2\pi i}\Gamma\left(-z\right)\\
\left[\Gamma\left(z+\frac{\Delta_1 + \Delta_2-d-s_{12}}{2}\right) A\left(z,\frac{s_{12}+d-\Delta_1-\Delta_2}{2},s_{13}\right) \right. \\
\: \left. + \: \Gamma\left(z+\frac{\Delta_1 + \Delta_2+2-d-s_{12}}{2} \right) B\left(z,\frac{s_{12}+d-2-\Delta_1-\Delta_2}{2},s_{13}\right)\right], \nonumber
\end{multline}}isolating the poles \eqref{spin1exchs12poles}, \eqref{z_Delta_34_poles} and \eqref{z_n_poles} concerned, where functions $A$ and $B$ are implicitly defined from \eqref{mass_spin_1_exch_integr}. Closing the $z$-integration contour to the right encloses the poles at $z=p$ along the positive real axis and gives the residue expansion: 
\begin{multline}\label{resexpspin1}
    M^{{\cal V}_{12 A}{\cal V}_{34 A}}_{\Delta_1 \Delta_2\Delta_3\Delta_4}\left(s_{12},s_{13}\right)=- \frac{1}{4}g_{12}g_{34}\,\pi^{\frac{d+2}{2}}\,\left(\Delta_1-\Delta_2\right)\left(\Delta_3-\Delta_4\right) \prod^4_{i=1}\frac{\Gamma\left(\tfrac{d}{2}-\Delta_i\right)}{4\pi^{\frac{d+2}{2}}}\\
\times\left(\frac{m}{2}\right)^{-2+3d-\sum_i\Delta_i}\Gamma\left(\frac{2-3d+\sum_i\Delta_i}{2}\right)\sum_{p=0}^{+\infty}\frac{(-1)^p}{p!}\\
\times \left[\sum_{q=0}^{+ \infty}\frac{(-1)^q}{q!} \frac{A(p,p+q,s_{13})}{s_{12}-(\Delta_1+\Delta_2-d+2(p+q))} + \: \sum_{q=0}^{+ \infty}\frac{(-1)^q}{q!} \frac{B(p,p+q,s_{13})}{s_{12}-(\Delta_1+\Delta_2-d+2(p+q+1))}\right] \\
 + \Delta_{1,2} \leftrightarrow \Delta_{3,4}.
\end{multline}
The residues at each twist can then be read off as linear combinations of the functions $A$ and $B$. From the conformal block expansion \eqref{cbespin1}, the leading twist contributions \eqref{twistfamspin1} with $n=0$ are given by the residue:
\begin{multline}
    \text{Res}\left[\rho\left(s_{12},s_{13}\right)M^{{\cal V}_{12 A}{\cal V}_{34 A}}_{\Delta_1 \Delta_2\Delta_3\Delta_4}\left(s_{12}, s_{13}\right)\right]_{s_{12}=\tau^{(0)}_{12}}\\ = {\tilde \rho}(\tau^{(0)}_{12},s_{13}) \left[a^{(m)}_{\tau^{(0)}_{12},1}{\cal Q}_{\tau^{(0)}_{12},1,0}\left(s_{13}\right)+a^{(m)}_{\tau^{(0)}_{12},0}{\cal Q}_{\tau^{(0)}_{12},0,0}\left(s_{13}\right)\right],
\end{multline}
and likewise for the residue at $s_{12}=\tau^{(0)}_{34}$. By comparing with the residue expansion \eqref{resexpspin1} above and identifying the kinematic polynomials \eqref{kinematic_pol} from the highest power of $s_{13}$, one finds:
{\allowdisplaybreaks\begin{subequations}\label{eq: massive spin one leading twist coef}
    \begin{align}
      a^{(m)}_{\tau^{(0)}_{12},1} & =  g_{12}g_{34}\,\pi^{\tfrac{d+2}{2}}\, \left(\frac{m}{2}\right)^{-2+3d-\sum_i \Delta_i}\Gamma\left(\frac{2-3d+\sum_{i} \Delta_i}{2}\right)\prod^4_{i=1}\frac{\Gamma\left(\tfrac{d}{2}-\Delta_i\right)}{4\pi^{\frac{d+2}{2}}}\\ \nonumber
        & \hspace*{1cm} \times 2 \frac{
\Gamma\!\left(\frac{d}{2}\right)
\Gamma\!\left(1 - \frac{d}{2} + \Delta_1\right)
\Gamma\!\left(1 - \frac{d}{2} + \Delta_2\right)
}
{
\Gamma(d + 1 - \Delta_1 - \Delta_2)\,
\Gamma(2 - d + \Delta_1 + \Delta_2)\,
\Gamma\!\left(\frac{2d - \Delta_1 - \Delta_2 - \Delta_3 - \Delta_4}{2}\right)} \\
& \hspace*{1cm} \times 
\Gamma\!\left(\frac{-\Delta_1 - \Delta_2 + \Delta_3 + \Delta_4}{2}\right)
\Gamma\!\left(\frac{d - \Delta_1 - \Delta_2 + \Delta_3 + \Delta_4}{2}\right)\nonumber \\
& \hspace*{1cm} \times \Gamma\!\left(\frac{2 - d + \Delta_1 + \Delta_2 - \Delta_3 + \Delta_4}{2}\right)\Gamma\!\left(\frac{2 - d + \Delta_1 + \Delta_2 + \Delta_3 - \Delta_4}{2}\right), \nonumber \\
a^{(m)}_{\tau^{(0)}_{12},0}  &= 0,
    \end{align}
\end{subequations}}and likewise for $\tau^{(0)}_{34}$. At leading twists $\tau^{(0)}_{12}$ and $\tau^{(0)}_{34}$ there is only a spin-1 operator contribution.

\vskip 4pt
One proceeds iteratively for higher twist contributions. In these cases, there is a mixing with the descendants of lower twist operators and their contributions must be subtracted. The subleading twist contributions are encoded in the residue
\begin{multline}
    \text{Res}\left[\rho(s_{12},s_{13})M^{{\cal V}_{12 A}{\cal V}_{34 A}}_{\Delta_1 \Delta_2\Delta_3\Delta_4}\left(s_{12},s_{13}\right) \right]_{s_{12} = \tau^{(1)}_{12}} = \tilde \rho(\tau^{(1)}_{12},s_{13}) \left[a^{(m)}_{\tau^{(0)}_{12},1}{\cal Q}_{\tau^{(0)}_{12},1,1}(s_{13}) \right.\\
   \left. + a^{(m)}_{\tau^{(1)}_{12},1}{\cal Q}_{\tau^{(1)}_{12},1,0}(s_{13}) + a^{(m)}_{\tau^{(1)}_{12},0}{\cal Q}_{\tau^{(1)}_{12},0,0}(s_{13})\right],
\end{multline}
where the first line is the contribution from the first descendant of the leading twist $\tau^{(0)}_{12}$ operator. After subtracting the latter one proceeds as for the leading twist contributions. This gives:
{\allowdisplaybreaks\begin{subequations}
    \begin{align}
      a_{\tau^{(1)}_{12},1}^{(m)} &= -(2-3d+\sum_{i=1}^4 \Delta_i)\left(\frac{m}{2}\right)^{-2+3d-\sum_{i=1}^4 \Delta_i}\Gamma\left(\frac{2-3d+\sum_{i=1}^4 \Delta_i}{2}\right) \\ \nonumber
      & \times g_{12}g_{34}\,\pi^{\tfrac{d+2}{2}}\,\prod^4_{i=1}\frac{\Gamma\left(\tfrac{d}{2}-\Delta_i\right)}{4\pi^{\frac{d+2}{2}}}\\ \nonumber
        & \times \Gamma\!\left(\frac{-2 + d - \Delta_1 - \Delta_2 + \Delta_3 + \Delta_4}{2}\right)
\sin\!\big(\pi\,(d - \Delta_1 - \Delta_2)\big)\\ \nonumber
        & \times \frac{
2
\Gamma\!\left( \tfrac{d}{2}-1 \right)
\Gamma\!\left(2 - \frac{d}{2} + \Delta_1\right)
\Gamma\!\left(2 - \frac{d}{2} + \Delta_2\right)}
{\pi\,(2 - d + \Delta_1 + \Delta_2)^2\,(d-3 - \Delta_1 - \Delta_2)\,(3d - 2(2 + \Delta_1 + \Delta_2))\,
}
\nonumber\\ 
&\nonumber\times \frac{\Gamma\!\left(\frac{4 - d + \Delta_1 + \Delta_2 + \Delta_3 - \Delta_4}{2}\right)
\Gamma\!\left(\frac{4 - d + \Delta_1 + \Delta_2 - \Delta_3 + \Delta_4}{2}\right)\Gamma\!\left(\frac{-2 - \Delta_1 - \Delta_2 + \Delta_3 + \Delta_4}{2}\right)}{\Gamma\!\left(\frac{-2 + 2d - \Delta_1 - \Delta_2 - \Delta_3 - \Delta_4}{2}\right)},
    \end{align}
      \begin{align}
      \label{eq: spin-o at n=1}
    a_{\tau^{(1)}_{12},0}^{(m)} & =-(2-3d+\sum_{i=1}^4 \Delta_i)\left(\frac{m}{2}\right)^{-2+3d-\sum_{i=1}^4 \Delta_i}\Gamma\left(\frac{2-3d+\sum_{i=1}^4 \Delta_i}{2}\right) \\ \nonumber
    & \times g_{12}g_{34}\,\pi^{\tfrac{d+2}{2}}\,\prod^4_{i=1}\frac{\Gamma\left(\tfrac{d}{2}-\Delta_i\right)}{4\pi^{\frac{d+2}{2}}}\\ 
    \nonumber
        & \times \Gamma\left( \frac{-2 + d - \Delta_{1} - \Delta_{2} + \Delta_{3} + \Delta_{4}}{2} \right)
\sin\big( \pi(d - \Delta_{1} - \Delta_{2})\big) \\
& \nonumber \times\frac{(\Delta_{1} - \Delta_{2})(\Delta_{3} - \Delta_{4})\,\Gamma\!\left( \tfrac{d}{2}-1 \right)\Gamma\!\left( 1 - \tfrac{d}{2} + \Delta_{1} \right)\Gamma\!\left( 1 - \frac{d}{2} + \Delta_{2} \right)}{4\pi\,(2d - 2 - \Delta_{1} - \Delta_{2})(1 - d + \Delta_{1} + \Delta_{2})(2 - d + \Delta_{1} + \Delta_{2})}\\ 
&\times\frac{\Gamma\!\left( \frac{2 - d + \Delta_{1} + \Delta_{2} + \Delta_{3} - \Delta_{4}}{2} \right)\Gamma\!\left( \frac{2 - d + \Delta_{1} + \Delta_{2} - \Delta_{3} + \Delta_{4}}{2} \right) \Gamma\!\left( \frac{-\Delta_{1} - \Delta_{2} + \Delta_{3} + \Delta_{4}}{2} \right)}{\Gamma\left( \frac{2d - \Delta_{1} - \Delta_{2} - \Delta_{3} - \Delta_{4}}{2} \right)}, \nonumber
\end{align}
\end{subequations}}and likewise for $\tau^{(1)}_{34}$. At subleading twists $\tau^{(1)}_{12}$ and $\tau^{(1)}_{34}$ we see that there are both spin-1 and spin-0 contributions.

\vskip 4pt
One proceeds similarly for all higher twists $\tau^{(n)}_{12}$ and $\tau^{(n)}_{34}$ with $n\geq2$ where, upon subtracting the contributions from descendants of lower twist operators, one uncovers new  spin-1 and spin-0 contributions of twists $\tau^{(n)}_{12}$ and $\tau^{(n)}_{34}$.

\vskip 4pt
In summary, we find that the exchange of a massive spin-1 field between massless scalars in $\mathbb{M}^{d+2}$ is encoded on the celestial sphere by the exchange of two infinite families of spin-1 operators with scaling dimensions:
\begin{subequations}\label{s1rdt}
 \begin{align}
    \Delta^{(n,1)}_{12}&=\Delta_1+\Delta_2+(1-d)+2n^{(1)}_{12}, \qquad n^{(1)}_{12}=0,1,2,\ldots,\\
   \Delta^{(n,1)}_{34}&=\Delta_3+\Delta_4+(1-d)+2n^{(1)}_{34}, \qquad n^{(1)}_{34}=0,1,2,\ldots\,,
\end{align}   
\end{subequations}
and two infinite families of scalar operators \eqref{scfam}:
\begin{subequations}
  \begin{align}
    \Delta^{(n,0)}_{12}&=\Delta_1+\Delta_2+(2-d)+2n^{(0)}_{12}, \qquad n^{(0)}_{12}=0,1,2,\ldots,\\
   \Delta^{(n,0)}_{34}&=\Delta_3+\Delta_4+(2-d)+2n^{(0)}_{34}, \qquad n^{(0)}_{34}=0,1,2,\ldots\,.
\end{align}  
\end{subequations}

\paragraph{Gauge boson.} When the exchanged spin-1 field is massless, the infinite families of operator contributions in the massive case collapse to a single pair of leading twist spin-1 operators with scaling dimensions:
\begin{align}\label{exchopml1}
    \Delta^{(0,1)}_{12}=\Delta_1+\Delta_2+(1-d), \qquad 
   \Delta^{(0,1)}_{34}=\Delta_3+\Delta_4+(1-d),
\end{align}
which are a shadow pair by virtue of the constraint \eqref{spin1mlconstr} on the external scaling dimensions $\Delta_i$:
\begin{equation}
    2-3d+\sum_i\Delta_i = 0.
\end{equation}
As we saw in section \ref{subsec::gaugeboson}, in this case the exchange is given, up to contact terms, by a conformal partial wave \eqref{mlspin1MA} encoding the contributions from the shadow pair of operators---from which one can read off the OPE data.

\vskip 4pt
Recall that the boundary two-point function \eqref{ml2pt} for massless scalars in Minkowski space only has support for $\Delta_{i} = \frac{d}{2}$. Setting $\Delta_1=\Delta_2=\frac{d}{2}$, the exchanged operators \eqref{exchopml1} correspond precisely to a boundary spin-1 current and boundary gauge boson:
\begin{equation}\label{bdcgb}
    \Delta^{(0,1)}_{12} = 1, \qquad \Delta^{(0,1)}_{34} = d-1,
\end{equation}
and likewise for $\Delta_3=\Delta_4=\frac{d}{2}$.

\vskip 4pt
The OPE data in the massless case can also be obtained from the massless limit of the massive OPE coefficients, where
\begin{subequations}
 \begin{align}
    a^{(0)}_{\tau^{(0)}_{12},1} & = \lim_{m \to 0}  a^{(m)}_{\tau^{(0)}_{12},1} \\
    &=2\pi i\, \delta\left(\tfrac{2-3d+\sum_i\Delta_i}{2}\right) a^{\prime\,(m=0)}_{\Delta^{(0,1)}_{12}, 1},
\end{align}   
\end{subequations}
where $a^{\prime\,(m=0)}_{\Delta^{(0,1)}_{12}, 1}$ is the coefficient of the CPW given in \eqref{spin1mlOPE} and it is straightforward to check that:\footnote{To obtain this we used: 
\begin{multline}
   \lim_{m\rightarrow0}\left(\frac{m}{2}\right)^{-2+3d-\sum_i\Delta_i}\left(-2+3d-\sum_i\Delta_i\right)\Gamma\left(\frac{2-3d+\sum_i\Delta_i}{2}\right) \\ = \left(-2+3d-\sum_i\Delta_i\right) 2\pi i \delta\left(\frac{2-3d+\sum_i\Delta_i}{2}\right). 
\end{multline}}
\begin{align}
    \lim_{m\to 0} a_{\tau^{(n>0)}_{12},1}^{(m)} = \lim_{m\to 0} a_{\tau^{(n>0)}_{12},0}^{(m)} = 0.
\end{align}
Likewise for $\tau^{(n)}_{12} \leftrightarrow \tau^{(n)}_{34}$.

\section{Spin-2 exchange}
\label{sec::spin2}

In this section we consider the celestial Mellin amplitude for the four-point exchange of massive (section \ref{subsec::massivespin2}) and massless (section \ref{subsec::graviton}) spin-2 fields between massless scalars in $\mathbb{M}^{d+2}$, extracting the direct channel OPE data on the celestial sphere in section \ref{subsec::OPEdataspin2}.

\subsection{Massive spin-2}
\label{subsec::massivespin2}

The equation for the propagator of a massive spin-2 field in Minkowski space is
\begin{equation}
    \left(\partial^2-m^2\right)G^{(m)\,\mu_1 \mu_2 \nu_1 \nu_2}\left(X,Y\right) = - \mathcal{P}_{m^2}^{\mu_1 \mu_2 \nu_1 \nu_2}\, \delta^{d+2}\left(X-Y\right),
\end{equation}
with symmetric and traceless projector:
\begin{subequations}
 \begin{align}\label{mspin2stp}
 \mathcal{P}_{m^2}^{\mu_1 \mu_2 \nu_1 \nu_2} &= \left(\mathcal{P}_{m^2}\right)^{\mu_1}{}_{\mu^\prime_1}\left(\mathcal{P}_{m^2}\right)^{\mu_2}{}_{\mu^\prime_2} \Pi^{\mu^\prime_1 \mu^\prime_2 \nu^\prime_1 \nu^\prime_2}_{d+1} \left(\mathcal{P}_{m^2}\right)_{\nu^\prime_1}{}^{\nu_1}\left(\mathcal{P}_{m^2}\right)_{\nu^\prime_2}{}^{\nu_2}, \\
    \Pi^{\mu_1 \mu_2 \nu_1 \nu_2}_{d+1} &=  \frac{1}{2}\left(\eta^{\mu_1 \nu_1} \eta^{\mu_2 \nu_2}+\eta^{\mu_1 \nu_2} \eta^{\mu_2 \nu_1}\right)-\frac{1}{d+1} \eta^{\mu_1 \mu_2} \eta^{\nu_1 \nu_2}. \label{stpdp1}
\end{align}   
\end{subequations}
The Feynman propagator for a massive spin-2 can then be constructed in terms of the scalar propagator as: 
\begin{equation}\label{spin_2_propagator}
    G^{(m)\,\mu_1\mu_2 \nu_1\nu_2}_T\left(X,Y\right)=\mathcal{P}_{m^2}^{\mu_1 \mu_2 \nu_1 \nu_2}G^{(m)}_T\left(X,Y\right).   \end{equation}
As for the spin-1 field, this allows to express Feynman diagrams involving massive spin-2 fields as differential operators in the external points acting on their scalar counterparts.

\vskip 4pt
We consider the four-point exchange of a massive spin-2 field $h_{\mu \nu}$ between massless scalars $\phi_i$ interacting via the cubic vertices
\begin{equation}
    \mathcal{V}_{12h} = g_{12} \left(\phi_1 \overleftrightarrow{\partial_\mu}\overleftrightarrow{\partial_\nu} \phi_2 \right) h^{\mu \nu}, \qquad \mathcal{V}_{34h} = g_{34} \left(\phi_3 \overleftrightarrow{\partial_\mu}\overleftrightarrow{\partial_\nu} \phi_4 \right) h^{\mu \nu},
\end{equation}
which reads
\begin{multline}\label{ampl_spin_2_massive}
    \mathcal{A}^{\mathcal{V}_{12h}\mathcal{V}_{34h}}(Y_1,Y_2,Y_3,Y_4) = -g_{12}g_{34} \int {\rm d}^{d+2}X_1 {\rm d}^{d+2}X_2 \left(G^{(0)}_T(Y_1,X_1)\overleftrightarrow{\partial_{\mu_1}}\overleftrightarrow{\partial_{\mu_2}}G^{(0)}_T(X_1,Y_2) \right) \\\times  G^{(m)\,\mu_1\mu_2 \nu_1\nu_2}_T\left(X_1,X_2\right)\left(G^{(0)}_T(Y_3,X_2)\overleftrightarrow{\partial_{\nu_1}}\overleftrightarrow{\partial_{\nu_2}}G^{(0)}_T(X_2,Y_4) \right).
\end{multline}
By translation invariance, all derivatives in the exchange integrand can be chosen to act on the external points $Y_i$. As for the spin-1 exchange, up to contact terms this gives:\footnote{Further details are given in section \ref{sec::higherspin}, where we present the extension to general spin $J$.}
\begin{equation}\label{spin2exchfromscalar}
    \mathcal{A}^{\mathcal{V}_{12h}\mathcal{V}_{34h}} 
    \approx {\cal D}^2\mathcal{A}^{\mathcal{V}_{12m}\mathcal{V}_{34m}}-\frac{m^4}{d+1}\mathcal{A}^{\mathcal{V}_{12m}\mathcal{V}_{34m}},
\end{equation}
where the first term on the rhs is two applications of the differential operator \eqref{D} on the massive scalar exchange \eqref{exchdef}. In Mellin space this corresponds to two iterations of the finite difference operation \eqref{recursion}:
\begin{equation}
     M^{\mathcal{V}_{12h}\mathcal{V}_{34h}}\left(\delta_{ij}\right) 
    \approx \Delta_{{\cal D}} \circ \Delta_{{\cal D}} \left[M^{\mathcal{V}_{12m}\mathcal{V}_{34m}}\left(\delta_{ij}\right)\right]-\frac{m^4}{d+1}M^{\mathcal{V}_{12m}\mathcal{V}_{34m}}\left(\delta_{ij}\right).
\end{equation}
 Iterating the spin-1 case, the corresponding celestial Mellin amplitude \eqref{MAcc1} is then a degree 2 polynomial in $s_{13}$ with poles in $s_{12}$ at:
\begin{equation}\label{massivespin2poles}
s_{12} = \Delta_1 + \Delta_2 -d-2+2n, \qquad s_{12} = \Delta_3 + \Delta_4 - d-2 + 2n.
\end{equation}
These correspond to operators on the co-dimension 2 celestial sphere of spins 2, 1 and 0, and twists:
\begin{subequations}
 \begin{align}
    \tau^{(n)}_{12} &= \Delta_1 + \Delta_2-d-2+2n_{12}, \qquad n_{12}=0,1,2,\ldots,\\ \tau^{(n)}_{34} &= \Delta_3 + \Delta_4 -d-2+2n_{34}, \qquad n_{34}=0,1,2,\ldots,
\end{align}   
\end{subequations}
encoding the bulk massive spin-2 exchanged single particle state in the direct channel on the celestial sphere. In section \ref{subsec::OPEdataspin2} we extract the corresponding OPE coefficients.

\subsection{Graviton}
\label{subsec::graviton}

For the graviton we proceed as for the gauge boson exchange in section \ref{subsec::gaugeboson}. For simplicity, the graviton propagator is taken in the de Donder gauge, which writes as
\begin{equation}
    G^{\mu_1 \mu_2, \nu_1 \nu_2}_T(X,Y)= \Pi^{\mu_1 \mu_2 \nu_1 \nu_2}_dG_T^{(0)}(X,Y),
\end{equation}
with 
\begin{equation}\label{spin2stpd}
    \Pi^{\mu_1 \mu_2 \nu_1 \nu_2}_d=\frac{1}{2}\left(\eta^{\mu_1 \nu_1} \eta^{\mu_2 \nu_2}+\eta^{\mu_1 \nu_2} \eta^{\mu_2 \nu_1}\right)-\frac{1}{d} \eta^{\mu_1 \mu_2} \eta^{\nu_1 \nu_2}.
\end{equation}
Other gauges might have been picked for the graviton, but just like for the gauge boson, gauge-dependent terms only contribute as contact terms, hence focusing on the de Donder gauge does not affect the generality of the result.
\vskip 4pt
We consider the four-point graviton exchange between massless scalars mediated by the cubic vertices\footnote{As mentioned in footnote \ref{fnt:onshell3pt}, cubic vertices between two scalars and a spinning field are unique on-shell \cite{Metsaev:2005ar}. When the external scalars are equal, the exchange diagram computed using the vertices \eqref{3ptgrav} therefore differ by contact terms from that computed using the standard coupling of gravity to matter via the stress-energy tensor $T_{\mu \nu} = -\partial_\mu \phi \partial_\nu \phi + \frac{1}{2}\eta_{\mu \nu} \partial^\sigma \phi \partial_\sigma \phi$.}
\begin{equation}\label{3ptgrav}
    \mathcal{V}_{12h} = g_{12} \left(\phi_1 \overleftrightarrow{\partial_\mu}\overleftrightarrow{\partial_\nu} \phi_2 \right) h^{\mu \nu},  \qquad \mathcal{V}_{34h} = g_{34} \left(\phi_3 \overleftrightarrow{\partial_\mu}\overleftrightarrow{\partial_\nu} \phi_4 \right) h^{\mu \nu},
\end{equation}
which reads
\begin{multline}
    \mathcal{A}^{\mathcal{V}_{12h}\mathcal{V}_{34h}}(Y_1,Y_2,Y_3,Y_4) = -g_{12}g_{34} \int {\rm d}^{d+2}X_1 {\rm d}^{d+2}X_2 \left(G^{(0)}_T(Y_1,X_1)\overleftrightarrow{\partial_{\mu_1}}\overleftrightarrow{\partial_{\mu_2}}G^{(0)}_T(X_1,Y_2) \right)\\
     \times G^{\mu_1\mu_2,\nu_1\nu_2}_T(X_1,X_2)\left(G^{(0)}_T(Y_3,X_2)\overleftrightarrow{\partial_{\nu_1}}\overleftrightarrow{\partial_{\nu_2}}G^{(0)}_T(X_2,Y_4) \right).
\end{multline}
As for the gauge boson exchange in section \ref{subsec::gaugeboson}, this can be expressed as the differential operator \eqref{D} in the external points acting on the exchange diagram \eqref{exchdef} of a massless scalar:
\begin{equation}\label{gravitonfromspin0}
    \mathcal{A}^{\mathcal{V}_{12h}\mathcal{V}_{34h}} \approx {\cal D}^2\mathcal{A}^{\mathcal{V}_{12m=0}\mathcal{V}_{34m=0}},
\end{equation}
where the trace term in \eqref{spin2stpd} contributes only contact terms (see section \ref{sec::higherspin}). Note that this is the square of the differential operator \eqref{gbexchfrsc} relating the gauge boson exchange to the massless scalar exchange, as expected from the double-copy structure of gravity.

 \vskip 4pt
At the level of the Mellin amplitudes the differential relation translates into \eqref{gravitonfromspin0}:
 \begin{equation}
     M^{\mathcal{V}_{12h}\mathcal{V}_{34h}}\left(\delta_{ij}\right) \approx \Delta_{{\cal D}} \circ \Delta_{{\cal D}} \left[M^{\mathcal{V}_{12m=0}\mathcal{V}_{34m=0}}\left(\delta_{ij}\right) \right],
 \end{equation}
 in terms of the Mellin amplitude of the massless scalar exchange, see \eqref{Mlbulkexch}. For integer $d \geq 3$ the latter simplifies \eqref{Mlbulkexchsimp} to a ratio of $\Gamma$-functions and the action of the difference operator \eqref{recursion} is straightforwardly implemented in Mathematica. Imposing the conformal constraints \eqref{confcon}, one finds that also for the graviton exchange the corresponding celestial Mellin amplitude is proportional to a spin-2 conformal partial wave (defined in appendix \ref{app::CPW}):
\begin{equation}\label{graviton_exch_mellin_ampl_first_coupling}
    M^{\mathcal{V}_{12h}\mathcal{V}_{34h}}_{\Delta_1 \Delta_2\Delta_3\Delta_4}\left(s_{12},s_{13}\right) = 2\pi i\,\delta\left(\frac{-3d + \sum_i \Delta_i}{2} \right)a^{\prime\,(m=0)}_{\Delta^{(0,2)}_{12}, J=2} \, \mathcal{F}_{\Delta_1 + \Delta_2 -d, 2}(s_{12}, s_{13}),
\end{equation}
with 
\begin{multline}\label{spin2mlOPE}
    a^{\prime\,(m=0)}_{\Delta^{(0,2)}_{12}, 2}=g_{12}g_{34}\,\pi^{\frac{d+2}{2}} \prod^4_{i=1}\left(\frac{\Gamma\left(\tfrac{d}{2}-\Delta_i\right)}{4\pi^{\frac{d+2}{2}}}\right)\frac{32  \Gamma \left(\frac{d}{2}+1\right) \Gamma\left(1-\frac{d}{2}+\Delta_1\right)\Gamma\left(1-\frac{d}{2}+\Delta_2\right) }{\Gamma (-d+\Delta_1+\Delta_2+2)}\\
    \times(-1)^{-d}\frac{ \Gamma (d+1-\Delta_3) \Gamma (d+1-\Delta_4) \Gamma (-d-1+\Delta_1+\Delta_2)\Gamma \left(\frac{3 d}{2}-\Delta_1-\Delta_2\right)}{\Gamma \left(-\frac{d}{2}\right)  \Gamma (-2 d+\Delta_1+\Delta_2)}.  
\end{multline}
This corresponds to the exchange of two (shadow) spin-2 operators with scaling dimensions:
\begin{align}\label{exchopml2}
    \Delta^{(0,2)}_{12}=\Delta_1+\Delta_2-d, \qquad 
   \Delta^{(0,2)}_{34}=\Delta_3+\Delta_4-d.
\end{align}
These are shadow of one another by virtue of the constraint \eqref{graviton_exch_mellin_ampl_first_coupling} on the external scaling dimensions.

\paragraph{Comment on the vDVZ discontinuity.} Notice that the expression \eqref{spin2exchfromscalar} for the massive spin-2 exchange reduces to the massless result \eqref{gravitonfromspin0} in the massless limit $m^2 \to 0$. In Minkowski space it is well known that the massless limit of Fierz-Pauli massive gravity does not reproduce the predictions of Einstein-Hilbert gravity---the vDVZ discontinuity \cite{vanDam:1970vg,Zakharov:1970cc}. It originates from the extra helicity-0 mode of a massive spin-2 field, which couples to the trace of the source. At the level of the propagator, this is the difference in the coefficient of the trace terms in the projectors \eqref{stpdp1} and \eqref{spin2stpd} between the massive and massless cases, and that difference survives as $m^2 \to 0$. As shown explicitly in section \ref{sec::higherspin}, the trace term is purely contact in the massless case (see equation \eqref{trcontr}). In other words, the vDVZ discontinuity lies in the contact terms, which we did not keep track of in the above. The discontinuity in fact persists for all higher spin $J \geq 2$ \cite{Francia:2007qt}, and for the same reason goes undetected when we consider the spin-$J$ exchange in section \ref{sec::higherspin}.

\subsection{OPE data}
\label{subsec::OPEdataspin2}

In this section we determine the direct channel OPE data for the four-point (massive and massless) spin-2 exchange.

\paragraph{Massive spin-2.} This follows the same steps as for the massive spin-1 in section \ref{subsec::opedataspin1}. The poles in $s_{12}$ in this case are at:
\begin{equation}\label{spin2exchs12poles}
    \qquad s_{12} = \Delta_1 + \Delta_2 -2-d+2(p+q), \qquad s_{12} = \Delta_3 + \Delta_4 -2 - d + 2(p+q), \qquad p, q \in \mathbb{N},
\end{equation}
which correspond to spin-2, spin-1 and spin-0 operators of twists
\begin{equation}\label{twistfamspin2}
    \tau^{(n)}_{12}= \Delta_1+\Delta_2-d-2+2n, \qquad \tau^{(n)}_{34}= \Delta_3+\Delta_4-d-2+2n, \qquad n \in \mathbb{N},
\end{equation}
and the conformal block expansion \eqref{MAcbe} of the Mellin amplitude takes the form:\footnote{Note that we leave implicit the contributions from poles \eqref{DTpoles} encoding double-trace operators \eqref{DT}. In this work we focus on the contributions from the exchanged single particle state in Minkowski space.} 
\begin{multline}\label{cbespin2}
\rho\left(s_{12},s_{13}\right)M^{{\cal V}_{12 h}{\cal V}_{34 h}}_{\Delta_1 \Delta_2\Delta_3\Delta_4}\left(s_{12}, s_{13}\right) = {\tilde \rho}\left(s_{12},s_{13}\right) \left[ \sum^\infty_{n=0} a^{(m)}_{\tau^{(n)}_{12},2}\sum^\infty_{m=0}\frac{{\cal Q}_{\tau^{(n)}_{12},2,m}\left(s_{13}\right)}{s_{12}-\tau^{(n)}_{12}-2m} \right. \\
\left. +\sum^\infty_{n=0} a^{(m)}_{\tau^{(n)}_{12},1}\sum^\infty_{m=0}\frac{{\cal Q}_{\tau^{(n)}_{12},1,m}\left(s_{13}\right)}{s_{12}-\tau^{(n)}_{12}-2m} +\sum^\infty_{n=0} a^{(m)}_{\tau^{(n)}_{12},0}\sum^\infty_{m=0}\frac{{\cal Q}_{\tau^{(n)}_{12},0,m}\left(s_{13}\right)}{s_{12}-\tau^{(n)}_{12}-2m}  + \left(\Delta_{1,2} \leftrightarrow \Delta_{3,4}\right) \right],
\end{multline}
where, as for the massive spin-1 exchange, there is mixing among spin-2, spin-1 and spin-0 operators (and their descendants). 

\vskip 4pt
Following the same procedure as for the spin-1 exchange one can determine the residue expansion of the spin-2 exchange Mellin amplitude on the poles \eqref{spin2exchs12poles}, which is straightforward to carry out in Mathematica. Its form is rather cumbersome so we do not report it explicitly here. The leading twist contributions \eqref{twistfamspin2} with $n=0$ are given by the residue:
\begin{multline}
    \text{Res}\left[\rho\left(s_{12},s_{13}\right)M^{{\cal V}_{12 h}{\cal V}_{34 h}}_{\Delta_1 \Delta_2\Delta_3\Delta_4}\left(s_{12}, s_{13}\right)\right]_{s_{12}=\tau^{(0)}_{12}}\\ = {\tilde \rho}\left(\tau^{(0)}_{12},s_{13}\right) \left[a^{(m)}_{\tau^{(0)}_{12},2}{\cal Q}_{\tau^{(0)}_{12},2,0}\left(s_{13}\right)+a^{(m)}_{\tau^{(0)}_{12},1}{\cal Q}_{\tau^{(0)}_{12},1,0}\left(s_{13}\right)+a^{(m)}_{\tau^{(0)}_{12},0}{\cal Q}_{\tau^{(0)}_{12},0,0}\left(s_{13}\right)\right],
\end{multline}
and likewise for the residue at $s_{12}=\tau^{(0)}_{34}$. By extracting this residue of the Mellin amplitude and identifying the kinematic polynomials \eqref{kinematic_pol} from the powers of $s_{13}$,\footnote{The highest power of $s_{13}$ comes from the spin-2 kinematic polynomial. Subtracting the contribution from the latter, the remaining linear term in $s_{13}$ is the contribution from the spin-1 kinematic polynomial etc.} one finds: 
{\allowdisplaybreaks\begin{subequations}\label{leadopespin2}\begin{align}
        a^{(m)}_{\tau^{(0)}_{12},2}  =& -32 g_{12}g_{34} \pi^{\tfrac{d+2}{2}}\left(\frac{m}{2}\right)^{3d-\sum_{i=1}^4 \Delta_i}\Gamma\left(\frac{-3d+\sum_i\Delta_i}{2}\right)\prod_i \frac{\Gamma\left(\frac{d}{2}- \Delta_i\right)}{4\pi^{\frac{d+2}{2}}}    \\ & \hspace*{1cm}
    \times \frac{ \Gamma \left(\frac{d}{2}+1\right)\Gamma\left(1-\frac{d}{2}+\Delta_1\right)\Gamma\left(1-\frac{d}{2}+\Delta_2\right)}{\Gamma (d+2-\Delta_1-\Delta_2) \Gamma (2-d+\Delta_1+\Delta_2) \Gamma \left(\frac{2d-\Delta_1-\Delta_2-\Delta_3-\Delta_4}{2}\right)} \nonumber \\  & \hspace*{1cm}
    \times  \Gamma \left(\frac{-\Delta_1-\Delta_2+\Delta_3+\Delta_4}{2}\right)\Gamma \left(\frac{2+d-\Delta_1-\Delta_2+\Delta_3+\Delta_4}{2}\right)  \nonumber \\ &  \hspace*{1cm}
    \times \Gamma\left(\frac{2-d+\Delta_1+\Delta_2-\Delta_3+\Delta_4}{2}\right)\Gamma\left(\frac{2-d+\Delta_1+\Delta_2+\Delta_3-\Delta_4}{2}\right),\nonumber \\
    a^{(m)}_{\tau^{(0)}_{12},1}&=0,\\
    a^{(m)}_{\tau^{(0)}_{12},0}&=0,
\end{align}
\end{subequations}}and likewise for $\tau^{(0)}_{34}$. At leading twists $\tau^{(0)}_{12}$ and $\tau^{(0)}_{34}$ there is only a spin-2 operator contribution.

\vskip 4pt
The subleading twist $\tau^{(1)}_{12}$ contributions are encoded in the residue:
\begin{multline}
    \text{Res}\left[\rho(s_{12},s_{13})M^{{\cal V}_{12 h}{\cal V}_{34 h}}_{\Delta_1 \Delta_2\Delta_3\Delta_4}\left(s_{12},s_{13}\right) \right]_{s_{12} = \tau^{(1)}_{12}} = \tilde \rho(\tau^{(1)}_{12},s_{13}) \left[a^{(m)}_{\tau^{(0)}_{12},2}{\cal Q}_{\tau^{(0)}_{12},2,1}(s_{13}) \right.\\
   \left. + a^{(m)}_{\tau^{(1)}_{12},2}{\cal Q}_{\tau^{(1)}_{12},2,0}(s_{13}) + a^{(m)}_{\tau^{(1)}_{12},1}{\cal Q}_{\tau^{(1)}_{12},1,0}(s_{13})+ a^{(m)}_{\tau^{(1)}_{12},0}{\cal Q}_{\tau^{(1)}_{12},0,0}(s_{13})\right],
\end{multline}
where the first line is the contribution from the first descendant of the leading twist $\tau^{(0)}_{12}$ contribution. Subtracting the latter descendant contribution one finds:
{\allowdisplaybreaks
\begin{subequations}
    \begin{align}           &a^{(m)}_{\tau^{(1)}_{12},2}=16g_{12}g_{34}\pi^{\tfrac{d+2}{2}}\left(\frac{m}{2}\right)^{3d-\sum_i\Delta_i}\Gamma\left(\frac{2-3d+\sum^4_{i=1}\Delta_i}{2}\right)\prod^4_{i=1}\frac{\Gamma\left(\tfrac{d}{2}-\Delta_i\right)}{4\pi^{\frac{d+2}{2}}}\\
    &\nonumber\phantom{xxxxxxx} \times (d-2 \Delta_1-2) (d-2\Delta_2-2)\frac{\Gamma\left(\frac{d - \Delta_1 - \Delta_2 + \Delta_3 + \Delta_4}{2}\right)}{\Gamma \left(\frac{2 d-2-\Delta_1-\Delta_2-\Delta_3-\Delta_4}{2} \right)} \\
    &\nonumber\phantom{xxxxxxx}\times
    \Gamma\left(\frac{4 - d + \Delta_1 + \Delta_2 + \Delta_3 - \Delta_4}{2}\right)
    \Gamma\left(\frac{4 - d + \Delta_1 + \Delta_2 - \Delta_3 + \Delta_4}{2}\right)\\
    &\nonumber\phantom{xxxxxxx}\times\frac{\Gamma \left(\frac{d}{2}\right) \Gamma \left(1-\frac{d}{2}+\Delta_1\right) \Gamma \left(1-\frac{d}{2}+\Delta_2\right)\Gamma\left(\frac{-2 - \Delta_1 - \Delta_2 + \Delta_3 + \Delta_4}{2}\right)}{\Gamma\left(d - \Delta_1 - \Delta_2\right)\Gamma\left(4-d+ \Delta_1 +\Delta_2\right) (-d+\Delta_1+\Delta_2+2)  (3 d-2 (\Delta_1+\Delta_2+1))},
    \end{align}
    \begin{align}&a^{(m)}_{\tau^{(1)}_{12},1}=-16g_{12}g_{34}\pi^{\tfrac{d+2}{2}}\left(\frac{m}{2}\right)^{3d-\sum_i\Delta_i}\Gamma\left(\frac{2-3d+\sum^4_{i=1}\Delta_i}{2}\right) \prod^4_{i=1}\frac{\Gamma\left(\tfrac{d}{2}-\Delta_i\right)}{4\pi^{\frac{d+2}{2}}} \\
    &\nonumber\phantom{xxxxxxx} \times ( \Delta_1-\Delta_2)(\Delta_3-\Delta_4)\frac{\Gamma\left(\frac{d - \Delta_1 - \Delta_2 + \Delta_3 + \Delta_4}{2}\right)}{\Gamma \left(\frac{2 d-\Delta_1-\Delta_2-\Delta_3-\Delta_4}{2} \right)} \\
    &\nonumber\phantom{xxxxxxx}\times
    \Gamma\left(\frac{2 - d + \Delta_1 + \Delta_2 + \Delta_3 - \Delta_4}{2}\right)
    \Gamma\left(\frac{2 - d + \Delta_1 + \Delta_2 - \Delta_3 + \Delta_4}{2}\right)\\
    &\nonumber\phantom{xxxxxxx}\times\frac{\Gamma \left(\frac{d}{2}\right) \Gamma \left(1-\frac{d}{2}+\Delta_1\right) \Gamma \left(1-\frac{d}{2}+\Delta_2\right)\Gamma\left(\frac{ - \Delta_1 - \Delta_2 + \Delta_3 + \Delta_4}{2}\right)}{\Gamma\left(d - \Delta_1 - \Delta_2+1\right)\Gamma\left(3-d+ \Delta_1 +\Delta_2\right) (-2d+\Delta_1+\Delta_2)},\\
    &a^{(m)}_{\tau^{(1)}_{12},0}=0,
\end{align}
\end{subequations}}and likewise for $\tau^{(1)}_{34}$. At subleading twists $\tau^{(1)}_{12}$ and $\tau^{(1)}_{34}$ one therefore finds spin-2 and spin-1 operator contributions.

\vskip 4pt
Moving to sub-sub-leading twists $\tau^{(2)}_{12}$ and $\tau^{(2)}_{34}$, we have:
\begin{multline}
    \text{Res}\left[\rho(s_{12},s_{13})M^{{\cal V}_{12 h}{\cal V}_{34 h}}_{\Delta_1 \Delta_2\Delta_3\Delta_4}\left(s_{12},s_{13}\right) \right]_{s_{12} = \tau^{(2)}_{12}}  = \tilde \rho(\tau^{(2)}_{12},s_{13}) \left[a^{(m)}_{\tau^{(0)}_{12},2}{\cal Q}_{\tau^{(0)}_{12},2,2}(s_{13}) \right.\\
   \left. + a^{(m)}_{\tau^{(1)}_{12},2}{\cal Q}_{\tau^{(1)}_{12},2,1}(s_{13}) + a^{(m)}_{\tau^{(1)}_{12},1}{\cal Q}_{\tau^{(1)}_{12},1,1}(s_{13})\right.\\
   \left. + a^{(m)}_{\tau^{(2)}_{12},2}{\cal Q}_{\tau^{(2)}_{12},2,0}(s_{13}) + a^{(m)}_{\tau^{(2)}_{12},1}{\cal Q}_{\tau^{(2)}_{12},1,0}(s_{13})+ a^{(m)}_{\tau^{(2)}_{12},0}{\cal Q}_{\tau^{(2)}_{12},0,0}(s_{13})\right],
\end{multline}
where the first line is the contribution from the second descendant of the leading twist $\tau^{(0)}_{12}$ operators, the second line is the contribution from the first descendant of the subleading twist $\tau^{(1)}_{12}$ operators. The corresponding OPE coefficients are given above. Subtracting the descendant contributions one finds:
{\allowdisplaybreaks 
\begin{subequations}
    \begin{align}
     &a^{(m)}_{\tau^{(2)}_{12},2}=-64g_{12}g_{34}\pi^{\tfrac{d+2}{2}}\left(\frac{m}{2}\right)^{3d-\sum_i\Delta_i}\Gamma\left(\frac{4-3d+\sum^4_{i=1}\Delta_i}{2}\right)\prod^4_{i=1}\frac{\Gamma\left(\tfrac{d}{2}-\Delta_i\right)}{4\pi^{\frac{d+2}{2}}}\label{ope_subsub_massive_spin_2}\\
    &\nonumber\phantom{xxxxxxx} \times\frac{1}{(-d+\Delta_1+\Delta_2+4)^2}\frac{ \Gamma\left(\frac{-2+d - \Delta_1 - \Delta_2 + \Delta_3 + \Delta_4}{2}\right)}{\Gamma \left(\frac{2 d-4-\Delta_1-\Delta_2-\Delta_3-\Delta_4}{2}\right)} \\
    &\nonumber\phantom{xxxxxxx}\times
    \Gamma\left(\frac{6 - d + \Delta_1 + \Delta_2 + \Delta_3 - \Delta_4}{2}\right)
    \Gamma\left(\frac{6 - d + \Delta_1 + \Delta_2 - \Delta_3 + \Delta_4}{2}\right)\\
    &\nonumber\phantom{xxxxxxx}\times\frac{\Gamma \left(-1+\frac{d}{2}\right) \Gamma \left(3-\frac{d}{2}+\Delta_1\right) \Gamma \left(3-\frac{d}{2}+\Delta_2\right)\Gamma\left(\frac{-4 - \Delta_1 - \Delta_2 + \Delta_3 + \Delta_4}{2}\right)}{\Gamma\left(d - \Delta_1 - \Delta_2\right)\Gamma\left(1-d+ \Delta_1 +\Delta_2\right)(-d+\Delta_1+\Delta_2+3)^2}\\  
    &\nonumber\phantom{xxxxxxx}\times\frac{1}{(-d+\Delta_1+\Delta_2+5)  (3 d-2 (\Delta_1+\Delta_2+2))(3 d-2 (\Delta_1+\Delta_2+3))},
    \end{align}
    \begin{align}
     &a^{(m)}_{\tau^{(2)}_{12},1}=-32g_{12}g_{34}\pi^{\tfrac{d+2}{2}}\left(\frac{m}{2}\right)^{3d-\sum_i\Delta_i}\Gamma\left(\frac{4-3d+\sum^4_{i=1}\Delta_i}{2}\right) \prod^4_{i=1}\frac{\Gamma\left(\tfrac{d}{2}-\Delta_i\right)}{4\pi^{\frac{d+2}{2}}} \label{ope_subsub_massive_spin_1} \\
    &\nonumber\phantom{xxxxxxx} \times ( \Delta_1-\Delta_2)(\Delta_3-\Delta_4)\frac{\Gamma\left(\frac{-2+d - \Delta_1 - \Delta_2 + \Delta_3 + \Delta_4}{2}\right)}{\Gamma \left(\frac{2 d-2-\Delta_1-\Delta_2-\Delta_3-\Delta_4}{2} \right)} \\
    &\nonumber\phantom{xxxxxxx}\times
    \Gamma\left(\frac{4 - d + \Delta_1 + \Delta_2 + \Delta_3 - \Delta_4}{2}\right)
    \Gamma\left(\frac{4 - d + \Delta_1 + \Delta_2 - \Delta_3 + \Delta_4}{2}\right)\\
    &\nonumber\phantom{xxxxxxx}\times\frac{\Gamma \left(-1+\frac{d}{2}\right) \Gamma \left(2-\frac{d}{2}+\Delta_1\right) \Gamma \left(2-\frac{d}{2}+\Delta_2\right)\Gamma\left(\frac{-2 - \Delta_1 - \Delta_2 + \Delta_3 + \Delta_4}{2}\right)}{\Gamma\left(d - \Delta_1 - \Delta_2\right)\Gamma\left(1-d+ \Delta_1 +\Delta_2\right) (2-2d+\Delta_1+\Delta_2)(3 d-2 (\Delta_1+\Delta_2+2))}\\
    &\nonumber\phantom{xxxxxxx}\times\frac{1}{(-d+\Delta_1+\Delta_2+2)^2 (-d+\Delta_1+\Delta_2+3)(-d+\Delta_1+\Delta_2+4) },
    \end{align}
    \begin{align}
     &a^{(m)}_{\tau^{(2)}_{12},0}=\label{ope_subsub_massive_spin_0}
     -g_{12}g_{34}\pi^{\tfrac{d+2}{2}}\left(\frac{m}{2}\right)^{3d-\sum_i\Delta_i}\Gamma\left(\frac{4-3d+\sum^4_{i=1}\Delta_i}{2}\right) \prod^4_{i=1}\frac{\Gamma\left(\tfrac{d}{2}-\Delta_i\right)}{4\pi^{\frac{d+2}{2}}} \\
    &\nonumber\phantom{xxxxxxx} \times  \frac{\alpha(\{\Delta_i\})}{(2-2d+\Delta_1+\Delta_2)}  \frac{\Gamma\left(\frac{-2+d - \Delta_1 - \Delta_2 + \Delta_3 + \Delta_4}{2}\right)}{\Gamma \left(\frac{2 d-\Delta_1-\Delta_2-\Delta_3-\Delta_4}{2} \right)} \\
    &\nonumber\phantom{xxxxxxx}\times
    \Gamma\left(\frac{2 - d + \Delta_1 + \Delta_2 + \Delta_3 - \Delta_4}{2}\right)
    \Gamma\left(\frac{2 - d + \Delta_1 + \Delta_2 - \Delta_3 + \Delta_4}{2}\right)\\
    &\nonumber\phantom{hellopaolo}\times\frac{\Gamma \left(-1+\frac{d}{2}\right) \Gamma \left(1-\frac{d}{2}+\Delta_1\right) \Gamma \left(1-\frac{d}{2}+\Delta_2\right)\Gamma\left(\frac{- \Delta_1 - \Delta_2 + \Delta_3 + \Delta_4}{2}\right)}{\Gamma\left(d - \Delta_1 - \Delta_2\right)\Gamma\left(1-d+ \Delta_1 +\Delta_2\right)(1-2d+\Delta_1+\Delta_2) }\\
    &\nonumber\phantom{xxxxxxx}\times\frac{1}{d(d+1)\left(1-d+ \Delta_1 +\Delta_2\right)\left(2-d+ \Delta_1 +\Delta_2\right)\left(3-d+ \Delta_1 +\Delta_2\right)},
 \end{align}
 \end{subequations}}
 where
 {\allowdisplaybreaks\begin{multline}
      \alpha(\{\Delta_i\})=2\left[2 d^2-d \left(\Delta_1^2+\Delta_1 (3-2 \Delta_2)+\Delta_2 (\Delta_2+3)+6\right)+4 (\Delta_1+1) (\Delta_2+1)\right]\\
      \times\left[2 d^2-d \left(3 \Delta_1+3 \Delta_2+(\Delta_3-\Delta_4)^2+6\right) \right. \\
      \left. +(\Delta_1+\Delta_2+\Delta_3-\Delta_4+2) (\Delta_1+\Delta_2-\Delta_3+\Delta_4+2)\right]. \nonumber
 \end{multline}}And likewise for $\tau^{(2)}_{34}$. At sub-sub-leading twists $\tau^{(2)}_{12}$ and $\tau^{(2)}_{34}$ we see that there spin-2, spin-1 and spin-0 contributions.

\vskip 4pt
For all higher twists $\tau^{(n)}_{12}$ and $\tau^{(n)}_{34}$ with $n\geq3$, proceeding in a similar way one uncovers new  spin-2, spin-1 and spin-0 contributions of twists $\tau^{(n)}_{12}$ and $\tau^{(n)}_{34}$ upon subtracting the contributions from descendants of lower twist operators.

\vskip 4pt
In summary, we find that the exchange of a massive spin-2 field between massless scalars in $\mathbb{M}^{d+2}$ is encoded on the celestial sphere by the exchange of the following operator families: Two infinite families of spin-2 operators with scaling dimensions:
\begin{subequations}
 \begin{align}
    \Delta^{(n,2)}_{12}&=\Delta_1+\Delta_2-d+2n^{(2)}_{12}, \qquad n^{(2)}_{12}=0,1,2,\ldots,\\
   \Delta^{(n,2)}_{34}&=\Delta_3+\Delta_4-d+2n^{(2)}_{34}, \qquad n^{(2)}_{34}=0,1,2,\ldots\,.
\end{align}   
\end{subequations}
Two infinite families of spin-1 operators \eqref{s1rdt}:
\begin{subequations}
 \begin{align}
    \Delta^{(n,1)}_{12}&=\Delta_1+\Delta_2+(1-d)+2n^{(1)}_{12}, \qquad n^{(1)}_{12}=0,1,2,\ldots,\\
   \Delta^{(n,1)}_{34}&=\Delta_3+\Delta_4+(1-d)+2n^{(1)}_{34}, \qquad n^{(1)}_{34}=0,1,2,\ldots\,.
\end{align}   
\end{subequations}
Two infinite families of scalar operators \eqref{scfam}:
\begin{subequations}
 \begin{align}
    \Delta^{(n,0)}_{12}&=\Delta_1+\Delta_2+(2-d)+2n^{(0)}_{12}, \qquad n^{(0)}_{12}=0,1,2,\ldots,\\
   \Delta^{(n,0)}_{34}&=\Delta_3+\Delta_4+(2-d)+2n^{(0)}_{34}, \qquad n^{(0)}_{34}=0,1,2,\ldots\,.
\end{align}   
\end{subequations}

\vskip 4pt
\paragraph{Graviton.} When the exchanged spin-2 field is massless the infinite families of operator contributions in the massive case collapse to a single pair of leading twist spin-2 operators
\begin{align}\label{exchopml2}
    \Delta^{(0,2)}_{12}=\Delta_1+\Delta_2-d, \qquad 
   \Delta^{(0,2)}_{34}=\Delta_3+\Delta_4-d,
\end{align}
which are shadow by virtue of the constraint \eqref{graviton_exch_mellin_ampl_first_coupling} on the external scaling dimensions:
\begin{equation}
    -3d+\sum_i\Delta_i = 0.
\end{equation}
As we saw in section \ref{subsec::graviton}, in this case the exchange is given, up to contact terms, by a conformal partial wave \eqref{graviton_exch_mellin_ampl_first_coupling} encoding the contributions from the shadow pair of operators---from which one can read off the OPE data.

\vskip 4pt
Recall that the boundary two-point function \eqref{ml2pt} for massless scalars in Minkowski space only has support for $\Delta_{i} = \frac{d}{2}$. Setting $\Delta_1=\Delta_2=\frac{d}{2}$, the exchanged operators \eqref{exchopml2} correspond precisely to a boundary spin-2 current and boundary graviton:
\begin{equation}\label{bdcgrv}
    \Delta^{(0,2)}_{12} = 0, \qquad \Delta^{(0,2)}_{34} = d,
\end{equation}
and likewise for $\Delta_3=\Delta_4=\frac{d}{2}$.

\vskip 4pt
The OPE data in the massless case can also be obtained from the massless limit of the massive OPE coefficients:
\begin{subequations}
 \begin{align}
    a^{(0)}_{\tau^{(0)}_{12},2} & = \lim_{m \to 0}  a^{(m)}_{\tau^{(0)}_{12},2} \\
    & = 2\pi i\,\delta\left(\frac{-3d+\sum_i\Delta_i}{2}\right) a^{\prime\,(m=0)}_{\Delta^{(0,2)}_{12}, 2},
\end{align}   
\end{subequations}
where $a^{\prime\,(m=0)}_{\Delta^{(0,2)}_{12}, 2}$ is the coefficient of the CPW given in \eqref{spin2mlOPE} and it is straightforward to check that:
\begin{equation}
    \lim_{m\to 0} a_{\tau^{(n>0)}_{12},2}^{(m)} = \lim_{m\to 0} a_{\tau^{(n>0)}_{12},1}^{(m)} =\lim_{m\to 0} a_{\tau^{(n>0)}_{12},0}^{(m)} = 0.
\end{equation}
Likewise for $\tau^{(n)}_{12} \leftrightarrow \tau^{(n)}_{34}$.

\section{Higher-spin exchanges}
\label{sec::higherspin}

\begin{figure}[t]
    \centering
    \includegraphics[width=\textwidth]{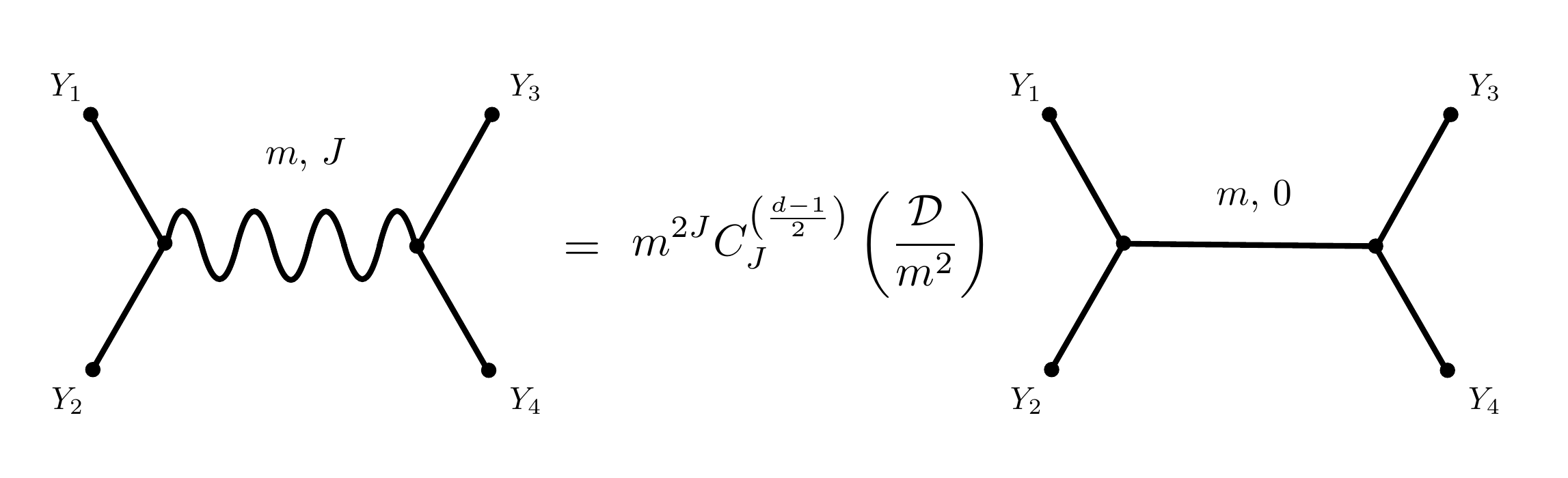}
    \caption{A spin-$J$ exchange can be expressed in terms of the differential operator ${\cal D}$, defined in \eqref{D}, acting on the spin-$0$ exchange of the same mass $m$, with Gegenbauer polynomial $C^{\left(\frac{d-1}{2}\right)}_J\left(z\right)$. See equation \eqref{spinJcjkspin0}.}
    \label{fig::spinjspin0}
\end{figure}

The approach naturally extends to particle exchanges of any spin. The equation for a massive integer spin-$J$ propagator in Minkowski space is
\begin{equation}
    \left(\partial^2-m^2\right)G^{(m)\,\mu_1 \ldots \mu_J \nu_1 \ldots \nu_J}\left(X,Y\right) = - \mathcal{P}_{m^2}^{\mu_1 \ldots \mu_J \nu_1 \ldots \nu_J}\, \delta^{d+2}\left(X-Y\right),
\end{equation}
with symmetric and traceless projector
\begin{equation}
    \mathcal{P}_{m^2}^{\mu_1 \dots \mu_J \nu_1\dots 
    \nu_J}=\frac{1}{J!}\sum_{\text{perm.} \nu_i}\prod_{i=1}^J\mathcal{P}_{m^2}^{\mu_i \nu_i} - \left(\mu_i\: \text{and}\: \nu_i\: \text{traces}\right),
    \label{spin_J_projector}
\end{equation}
where $\mathcal{P}_{m^2}^{\mu \nu}$ is given explicitly by
\begin{equation}
    \mathcal{P}_{m^2}^{\mu \nu}=\eta^{\mu \nu}+\frac{1}{m^2}\frac{\partial}{\partial X_{\mu}}\frac{\partial}{\partial Y_{\nu}}.
\end{equation}
The Feynman propagator can therefore be expressed in terms of the scalar propagator via
\begin{equation}\label{spinJfeyn}
    G^{(m)\,\mu_1 \ldots \mu_J \nu_1 \ldots \nu_J}_T\left(X,Y\right) = \mathcal{P}_{m^2}^{\mu_1 \dots \mu_J \nu_1\dots 
    \nu_J}G^{(m)}_T\left(X,Y\right),
\end{equation}
which is the generalisation of the spin-2 \eqref{spin_2_propagator} and spin-1 \eqref{spin_1_propagator} expressions to arbitrary integer spin $J$. This representation for the Feynman propagator allows to express correlators involving spinning fields in terms of differential operators acting on their scalar counterparts.

\vskip 4pt
Consider the four-point s-channel exchange of a massive spin-$J$ field $\varphi_{\mu_1 \ldots \mu_J}$ between massless scalars $\phi_i$ via the cubic vertices
\begin{equation}\label{cubicJ}
    \mathcal{V}_{12J} = g_{12} \left(\phi_1 \overleftrightarrow{\partial_{\mu_1}}\dots\overleftrightarrow{\partial_{\mu_J}} \phi_2 \right) \varphi^{\mu_1 \ldots \mu_J}, \qquad \mathcal{V}_{34J} = g_{34} \left(\phi_3 \overleftrightarrow{\partial_{\mu_1}}\dots\overleftrightarrow{\partial_{\mu_J}} \phi_4 \right) \varphi^{\mu_1 \ldots \mu_J}.
\end{equation}
Explicitly this reads
\begin{multline}\label{ampl_spin_J_massive}
    \mathcal{A}^{\mathcal{V}_{12J}\mathcal{V}_{34J}}(Y_1,Y_2,Y_3,Y_4) = -g_{12}g_{34} \int {\rm d}^{d+2}X_1 {\rm d}^{d+2}X_2 \left(G^{(0)}_T(Y_1,X_1)\overleftrightarrow{\partial_{\mu_1}}\dots\overleftrightarrow{\partial_{\mu_J}}G^{(0)}_T(X_1,Y_2) \right)\\
    \times \: G^{(m)\,\mu_1\dots\mu_J \nu_1\dots\nu_J}_T\left(X_1,X_2\right) \left(G^{(0)}_T(Y_3,X_2)\overleftrightarrow{\partial_{\nu_1}}\dots\overleftrightarrow{\partial_{\nu_J}}G^{(0)}_T(X_2,Y_4) \right).
\end{multline}
Using translation symmetry, the derivatives acting on the scalar propagators can be pulled out of the bulk integrand:
\begin{multline} 
    \mathcal{A}^{\mathcal{V}_{12J}\mathcal{V}_{34J}}(Y_1,Y_2,Y_3,Y_4) = -g_{12}g_{34} \left(\partial^{Y_2}_{\mu_1}-\partial^{Y_1}_{\mu_1}\right)\dots\left(\partial^{Y_2}_{\mu_J}-\partial^{Y_1}_{\mu_J}\right)\left(\partial^{Y_4}_{\nu_1}-\partial^{Y_3}_{\nu_1}\right)\dots\left(\partial^{Y_4}_{\nu_J}-\partial^{Y_3}_{\nu_J}\right)\\ \nonumber
    \hspace*{-0.5cm} \times  \int {\rm d}^{d+2}X_1 {\rm d}^{d+2}X_2\, G^{(0)}_T(Y_1,X_1) G^{(0)}_T(Y_2,X_1)  G^{(m)\,\mu_1\dots\mu_J \nu_1\dots\nu_J}_T\left(X_1,X_2\right) \\ \nonumber  \hspace*{6cm} \times G^{(0)}_T(Y_3,X_2)G^{(0)}_T(Y_4,X_2).
\end{multline}
For the contraction with the symmetric and traceless projector \eqref{spin_J_projector}, up to contact terms we have:
\begin{multline}\label{stconmJ}
   \mathcal{P}_{m^2}^{\mu_1 \dots \mu_J \nu_1\dots 
    \nu_J} \left(\partial^{Y_2}_{\mu_1}-\partial^{Y_1}_{\mu_1}\right)\dots\left(\partial^{Y_2}_{\mu_J}-\partial^{Y_1}_{\mu_J}\right)\left(\partial^{Y_4}_{\nu_1}-\partial^{Y_3}_{\nu_1}\right)\dots\left(\partial^{Y_4}_{\nu_J}-\partial^{Y_3}_{\nu_J}\right)\\
    \approx \frac{1}{J!}\left[\sum_{\sigma\in\Sigma_J}\left(\eta_{\mu_1\nu_{\sigma(1)}}\dots\eta_{\mu_J\nu_{\sigma(J)}}\right)-\text{traces}\right]\\ \times \left(\partial^{Y_2}_{\mu_1}-\partial^{Y_1}_{\mu_1}\right)\dots\left(\partial^{Y_2}_{\mu_J}-\partial^{Y_1}_{\mu_J}\right)\left(\partial^{Y_4}_{\nu_1}-\partial^{Y_3}_{\nu_1}\right)\dots\left(\partial^{Y_4}_{\nu_J}-\partial^{Y_3}_{\nu_J}\right).
\end{multline}
This expresses the spin-$J$ exchange as a differential operator in the external points acting on the corresponding massive scalar exchange diagram \eqref{exchdef}: 
\begin{multline}\label{Jexchfr0nonsimpl}
  \mathcal{A}^{\mathcal{V}_{12J}\mathcal{V}_{34J}}(Y_1,Y_2,Y_3,Y_4)
   \approx  \left[\left(\partial_{Y_2}-\partial_{Y_1}\right)^{2}\left(\partial_{Y_4}-\partial_{Y_3}\right)^{2}\right]^{J/2} \\ \times C^{\left(\frac{d-1}{2}\right)}_J\left(\frac{\left(\partial_{Y_1}-\partial_{Y_2}\right)\cdot \left(\partial_{Y_3}-\partial_{Y_4}\right)}{\sqrt{\left(\partial_{Y_2}-\partial_{Y_1}\right)^{2}\left(\partial_{Y_4}-\partial_{Y_3}\right)^{2}}}\right)\mathcal{A}^{\mathcal{V}_{12m}\mathcal{V}_{34m}}(Y_1,Y_2,Y_3,Y_4),
\end{multline}
where the Gegenbauer polynomial implements the symmetric and traceless contraction \eqref{stconmJ} and is given by:
\begin{subequations}\label{gegenbauer}
 \begin{align}
    C^{\left(\frac{d-1}{2}\right)}_J\left(z\right) &= \sum^{\lfloor J/2 \rfloor}_{k=0}c^{(d+1)}_{J,k}\, z^{J-2k}, \\
    c^{(d+1)}_{J,k}&=\left(-4\right)^{-k}\frac{J!}{k! \left(J-2k\right)!} \frac{\Gamma\left(J-k+\frac{d-1}{2}\right)}{\Gamma\left(\frac{d-1}{2}+J\right)}.
\end{align}   
\end{subequations}
The expression \eqref{Jexchfr0nonsimpl} can be simplified further up to contact terms, noting that each trace in the contraction \eqref{stconmJ} contributes as\footnote{In the first equality we used translation invariance in the form: $\partial_{X_1}=\partial_{Y_1}+\partial_{Y_2}$ and $\partial_{X_2}=\partial_{Y_3}+\partial_{Y_4}$, and the equation of motion for the external massless scalar fields $\partial^2_{Y_i} \approx 0$. The second equality uses the equation of motion for the exchanged massive scalar field $\partial^2_{X_i} \approx m^2$.}
\begin{align}\label{trcontr}
    \left(\partial_{Y_2}-\partial_{Y_1}\right)^2\left(\partial_{Y_4}-\partial_{Y_3}\right)^2 \approx \partial^2_{X_1} \partial^2_{X_2} \approx m^4,
\end{align}
which gives:
\begin{equation}\label{Jexchfr0}
  \mathcal{A}^{\mathcal{V}_{12J}\mathcal{V}_{34J}}(Y_1,Y_2,Y_3,Y_4)
  \approx  m^{2J} C^{\left(\frac{d-1}{2}\right)}_J\left(\frac{{\cal D}}{m^2}\right)\mathcal{A}^{\mathcal{V}_{12m}\mathcal{V}_{34m}}(Y_1,Y_2,Y_3,Y_4),
\end{equation}
where ${\cal D}$ is the shorthand for the differential operator \eqref{D}. This is depicted in figure \ref{fig::spinjspin0}. Equivalently, using the following recursion relation for Gegenbauer polynomials:
\begin{equation}
    C^{\left(\frac{d-1}{2}\right)}_{J+1}\left(z\right)=z\,C^{\left(\frac{d-1}{2}\right)}_{J}\left(z\right)-\frac{J(d+J-2)}{\left(d+2J-1\right)\left(d+2J-3\right)}C^{\left(\frac{d-1}{2}\right)}_{J-1}\left(z\right),
\end{equation}
the spin$-J$ exchange can be written in terms of the spin-$(J-1)$ and spin-$(J-2)$ exchanges as:
\begin{multline}\label{JexchfrJ-1}
  \mathcal{A}^{\mathcal{V}_{12J}\mathcal{V}_{34J}}(Y_1,Y_2,Y_3,Y_4)
  \approx {\cal D}\mathcal{A}^{\mathcal{V}_{12(J-1)}\mathcal{V}_{34(J-1)}}(Y_1,Y_2,Y_3,Y_4)\\
  -m^4\frac{(J-1)(J+d-3)}{\left(d+2J-3\right)\left(d+2J-5\right)}\mathcal{A}^{\mathcal{V}_{12(J-2)}\mathcal{V}_{34(J-2)}}(Y_1,Y_2,Y_3,Y_4),
\end{multline}
which reduces to the expression \eqref{spin2exchfromscalar} for the spin-2 exchange for $J=2$ and \eqref{spin-1exchexpansion} for the spin-1 exchange setting $J=1$.

\vskip 4pt
In practice, to evaluate the action of the derivatives it is convenient to work with the expression:
\begin{equation}\label{spinJcjkspin0}
  \mathcal{A}^{\mathcal{V}_{12J}\mathcal{V}_{34J}}(Y_1,Y_2,Y_3,Y_4)
   \approx  \sum^{\lfloor J/2 \rfloor}_{k=0}c^{(d+1)}_{J,k}\,m^{4k}{\cal D}^{J-2k} \mathcal{A}^{\mathcal{V}_{12m}\mathcal{V}_{34m}}(Y_1,Y_2,Y_3,Y_4),
\end{equation}
in terms of the Gegenbauer coefficients $c^{(d+1)}_{J,k}$ in \eqref{gegenbauer}. These relations can also be extended straightforwardly to massive external scalar fields, which is discussed in appendix \ref{app::extmass}.

\vskip 4pt
We see that a spin-$J$ exchange can be generated from the corresponding scalar exchange \eqref{exchdef} through iterative applications of the differential operator \eqref{spin-1exchexpansion}. In Mellin space, this translates into iterations of the difference operation \eqref{recursion} on the Mellin amplitude \eqref{MAexchdij} for the massive scalar exchange:
\begin{equation}
  M^{\mathcal{V}_{12J}\mathcal{V}_{34J}}(\delta_{ij})
   \approx  \sum^{\lfloor J/2 \rfloor}_{k=0}c^{(d+1)}_{J,k}\,m^{4k} \underbrace{\Delta_{{\cal D}} \circ \dots \circ \Delta_{{\cal D}}}_{J-2k\, \text{applications}}
    M^{\mathcal{V}_{12m}\mathcal{V}_{34m}}(\delta_{ij}).
\end{equation}
This form is straightforward to implement in Mathematica for any given spin-$J$.

\vskip 4pt
The celestial Mellin amplitude is obtained by simply enforcing the conformal constraints \eqref{confcon}. Since the spin-$J$ celestial Mellin amplitude is obtained from up to $J$ applications of the difference operation \eqref{recursion}, it is a degree-$J$ polynomial in the Mellin variable $s_{13}$ with the following poles in $s_{12}$:
\begin{equation}
     s_{12} = \Delta_1 + \Delta_2 +2-d-2J+2n, \qquad s_{12} = \Delta_3 + \Delta_4+2 - d-2J + 2n,  \qquad n \in \mathbb{N},
\end{equation}
which are shifted by $-2J$ with respect to the scalar case \eqref{s_12_poles_scalar1}. These correspond to operators on the co-dimension 2 celestial sphere of all spins from $J$ down to $0$, with twists:
\begin{subequations}\label{spinJexchtwists}
 \begin{align}
    \tau^{(n)}_{12} &= \Delta_1 + \Delta_2+2 - d-2J+2n_{12}, \qquad n_{12}=0,1,2,\ldots,\\ \tau^{(n)}_{34} &= \Delta_3 + \Delta_4 +2 - d-2J+2n_{34}, \qquad n_{34}=0,1,2,\ldots,
\end{align}   
\end{subequations}
and they encode the massive spin-$J$ exchanged single particle state in Minkowski space. 

\vskip 4pt
One proceeds as for the spin-2 and spin-1 exchanges to extract the corresponding OPE data. At leading twists $\tau^{(0)}_{12}$ and  $\tau^{(0)}_{34}$ there is only a spin-$J$ operator contribution, with OPE coefficient:\footnote{This can be proven by induction using the difference operator \eqref{recursion}, which is detailed in appendix \ref{app::spinjproof}.} 
{\allowdisplaybreaks \begin{subequations}\label{ope_coeff_massive_spin_J}
 \begin{align}
             a^{(m)}_{\tau^{(0)}_{12}, J}  &= -2^{4J-3} g_{12}g_{34} \pi^{\tfrac{d+2}{2}}\prod_i \frac{\Gamma\left(\frac{d}{2}- \Delta_i\right)}{4\pi^{\frac{d+2}{2}}}\\ \nonumber
        & \times\left(\frac{m}{2}\right)^{-4+3d+2J-\sum^4_{i=1}\Delta_i}\Gamma\left(\frac{4-3d-2J+\sum_i\Delta_i}{2}\right)\\  \nonumber
    & \times \frac{\Gamma (\Delta_1+\Delta_2+1-d-J)  \Gamma \left(-1+\frac{d}{2}+J\right)\Gamma\left(1-\tfrac{d}{2}+\Delta_1\right)\Gamma\left(1-\tfrac{d}{2}+\Delta_2\right)}{\Gamma (2-d+\Delta_1+\Delta_2)\Gamma (d-\Delta_1-\Delta_2) \Gamma (1-d+\Delta_1+\Delta_2) \Gamma \left(\frac{2d-\Delta_3-\Delta_4-\Delta_1-\Delta_2 }{2}\right)}\\
    \nonumber
    & \times \Gamma \left(\frac{-\Delta_1-\Delta_2+\Delta_3+\Delta_4}{2}\right) \Gamma \left(\frac{2J-2+d-\Delta_1-\Delta_2+\Delta_3+\Delta_4}{2}\right)
    \\  \nonumber
   & \times \Gamma\left(\frac{2-d+\Delta_1+\Delta_2-\Delta_3+\Delta_4}{2}\right)\Gamma\left(\frac{2-d+\Delta_1+\Delta_2+\Delta_3-\Delta_4}{2}\right),\\
    a^{(m)}_{\tau^{(0)}_{12}, J-1}  &= 0,\\
    & \vdots\\ \nonumber
     a^{(m)}_{\tau^{(0)}_{12}, 0}  &= 0,
\end{align}   
\end{subequations}}and likewise for $\tau^{(0)}_{34}$. At higher twists $\tau^{(n>0)}_{12}$ and $\tau^{(n>0)}_{34}$, as for the spin-2 and spin-1 exchanges it is necessary to subtract contributions from descendants of lower twist, where at twist $n \leq J $ contributions of spin $(J-n)$ emerge together with new contributions from all higher spins up to $J$.

\vskip 4pt
When the exchanged spin-$J$ particle is massless, we have the recursion:
\begin{align}
    \mathcal{A}^{\mathcal{V}_{12J}\mathcal{V}_{34J}}(Y_i)\approx {\cal D}\mathcal{A}^{\mathcal{V}_{12(J-1)}\mathcal{V}_{34(J-1)}}(Y_i) \approx {\cal D}^{J}\mathcal{A}^{\mathcal{V}_{12m=0}\mathcal{V}_{34m=0}}(Y_i),\label{mlJfr0}
\end{align}
which is given by $J$ applications of the differential operator \eqref{D} on the massless scalar exchange \eqref{Mlbulkexch}. In Mellin space this reads:
\begin{align} \nonumber
     M^{\mathcal{V}_{12J}\mathcal{V}_{34J}}(\delta_{ij}) &\approx \Delta_{{\cal D}} M^{\mathcal{V}_{12(J-1)}\mathcal{V}_{34(J-1)}}(\delta_{ij})  \\ &\approx \underbrace{\Delta_{{\cal D}} \circ \dots \circ \Delta_{{\cal D}}}_{J\, \text{applications}}  M^{\mathcal{V}_{12m=0}\mathcal{V}_{34m=0}}(\delta_{ij}),  
\end{align}
which can be deduced also from the massless limit of \eqref{JexchfrJ-1}. For integer $d \geq 3$ the Mellin amplitude \eqref{Mlbulkexchsimp} for the massless scalar exchange is a simple ratio of $\Gamma$-functions. Consequently, as for the gauge boson and graviton exchanges, one expected that the celestial Mellin amplitude for the massless spin-$J$ exchange is proportional to a spin-$J$ conformal partial wave $\mathcal{F}_{\Delta, J}$ (defined in appendix \ref{app::CPW}):
\begin{multline}\label{mass_lim_exch_res_sum_spin_J_2}
    M^{{\cal V}_{12J}{\cal V}_{34 J}}_{\Delta_1 \Delta_2\Delta_3\Delta_4}\left(s_{12},s_{13}\right)
     \\
    =  2\pi i\, \delta\left(\frac{-3d +4-2J + \sum_i \Delta_i}{2} \right)a^{\prime\,(m=0)}_{\tau^{(0)}_{12}, J} \mathcal{F}_{\Delta_1 + \Delta_2+2-d-J, J}(s_{12}, s_{13}),
\end{multline}
where
\begin{multline}
    a^{\prime\,(m=0)}_{\tau^{(0)}_{12}, J}=-2^{4 J-3}g_{12}g_{34}\,\pi^{\frac{d+2}{2}} \prod^4_{i=1}\frac{\Gamma\left(\tfrac{d}{2}-\Delta_i\right)}{4\pi^{\frac{d+2}{2}}} \\ \times \frac{(-1)^{-d}  \Gamma \left(\frac{d}{2}\right) \Gamma \left(1-\frac{d}{2}\right)  \Gamma\left(1-\frac{d}{2}+\Delta_1\right)\Gamma\left(1-\frac{d}{2}+\Delta_2\right) }{\Gamma (-d+\Delta_1+\Delta_2+2)}\\
    \times\frac{ \Gamma (d+J-\Delta_3-1) \Gamma (d+J-\Delta_4-1) \Gamma (-d-J+\Delta_1+\Delta_2+1)\Gamma \left(\frac{3 d}{2}-\Delta_1-\Delta_2\right)}{\Gamma \left(-\frac{d}{2}-J+2\right)^2  \Gamma (-2 d-2 J+\Delta_1+\Delta_2+4) \left(-\frac{3 d}{2}+\Delta_1+\Delta_2+1\right)_{2-J}}.
\end{multline}
This corresponds to the exchange of a pair of spin-$J$ operators with scaling dimensions:
\begin{align}\label{exchopmlJ}
    \Delta^{(0,J)}_{12}=\Delta_1 + \Delta_2 -d +2-J, \qquad 
   \Delta^{(0,J)}_{34}=\Delta_3 + \Delta_4 -d +2-J,
\end{align}
which are shadow of one another by virtue of the constraint \eqref{mass_lim_exch_res_sum_spin_J_2} on the scaling dimensions:
\begin{equation}
   -3d +4-2J + \sum_i \Delta_i = 0. 
\end{equation}
The result can be verified from the massless limit of the massive exchange, where from \eqref{ope_coeff_massive_spin_J} we have
\begin{equation}
      \lim_{m \rightarrow 0}a^{(m)}_{\tau^{(0)}_{12}, J} = 2\pi i\, \delta\left(\frac{-3d +4-2J + \sum_i \Delta_i}{2} \right) a^{\prime\,(m=0)}_{\tau^{(0)}_{12}, J}.
\end{equation}
There are no contributions from lower spin operators since massless particles in $\mathbb{M}^{d+2}$ are labeled by $SO(d)$ spin, as are irreducible representations of the conformal group on the $d$-dimensional celestial sphere.

\vskip 4pt
Recall that the boundary two-point function \eqref{ml2pt} for massless scalars in Minkowski space only has support for $\Delta_{i} = \frac{d}{2}$. Setting $\Delta_1=\Delta_2=\frac{d}{2}$, the exchanged operators \eqref{exchopmlJ} correspond precisely to a boundary spin-$J$ current and boundary spin-$J$ gauge boson:
\begin{equation}
    \Delta^{(0,J)}_{12} = 2-J, \qquad \Delta^{(0,J)}_{34} = J+d-2,
\end{equation}
and likewise setting $\Delta_3=\Delta_4=\frac{d}{2}$.

\section*{Acknowledgements} 

This research was supported by the European Union (ERC grant ``HoloBoot'', project number 101125112),\footnote{Views and opinions expressed are however those of the author(s) only and do not necessarily reflect those of the European Union or the European Research Council. Neither the European Union nor the granting authority can be held responsible for them.} by the MUR-PRIN grant No. PRIN2022BP52A (European Union - Next Generation EU) and by the INFN initiative STEFI.

\newpage

\begin{appendix}

\section{External massive scalars}
\label{app::extmass}

While in this work we focused for simplicity on processes with massless external scalar fields, the tools we present to study spinning exchange processes apply equally well for massive external scalars.  

\vskip 4pt
Consider exchange of a massive spin-$J$ field mediated by the cubic vertices \eqref{cubicJ} but now with scalars $\phi_i$ of mass $m_i$,
\begin{multline}
    \mathcal{A}^{\mathcal{V}_{12J}\mathcal{V}_{34J}}(Y_1,Y_2,Y_3,Y_4) = -g_{12}g_{34} \int {\rm d}^{d+2}X_1 {\rm d}^{d+2}X_2 \left(G^{(m_1)}_T(Y_1,X_1)\overleftrightarrow{\partial_{\mu_1}}\dots\overleftrightarrow{\partial_{\mu_J}}G^{(m_2)}_T(X_1,Y_2) \right)\\
    \times \: G^{(m)\,\mu_1\dots\mu_J \nu_1\dots\nu_J}_T\left(X_1,X_2\right) \left(G^{(m_3)}_T(Y_3,X_2)\overleftrightarrow{\partial_{\nu_1}}\dots\overleftrightarrow{\partial_{\nu_J}}G^{(m_4)}_T(X_2,Y_4) \right),
\end{multline}
where as before the Feynman propagator for the spin-$J$ field can be expressed as the symmetric traceless projector \eqref{spin_J_projector} acting on its spin-$0$ counterpart \eqref{spinJfeyn}, and we have:
\begin{multline} 
    \mathcal{A}^{\mathcal{V}_{12J}\mathcal{V}_{34J}}(Y_1,Y_2,Y_3,Y_4) = -g_{12}g_{34} \left(\partial^{Y_2}_{\mu_1}-\partial^{Y_1}_{\mu_1}\right)\dots\left(\partial^{Y_2}_{\mu_J}-\partial^{Y_1}_{\mu_J}\right)\left(\partial^{Y_4}_{\nu_1}-\partial^{Y_3}_{\nu_1}\right)\dots\left(\partial^{Y_4}_{\nu_J}-\partial^{Y_3}_{\nu_J}\right)\\ \nonumber
    \hspace*{-0.5cm} \times  \int {\rm d}^{d+2}X_1 {\rm d}^{d+2}X_2\, G^{(m_1)}_T(Y_1,X_1) G^{(m_2)}_T(Y_2,X_1)  G^{(m)\,\mu_1\dots\mu_J \nu_1\dots\nu_J}_T\left(X_1,X_2\right) \\ \nonumber  \hspace*{6cm} \times G^{(m_3)}_T(Y_3,X_2)G^{(m_4)}_T(Y_4,X_2).
\end{multline}
The difference with respect to the massless case lies in the contact terms, where now we have\footnote{This is obtained using the equations of motion $\partial^2_{Y_i} \approx m^2_i$ and $\partial^2_{X_i} \approx m^2$.} 
\begin{multline}
    \left(\partial^{Y_2}_{\mu_1}-\partial^{Y_1}_{\mu_1}\right)\left(\partial^{Y_2}_{\mu_2}-\partial^{Y_1}_{\mu_2}\right)\mathcal{P}_{m^2}^{\mu_1 \mu_2}\mathcal{P}_{m^2}^{\nu_1 \nu_2}\left(\partial^{Y_4}_{\nu_1}-\partial^{Y_3}_{\nu_1}\right)\left(\partial^{Y_4}_{\nu_2}-\partial^{Y_3}_{\nu_2}\right) \\
    \approx \left[m^2-\frac{\left(m^2_1-m^2_2\right)^2}{m^2}\right]\left[m^2-\frac{\left(m^2_3-m^2_4\right)^2}{m^2}\right],
\end{multline}
and
\begin{multline}
    \left(\partial^{Y_2}_{\mu_1}-\partial^{Y_1}_{\mu_1}\right)\mathcal{P}_{m^2}^{\mu_1 \nu_1}\left(\partial^{Y_4}_{\nu_1}-\partial^{Y_3}_{\nu_1}\right) \\
    \approx \left(\partial_{Y_1}-\partial_{Y_2}\right) \cdot \left(\partial_{Y_3}-\partial_{Y_4}\right) +\frac{\left(m^2_1-m^2_2\right)\left(m^2_3-m^2_4\right)}{m^2}.
\end{multline}
Together, these generalise to external massive scalars the representation \eqref{Jexchfr0} for the massive spin-$J$ exchange in terms of the corresponding scalar exchange diagram:
\begin{multline}
  \mathcal{A}^{\mathcal{V}_{12J}\mathcal{V}_{34J}}(Y_1,Y_2,Y_3,Y_4)
   \approx  \left[m^4-\left(m^2_1-m^2_2\right)^2\right]^{J/2}\left[m^4-\left(m^2_3-m^2_4\right)^2\right]^{J/2}m^{-2J} \\ \times C^{\left(\frac{d-1}{2}\right)}_J\left(\tfrac{m^2\left(\partial_{Y_1}-\partial_{Y_2}\right)\cdot \left(\partial_{Y_3}-\partial_{Y_4}\right)+\left(m^2_1-m^2_2\right)\left(m^2_3-m^2_4\right)}{\sqrt{\left(m^4-\left(m^2_1-m^2_2\right)^2\right)\left(m^4-\left(m^2_3-m^2_4\right)^2\right)}}\right)\mathcal{A}^{\mathcal{V}_{12m}\mathcal{V}_{34m}}(Y_1,Y_2,Y_3,Y_4).
\end{multline}
As a consistency check this reduces to \eqref{Jexchfr0} upon setting $m^2_i =0$.

\vskip 4pt
The consistent cubic coupling of a massless spin-$J$ field to scalar matter requires the scalars to have equal mass \cite{Berends:1984rq,Berends:1985xx}. This implies the representation \eqref{mlJfr0} for the massless spin-$J$ exchange continues to hold for massive external scalars:
\begin{multline}
     \mathcal{A}^{\mathcal{V}_{12J}\mathcal{V}_{34J}}(Y_1,Y_2,Y_3,Y_4)
  \\ \approx \left[\left(\partial_{Y_1}-\partial_{Y_2}\right)\cdot \left(\partial_{Y_3}-\partial_{Y_4}\right)\right]^{J}\mathcal{A}^{\mathcal{V}_{12m=0}\mathcal{V}_{34m=0}}(Y_1,Y_2,Y_3,Y_4),
\end{multline}
setting $m^2_1 = m^2_2$ and $m^2_3 = m^2_4$.

\vskip 4pt
These expressions allow for exchange diagrams with massive external scalars to be treated similarly to the massless case considered in the main text. The Mellin amplitude for a four-point scalar exchange with external massive scalar fields was given in appendix B of \cite{Pacifico:2024dyo}, where the dependence on each external mass $m_i$ is encoded in a Barnes integral \eqref{SPm} from the corresponding Feynman propagator---which is a modified Bessel-$K$ function. Compared to the power-law form \eqref{MLschw} of the massless Feynman propagator, this leads to a significantly more involved pole structure in the Mellin variable $s_{12}$.

\section{Spin-$J$ OPE coefficients}\label{app::spinjproof}
In this appendix we outline a derivation of the expression \eqref{ope_coeff_massive_spin_J} for the OPE coefficients of the leading twist contribution to massive spin$-J$ exchange diagrams. The proof follows by induction. The base cases are given by the spin-$0, 1$ and $2$ exchanges. Let's suppose \eqref{ope_coeff_massive_spin_J} to hold for spin$-J$. The inductive step can be implemented using the expression \eqref{JexchfrJ-1} for the spin-$(J+1)$ exchange in terms of spin$-J$ and $J-1$ exchanges, which in Mellin space reads:
\begin{multline}\label{eq:recursionMellinlevel}
M^{\mathcal{V}_{12(J+1)}\mathcal{V}_{34(J+1)}}(\delta_{ij})\approx\Delta_{{\cal D}}\left[M^{\mathcal{V}_{12J}\mathcal{V}_{34J}}(\delta_{ij})\right]\\
  -m^4 \frac{J(d+J-2)}{\left(d+2J-1\right)\left(d+2J-3\right)}M^{\mathcal{V}_{12(J-1)}\mathcal{V}_{34(J-1)}}(\delta_{ij}).
\end{multline}
From the knowledge of the poles for the lower spin exchanges and the action of the difference operator \eqref{D}, one can conclude that the Mellin amplitude for the spin-$(J+1)$ exchange has simple poles at:
\begin{equation}
    s_{12} = \tau^{(n)}_{12}-2, \quad s_{12} = \tau^{(n)}_{34}-2,  \quad n \in \mathbb{N},
\end{equation}
where $\tau^{(n)}_{12}$ and $\tau^{(n)}_{34}$ are the twists \eqref{spinJexchtwists} associated to the spin-$J$ exchange. The leading twist contributions $\tau^{(0)}_{12}-2$ and $\tau^{(0)}_{34}-2$ to the spin-$(J+1)$ exchange can only originate from the action of $\Delta_{{\cal D}}$ on the leading twist contribution to the spin-$J$ exchange on the first line of \eqref{eq:recursionMellinlevel}, which induces shifts in $s_{12}, s_{13}$ (and hence also in the $s_{12}$ poles) and increases the degree of the polynomial in $s_{13}$ by one unit. At the level of the residue expansion, this in particular would imply that:
    \begin{multline}\label{indstep}
           \Delta_{{\cal D}}\left[a^{(m)}_{\tau^{(0)}_{12}, J}\frac{{\cal Q}_{\tau_{12}^{(0)}, J,0}\left(s_{13}\right)}{s_{12}-\tau_{12}^{(0)}}\right]  \\ = \frac{1}{s_{12} - (\tau_{12}^{(0)}-2)}\left(a^{(m)}_{\tau^{(0)}_{12}-2,J+1}{\cal Q}_{\tau^{(0)}_{12}-2,J+1,0}\left(s_{13}\right) +O(s^J_{13}) \right)+ \,\ldots\,,
\end{multline}
where the $\ldots$ denote subleading twist contributions and $O(s^J_{13})$ potential lower spin contributions at leading twist. The spin-$(J+1)$ leading twist coefficient $a^{(m)}_{\tau^{(0)}_{12}-2,J+1}$ can be obtained from its counterpart $a^{(m)}_{\tau^{(0)}_{12},J}$ in the spin-$J$ exchange by reading off the coefficient of $s^{J+1}_{13}$. Plugging in \eqref{ope_coeff_massive_spin_J}, the hypothesis is confirmed for $a^{(m)}_{\tau^{(0)}_{12}-2,J+1}$ and we observe that the $O(s^J_{13})$ on the rhs of \eqref{indstep} vanishes, which concludes the proof.

\section{Schwinger parameter integrals}
\label{app::schwint}

We follow the approach outlined in \cite{Pacifico:2024dyo} to determine Mellin amplitudes for perturbative time-ordered correlators in Minkowski space, which reduces the calculation to certain integrals over the Schwinger parameters of the corresponding Feynman propagators. In this appendix we collect various results for the Schwinger parameter integrals appearing in contact and exchange diagrams, which were originally presented in \cite{Pacifico:2024dyo}.

\paragraph{Contact diagrams.} For contact diagrams \eqref{contnpt}, the Schwinger parameter integrals take the form
\begin{align}\label{contactschw}
     t_{\text{cont.}}\left(a_1,\ldots a_n;b\right)&=\int^{\infty}_0 \prod^n_{i=1} \frac{{\rm d}t_i}{t_i} t^{a_i}_i \left(t_1+\ldots + t_n\right)^{b}.
\end{align} 
This can be evaluated by writing 
\begin{equation}
    \left(t_1+\ldots + t_n\right)^{b}=\frac{1}{\Gamma\left(-b\right)}\int^\infty_0\frac{{\rm d}u}{u}u^{-b}e^{-u(t_1+\ldots + t_n)}.
\end{equation}
and performing the change of variables $t_i\to t_i/u$, which gives
\begin{align}
     t_{\text{cont.}}\left(a_1,\ldots a_n;b\right) &= \frac{1}{\Gamma\left(-b\right)} \left( \int^{\infty}_0 \prod^n_{i=1} \frac{{\rm d}t_i}{t_i} t^{a_i}_i e^{-t_i} \right)\left(\int^\infty_0 \frac{{\rm d}u}{u} u^{-b-\sum_i a_i}\right),\\
     &= 2 \pi i\, \delta (b+\sum_i a_i) \frac{1}{\Gamma\left(-b\right)} \prod^n_{i=1}\Gamma\left(a_i\right).
 \end{align}

\paragraph{Exchange diagrams.} For exchange diagrams \eqref{schwintexch}, the Schwinger parameter integrals take the form
\begin{multline}
t_{\text{exch}}\left(a_1,a_2,a_3,a_4;a;b;b_{12},b_{34}\right)=\int^\infty_0 \frac{{\rm d}{\bar t}}{{\bar t}} {\bar t}^{-(\delta_{13}+\delta_{14}+\delta_{23}+\delta_{24})+a}\\ \times \int^\infty_0 \prod^4_{i=1}\frac{{\rm d}t_i}{t_i} t^{a_i}_i\left[({\bar t}+t_1+t_2)({\bar t}+t_3+t_4)-{\bar t}^2\right]^{b}\\ \times \left({\bar t}+t_1+t_2\right)^{-\delta_{34}+b_{12}}\left({\bar t}+t_3+t_4\right)^{-\delta_{12}+b_{34}}.
\end{multline}
Using the generalisation of the Binomial expansion
\begin{equation}
    \left(x+y\right)^{b} = \frac{1}{\Gamma\left(-b\right)} \int^{+i\infty}_{-i\infty}\frac{{\rm d}z}{2\pi i} \Gamma\left(z-b\right)\Gamma\left(-z\right)x^z y^{b-z},
\end{equation}
with
\begin{equation}
    x = -{\bar t}^2, \qquad y = ({\bar t}+t_1+t_2)({\bar t}+t_3+t_4),
\end{equation}
one obtains
\begin{multline}
t_{\text{exch}}\left(a_1,a_2,a_3,a_4;a;b;b_{12},b_{34}\right)=\frac{1}{\Gamma\left(-b\right)} \int^{+i\infty}_{-i\infty}\frac{{\rm d}z}{2\pi i}\Gamma\left(z-b\right)\Gamma\left(-z\right)\left(-1\right)^z\\ \times \int^{\infty}_0 \frac{{\rm d}{\bar t}}{{\bar t}} {\bar t}^{-\left(\delta_{13}+\delta_{14}+\delta_{23}+\delta_{24}\right)+a+2z}\\ \times  \int^\infty_0 \frac{{\rm d}t_1}{t_1}\frac{{\rm d}t_2}{t_2} t^{a_1}_1t^{a_2}_2\left({\bar t}+t_1+t_2\right)^{b+b_{12}-\left(\delta_{34}+z\right)}\\ \times \int^\infty_0 \frac{{\rm d}t_3}{t_3}\frac{{\rm d}t_4}{t_4} t^{a_3}_3t^{a_4}_4\left({\bar t}+t_3+t_4\right)^{b+b_{34}-\left(\delta_{12}+z\right)},
\end{multline}
where the integrals over $t_{1,2}$ and $t_{3,4}$ factorise and take the form of the contact diagram integrals \eqref{contactschw}. The same steps give
\begin{align}\nonumber 
  \int^\infty_0 \frac{{\rm d}t_i}{t_i}\frac{{\rm d}t_j}{t_j} t^{a_i}_it^{a_j}_j ({\bar t}+t_i+t_j)^{c} ={\bar t}^{a_i+a_j+c} \frac{\Gamma\left(
    a_i\right)\Gamma\left(
    a_j\right)\Gamma\left(-a_i-a_j-c\right)}{\Gamma\left(-c\right)},
\end{align}
and the remaining integral over ${\bar t}$ a Dirac delta function, giving:
 \begin{multline}
t_{\text{exch}}\left(a_1,a_2,a_3,a_4;a;b;b_{12},b_{34}\right)=2 \pi i\, \delta\left(-\tfrac{d+2}{2}-\tfrac{\beta}{2}+a+a_1+a_2+a_3+a_4+2b+b_{12}+b_{34}\right)\\ \times \frac{\left(\prod^4_{i=1}\Gamma\left(a_i\right)\right)}{\Gamma\left(-b\right)} \int^{+i\infty}_{-i\infty}\frac{{\rm d}z}{2\pi i} \frac{\Gamma\left(z+\delta_{34}-b-b_{12}-a_1-a_2+z\right)\Gamma\left(z+\delta_{12}-b-b_{34}-a_3-a_4+z\right)}{\Gamma\left(z+\delta_{34}-b-b_{12}\right)\Gamma\left(z+\delta_{12}-b-b_{34}\right)}\\ \times \Gamma\left(z-b\right)\Gamma\left(-z\right)\left(-1\right)^z.
\end{multline}
This is the generalised ${}_3F_2$ hypergeometric function \eqref{MB3F2}.

\section{Conformal Partial Waves in Mellin space}

In this appendix we review the relevant aspects of conformal partial waves and their representation in Mellin space, where the latter was first introduced by Mack in \cite{Mack:2009mi,Mack:2009gy}.

\subsection{Conformal Partial Waves}\label{app::CPW}

Conformal Partial Waves (CPWs) ${\cal F}_{\Delta,l}$ are single-valued Eigenfunctions of the Casimir invariants of the conformal group. Four point functions can be rewritten in a conformally invariant manner as
\begin{multline}\label{CPW}
{\cal F}_{\Delta,l}(\vec{x}_1,\vec{x}_2,\vec{x}_3,\vec{x}_4) =\frac{1}{\left(x^2_{12}\right)^{\frac{\Delta_1+\Delta_2}{2}}\left(x^2_{34}\right)^{\frac{\Delta_3+\Delta_4}{2}}}\left(\frac{x^2_{24}}{x^2_{14}}\right)^{\frac{\Delta_1-\Delta_2}{2}}\left(\frac{x^2_{14}}{x^2_{13}}\right)^{\frac{\Delta_3-\Delta_4}{2}}{\cal F}_{\Delta,l}(u,v),
\end{multline}
which in turn has the Mellin representation
\begin{equation}\label{mell_rep_cpw}
    {\cal F}_{\Delta,l}(u,v)=\int^{+i\infty}_{-i\infty}\frac{{\rm d}s_{12}{\rm d}s_{13}}{\left(4\pi i\right)^2}\, u^{\frac{s_{12}}{2}}v^{-\left(\frac{s_{12}+s_{13}}{2}\right)}\,\rho\left(s_{12},s_{13}\right)\,{\cal F}_{\Delta,l}\left(s_{12},s_{13}\right),
\end{equation}
where $u$ and $v$ are the conformal invariant cross ratios defined in \eqref{crossratios}. The Mellin amplitude is given by: 
\begin{equation}\label{mell_rep_cpw_ii}
    {\cal F}_{\Delta,l}\left(s_{12},s_{13}\right)={\mathfrak C}_{l,\tau}(\Delta_i)\Omega_{l}(s_{12})P_{\tau,l}(s_{12},s_{13}),
\end{equation}
with twist $\tau=\Delta-l$, and ${\mathfrak C}_{l,\tau}$ and $\Omega_{l}$ are defined as
\begin{subequations}
\begin{align}
    {\mathfrak C}_{l,\tau}(\tau_i)&=-\frac{2^{-2l}\Gamma (2 l+\tau )(l+\tau-1)_l}{\Gamma \left(\tfrac{d}2- l-\tau \right)}\\
    &\nonumber\phantom{xx}\times\frac{1}{\Gamma \left(\tfrac{\tau +2l+\Delta_1-\Delta_2}{2}\right) \Gamma \left(\tfrac{\tau+2l -\Delta_1+\Delta_2}{2}\right) \Gamma \left(\tfrac{\tau +2l+\Delta_3-\Delta_4}{2}\right) \Gamma \left(\tfrac{\tau+2l -\Delta_3+\Delta_4}{2}\right)},\\ \label{omegal}
    \Omega_l(s_{12})&=\frac{\Gamma \left(\tfrac{\tau -s_{12}}{2}\right) \Gamma \left(\tfrac{d-2 l-s_{12}-\tau}{2}\right)}{\Gamma \left(\tfrac{-s_{12}+\Delta_1+\Delta_2}{2}\right) \Gamma \left(\tfrac{-s_{12}+\Delta_3+\Delta_4}{2}\right)}\,.
\end{align}
\end{subequations} 
$P_{\tau,l}(s_{12},s_{13})$ is a degree $l$ polynomial in both the Mellin variables $s_{12}$ and $s_{13}$ known as Mack-polynomials. In general, the explicit form of these polynomials is rather intricate, and are generally given in terms of nested sums. We choose the normalization of the Mack polynomials such that $P_{\tau,l}(s_{12},s_{13})\sim s_{13}^l+\ldots$ and we give their explicit form in Appendix \ref{a::mack_pol}. 

\vskip 4pt
The kinematic information inside ${\cal F}_{\Delta,l}$ can be decoded in more detail if we look at the poles in $s_{12}$ of the Mellin representation \eqref{mell_rep_cpw_ii}. Indeed, the function $\Omega_l$ in \eqref{omegal} has poles at
\begin{equation}\label{cpwpoles}
   s_{12}=\tau+2m, \qquad  s_{12}=d-\tau-2l+2m, \qquad m \in \mathbb{N},
\end{equation}
corresponding to the exchange of a pair of spin-$l$ operators with shadow scaling dimensions $\Delta$ and $d-\Delta$, where $\tau=\Delta-l$. The $\Gamma$-function factors $\Gamma\left(\frac{\Delta_3+\Delta_4-s_{12}}{2}\right)\Gamma\left(\frac{\Delta_1+\Delta_2-s_{12}}{2}\right)$ in the denominator of $\Omega_l$ cancel the double-trace operator poles \eqref{DT} of the Mellin measure \eqref{MMeasure}.

 \vskip 4pt
 The residues of the conformal partial wave \eqref{mell_rep_cpw_ii} on the poles of $\Omega_l$ are the kinematic polynomials ${\cal Q}_{\tau,l,m}$ of degree $l$ in $s_{13}$,
\begin{equation}\label{residue_mell_rep_cpw}
    \rho(s_{12},s_{13}){\cal F}_{\Delta,l}(s_{12},s_{13})=\tilde{\rho}(s_{12},s_{13})\sum^\infty_{m=0}\frac{{\cal Q}_{\tau,l,m}\left(s_{13}\right)}{s_{12}-\tau-2m} + \text{shadow},
\end{equation}
where shadow in last equation refers to the shadow sequence of poles at $s_{12}=d-\tau-2l+2m$. The kinematic polynomials ${\cal Q}_{\tau,l,m}(s_{13})$ are related to the Mack polynomials via \cite{Costa:2012cb}
\begin{equation}\label{kinematic_pol}
    {\cal Q}_{\tau,l,m}(s_{13})=\Xi^{\{\Delta_i\}}_{l,\tau,m}P_{\tau,l}(\tau+2m,s_{13}),
\end{equation}
where
\begin{multline}
\label{eq:xiMack}
    \Xi^{\{\Delta_i\}}_{l,\tau,m}=-\frac{2^{-2l+1}(l+\tau-1)_l\Gamma(2l+\tau)}{m! \left(-\frac{d}{2}+l+\tau +1\right)_m}\frac{1}{\Gamma\left(l+\tfrac{\tau+\Delta_1-\Delta_2}{2}\right)\Gamma\left(l+\tfrac{\tau-\Delta_1+\Delta_2}{2}\right)}\\
    \times\frac{1}{\Gamma\left(l+\tfrac{\tau+\Delta_3-\Delta_4}{2}\right)\Gamma\left(l+\tfrac{\tau-\Delta_3+\Delta_4}{2}\right)}.
\end{multline}
The primary contribution corresponds to the case $m=0$ and the descendants contributions $m>0$ are completely fixed by conformal symmetry. In \cite{Costa:2012cb}, it is shown that the kinematic polynomials \eqref{kinematic_pol} corresponding to the lowest (primary) twist, $m=0$, form an orthogonal basis and can be expressed in terms of the so-called continuous Hahn polynomials $Q_{\tau,l}(s_{13})$ \cite{andrews_askey_roy_1999} as
\begin{multline}
    {\cal Q}_{\tau,l,0}(s_{13})=-\frac{2^{-2l+1}(l+\tau-1)_l\Gamma(2l+\tau)}{\Gamma\left(l+\tfrac{\tau+\Delta_1-\Delta_2}{2}\right)\Gamma\left(l+\tfrac{\tau-\Delta_1+\Delta_2}{2}\right)}\\
    \times\frac{1}{\Gamma\left(l+\tfrac{\tau+\Delta_3-\Delta_4}{2}\right)\Gamma\left(l+\tfrac{\tau-\Delta_3+\Delta_4}{2}\right)}Q_{\tau,l}(s_{13}).
\end{multline}
The continuous Hahn polynomial $Q_{\tau,l}(s_{13})$ is also a degree $l$ polynomial in $s_{13}$ whose normalisation is also chosen to be unit normalisation, ie. $Q_{\tau,l}(s_{13})\sim s^l_{13}+\ldots$. As it has been derived explicitly in \cite{Costa:2012cb}, the continuous Hahn polynomials can be expressed in terms of ${}_3F_2$ hypergeometric functions as 
\begin{equation}\label{hahn_def}
\begin{split}
    Q_{\tau,l}(s_{13})&=\frac{2^l\left(\frac{\Delta_{12}+\tau}{2}\right)_l\left(\frac{\Delta_{34}+\tau}{2}\right)_l}{\left(\tau+l-1\right)_l}{}_3F_2\left(\begin{matrix}-l,l+\tau-1,\frac{\Delta_{34}-s_{13}}{2}\\\frac{\tau+\Delta_{12}}{2},\frac{\tau+\Delta_{34}}{2}\end{matrix};1\right)\\
    &=\frac{2^l\left(\frac{\Delta_{12}+\tau}{2}\right)_l\left(\frac{\Delta_{34}+\tau}{2}\right)_l}{\left(\tau+l-1\right)_l}\sum^l_{n=0}\frac{1}{n!}\frac{(-l)_n(\tau+l-1)_n(\frac{\Delta_{34}-s_{13}}{2})_n}{(\frac{\tau+\Delta_{12}}{2})_n(\frac{\tau+\Delta_{34}}{2})_n}.
\end{split}
\end{equation}
The unit normalisation for the leading order can be checked by considering the $n=l$ term in the sum in \eqref{hahn_def} given other terms correspond to lower degrees in $s_{13}$,
\begin{equation}\label{leading_term_hahn}
    Q_{\tau,l}(s_{13})=\frac{2^l(-l)_l\left(\frac{\Delta_{34}-s_{13}}{2}\right)_l}{l!}+\ldots,
\end{equation}
where $\ldots$ represents the terms $n<l$. From the properties of Pochhammer symbol
\begin{subequations}
    \begin{align}
       (-a)_m&=\frac{(-1)^m}{(a+1)_{-m}},\qquad m\in \mathbb{Z},\\
       (a+b)_m&=\sum^m_{j=0}\binom{m}{j}(a)_{m-j}(b)_j\;,
\end{align}
\end{subequations}
we have that
\begin{subequations}
    \begin{align}
        (-l)_l&=(-1)^l \,l\,!\;,\label{identities_for_hahn_norm_i}\\
       \left(\frac{\Delta_{34}-s_{13}}{2}\right)_l&=\frac{1}{2^l}\sum^l_{j=0}(-1)^j\binom{l}{j}(\Delta_{34})_{l-j}(s_{13})_j\label{identities_for_hahn_norm_ii}\\
       \nonumber&=\frac{(-1)^l s^l_{13}}{2^l}+O(s^{l-1}_{13})\;,
\end{align}
\end{subequations}
and in last sum we isolated the $\propto s^l_{13}$ contribution from the $j=l$ term as it gives the leading term in $Q_{l,\tau}(s_{13})$. Substituting \eqref{identities_for_hahn_norm_i} and \eqref{identities_for_hahn_norm_ii} in \eqref{leading_term_hahn}, we obtain that $Q_{\tau,l}(s_{13})=s^l_{13}+O(s^{l-1}_{13})$.

\subsection{Mack polynomials}\label{a::mack_pol}
In this appendix we recall the definition of Mack polynomials and their main properties that are useful in this work. We follow the normalisation presented in \cite{Sleight:2018epi} and define Mack polynomials as\footnote{To lighten notation we will report the Mack polynomials simply as $P_{\tau,l^\prime}(s_{12},s_{13})$, considering implicitly the dependence on $\Delta_1,\Delta_2,\Delta_3,\Delta_4$.}:
\begin{multline}\label{MackPscalar}
    P_{\tau,l^\prime}(s_{12},s_{13}|\Delta_1,\Delta_2,\Delta_3,\Delta_4)=\frac{4^{l^\prime} \left(\tfrac{\tau +\Delta_1-\Delta_2}{2}\right)_{l^\prime} \left(\tfrac{\tau -\Delta_1+\Delta_2}{2}\right)_{l^\prime}}{(d-l^\prime-\tau -1)_{l^\prime} (l^\prime+\tau -1)_{l^\prime}}\\
    \times\sum_{k=0}^{[l^\prime/2]}c^{(d)}_{l^\prime,k}\left[\sum_{\sum_ir_i=l^\prime-2k}p_{r_1,r_2,r_3,r_4}^{l^\prime,k}(s_{12},s_{13}|\Delta_1,\Delta_2,\Delta_3,\Delta_4)\right],
\end{multline}
 where we introduced:
\begin{multline}
    p_{r_i}^{l^\prime,k}(s_{12},s_{13}|\Delta_1,\Delta_2,\Delta_3,\Delta_4)=\left(\frac{\tau -s_{12}}{2}\right)_k \left(\frac{d-2 l^\prime-s_{12}-\tau}{2}\right)_k\\\times\,\frac{\left(\frac{-d+\tau +\Delta_3-\Delta_4+2}{2} \right)_{k+r_1+r_2} \left(\frac{-d+\tau -\Delta_3+\Delta_4+2}{2}\right)_{k+r_3+r_4}}{\left(\frac{\tau -\Delta_1+\Delta_2}{2}\right)_{k+r_1+r_3} \left(\frac{\tau +\Delta_1-\Delta_2}{2}\right)_{k+r_2+r_4}}\,\mathfrak{p}_{r_i}(s_{12},s_{13}|\Delta_1,\Delta_2,\Delta_3,\Delta_4),
\end{multline}
with\footnote{The multinomial coefficient $(r_1,r_2,r_3,r_4)!$ is defined as
\begin{equation}
    (r_1,r_2,r_3,r_4)!:=\frac{(r_1+r_2+r_3+r_4)!}{r_1!r_2!r_3!r_4!}.
\end{equation}
}
\begin{multline}
    \mathfrak{p}_{r_i}(s_{12},s_{13}|\Delta_1,\Delta_2,\Delta_3,\Delta_4)=\frac{(-1)^{r_1+r_4}}{2^{r_1+r_2+r_3+r_4}}(r_1,r_2,r_3,r_4)!\\\times\, \left(\frac{s_{12}+s_{13}}{2}\right)_{r_1} \left(\frac{-s_{13}+\Delta_3-\Delta_4}{2}\right)_{r_2}\\
    \times\left(\frac{-s_{13}-\Delta_1+\Delta_2}{2}\right)_{r_3} \left(\frac{s_{12}+s_{13}+\Delta_1-\Delta_2-\Delta_3+\Delta_4}{2}\right)_{r_4}.
\end{multline}
The coefficients $c^{(d)}_{l^\prime,k}$ are the Gegenbauer polynomial coefficients \eqref{gegenbauer}. 

\vskip 4pt
The definition of the Mack polynomials \eqref{MackPscalar} ensures a unit normalization for the leading term as  $P_{\tau,l}(s_{12},s_{13})\sim s_{13}^{l}+\ldots\,$\,. To see this, let us consider the term $k=0$ in \eqref{MackPscalar} as it will give the leading $s_{13}$ contribution to the polynomial
\begin{multline}\label{mack_norm_proof_sum}
    P_{\tau,l}(s_{12},s_{13})=\frac{s_{13}^{l}(-1)^{l}l! \left(\tfrac{\tau +\Delta_1-\Delta_2}{2}\right)_{l} \left(\tfrac{\tau -\Delta_1+\Delta_2}{2}\right)_{l}}{(d-l-\tau -1)_{l} (l+\tau -1)_{l}}\\
    \times\sum_{\sum{r_i}=l}\frac{\left(\frac{-d+\tau +\Delta_3-\Delta_4+2}{2} \right)_{r_1+r_2} \left(\frac{-d+\tau -\Delta_3+\Delta_4+2}{2}\right)_{r_3+r_4}}{\left(\frac{\tau -\Delta_1+\Delta_2}{2}\right)_{r_1+r_3} \left(\frac{\tau +\Delta_1-\Delta_2}{2}\right)_{r_2+r_4}}\frac{1}{r_1!r_2!r_3!r_4!}\,+\ldots\,,
\end{multline}
where $\ldots$ are the $k>0$ contributions. To perform the sum we change variables to $q_1=l-r_1-r_2$ and $q_2=l-r_1-r_3$, which imply $r_4=-l+r_1+q_1+q_2$:
\begin{multline}\label{mack_norm_proof_sum_i}
    \sum_{\sum{r_i}=l}\frac{\left(\frac{-d+\tau +\Delta_3-\Delta_4+2}{2} \right)_{r_1+r_2} \left(\frac{-d+\tau -\Delta_3+\Delta_4+2}{2}\right)_{r_3+r_4}}{\left(\frac{\tau -\Delta_1+\Delta_2}{2}\right)_{r_1+r_3} \left(\frac{\tau +\Delta_1-\Delta_2}{2}\right)_{r_2+r_4}}\frac{1}{r_1!r_2!r_3!r_4!}\\
    =\sum^l_{q_1=0}\sum^l_{q_2=0}\sum^{l-q_1}_{r_1=0}\frac{\left(\frac{-d+\tau +\Delta_3-\Delta_4+2}{2} \right)_{l-q_1} \left(\frac{-d+\tau -\Delta_3+\Delta_4+2}{2}\right)_{q_1}}{\left(\frac{\tau -\Delta_1+\Delta_2}{2}\right)_{l-q_2} \left(\frac{\tau +\Delta_1-\Delta_2}{2}\right)_{q_2}}\\\times\frac{1}{r_1!(l-r_1-q_1)!(l-r_1-q_2)!(r_1+q_1+q_2-l)!}\,.
\end{multline}
Now we can perform the sum over $r_1$ as
    \begin{align}
    &\sum^{l-q_1}_{r_1=0}\frac{1}{r_1!(l-r_1-q_1)!(l-r_1-q_2)!(r_1+q_1+q_2-l)!}\\ \nonumber & \hspace*{4cm}=\frac{1}{(l-q_1)!(l-q_2)!(q_1+q_2-l)!}\sum^{l-q_1}_{r_1=0}\frac{1}{r_1!}\frac{(q_1-l)_{r_1}(q_2-l)_{r_1}}{(q_1+q_2-l+1)_{r_1}}\\ \nonumber
    & \hspace*{4cm} =\frac{1}{(l-q_1)!(l-q_2)!(q_1+q_2-l)!}{}_2F_1\left(\begin{matrix}q_1-l,q_2-l\\q_1+q_2-l+1\end{matrix};1\right)\\ \nonumber
    & \hspace*{4cm}=\frac{l!}{q_1!q_2!(l-q_1)!(l-q_2)!}.
    \end{align}
In second line we used the Pochhammer symbol property
\begin{equation}
    (-1)^k\frac{(-a)_{k}}{a!}=\frac{1}{(a-k)!},
\end{equation}
in third line we identified the Gaussian hypergeometric function ${}_2F_1$ as
\begin{equation}
    {}_2F_1\left(\begin{matrix}-n,-m\\-n-m+l+1\end{matrix};1\right)=\sum^{n}_{k=0}\frac{1}{k!}\frac{(-n)_k(-m)_k}{(-n-m+l+1)_k},
\end{equation}
and in fourth line we used the identity
\begin{equation}
    {}_2F_1\left(\begin{matrix}a,b\\c\end{matrix};1\right)=\frac{\Gamma(c)\Gamma(c-a-b)}{\Gamma(c-a)\Gamma(c-b)},\qquad\text{Re}(c-a-b)>0\,.
\end{equation}
Plugging the result of the sum over $r_1$ in \eqref{mack_norm_proof_sum_i}, we are able to factorize the two sums over $q_1$ and $q_2$ as:
\begin{multline}
   \sum^l_{q_1=0}\sum^l_{q_2=0}\frac{\left(\frac{-d+\tau +\Delta_3-\Delta_4+2}{2} \right)_{l-q_1} \left(\frac{-d+\tau -\Delta_3+\Delta_4+2}{2}\right)_{q_1}}{\left(\frac{\tau -\Delta_1+\Delta_2}{2}\right)_{l-q_2} \left(\frac{\tau +\Delta_1-\Delta_2}{2}\right)_{q_2}}\frac{l!}{q_1!q_2!(l-q_1)!(l-q_2)!}\\ 
   =\frac{1}{l!}\frac{\left(-d+\tau+2\right)_{l}(l+\tau-1)_l}{\left(\frac{\tau -\Delta_1+\Delta_2}{2}\right)_{l}\left(\frac{\tau +\Delta_1-\Delta_2}{2}\right)_{l}}\,,
\end{multline}
and we performed these sums using the Pochhammer binomial identities
\begin{subequations}
    \begin{align}
        (a+b)_n&=\sum^n_{j=0}\binom{n}{j}(a)_{n-j}(b)_j,\\
        \frac{(a+b-1+n)_n}{(a)_n(b)_n}&=\sum^n_{j=0}\binom{n}{j}\frac{1}{(a)_{n-j}(b)_{j}}.
    \end{align}
\end{subequations}
Finally we can plug the results of the sums over $q_1$ and $q_2$ into \eqref{mack_norm_proof_sum} and using the identity of Pochhammer
\begin{equation}
    (-d+\tau+2)_l=(-1)^l(d-l-\tau-1)_l,
\end{equation}
we get indeed that $P_{\tau,l}(s_{12},s_{13})=s^l_{13}+O(s^{l-1}_{13})$.
Mack polynomials reduce to the kinematic polynomials ${\cal Q}_{\tau,l,m}(s_{13})$ controlling the OPE expansion for specific values of $s_{12}$ \cite{Costa:2012cb}:
\begin{equation}
    {\cal Q}_{\tau,l,m}(s_{13})=\Xi^{\{\Delta_i\}}_{l,\tau,m}P_{\tau,l}(\tau+2m,s_{13}),
\end{equation}
where
\begin{multline}
    \Xi^{\{\Delta_i\}}_{l,\tau,m}=-\frac{2^{-2l+1}(l+\tau-1)_l\Gamma(2l+\tau)}{m! \left(-\frac{d}{2}+l+\tau +1\right)_m}\frac{1}{\Gamma\left(l+\tfrac{\tau+\Delta_1-\Delta_2}{2}\right)\Gamma\left(l+\tfrac{\tau-\Delta_1+\Delta_2}{2}\right)}\\
    \times\frac{1}{\Gamma\left(l+\tfrac{\tau+\Delta_3-\Delta_4}{2}\right)\Gamma\left(l+\tfrac{\tau-\Delta_3+\Delta_4}{2}\right)},
\end{multline}
as defined in \eqref{eq:xiMack}.

\section{Contact terms}
\label{app::contactterms}

In this appendix we give further details on the various types of contact terms appearing in four-point exchange diagrams. Such terms are neglected in the main text, which focuses on the contribution from the exchanged single-particle state.

\subsection{Toy example: Scalar exchange diagram with derivative vertex}
\label{section: Exchange diagram with derivative scalar vertex}

Consider the exchange of a massive scalar field $\phi$ between massless scalars $\phi_i$ generated by the cubic vertices
\begin{equation}\label{derivcubic}
    \mathcal{V}_{\bar{1}\bar{2}m} = g_{12}\partial_\mu \phi_1 \partial^\mu \phi_2 \phi, \qquad \mathcal{V}_{m34} = g_{34} \phi \phi_3 \phi_4.
\end{equation}
The vertex $\mathcal{V}_{\bar{1}\bar{2}m}$ is on-shell equivalent to the non-derivative cubic vertex $\mathcal{V}_{12m}$, which implies the four-point exchange diagram generated by the cubic vertices \eqref{derivcubic} differs from that \eqref{exchdef} generated by non-derivative cubic vertices \eqref{cubicv} by contact terms. In this appendix we examine the form of such contact terms in more detail, which is a toy example of the remark in footnote \ref{foo::cubicimpr}.

\vskip 4pt
The four-point exchange diagram generated by the vertices \eqref{derivcubic} can be expressed as the following differential operator acting on its non-derivative counterpart \eqref{exchdef} (see also section 2.1 of \cite{Pacifico:2024dyo}):
\begin{equation}\label{eq: DY1DY2}
    \mathcal{A}^{\mathcal{V}_{\bar{1}\bar{2}m}\mathcal{V}_{m34}}(Y_1,Y_2,Y_3,Y_4) = \left(\partial_{Y_1}\cdot\partial_{Y_2}\right)\mathcal{A}^{\mathcal{V}_{12m}\mathcal{V}_{m34}}(Y_1,Y_2,Y_3,Y_4),
\end{equation}
where we recall the form of \eqref{exchdef} for convenience below
\begin{multline}\label{exchmasslessdefA}
   {\cal A}^{{\cal V}_{12 m}{\cal V}_{34 m}}\left(Y_1,Y_2,Y_3,Y_4\right) = -g_{12}g_{34} \int {\rm d}^{d+2}X {\rm d}^{d+2}Y\, G^{(0)}_{T}\left(X,Y_1\right)G^{(0)}_{T}\left(X,Y_2\right) \\ \times G^{(m)}_{T}\left(X,Y\right)G^{(0)}_{T}\left(Y,Y_3\right)G^{(0)}_{T}\left(Y,Y_4\right).
\end{multline}
By translation invariance we can write:
\begin{equation}
   \partial_{Y_1}\cdot\partial_{Y_2} = \frac{1}{2}(\partial^2_{X} - \partial^2_{Y_1} - \partial^2_{Y_2}),
\end{equation}
so that using the propagator equations for $\phi$ and $\phi_i$ we have
\begin{multline}\label{derivexchA}
   2 \mathcal{A}^{\mathcal{V}_{\bar{1}\bar{2}m}\mathcal{V}_{m34}}(Y_1,Y_2,Y_3,Y_4) = m^2 {\cal A}^{{\cal V}_{12 m}{\cal V}_{34 m}}\left(Y_1,Y_2,Y_3,Y_4\right) \\ + g_{12}g_{34} \int {\rm d}^{d+2}X \, G^{(0)}_{T}\left(X,Y_1\right)G^{(0)}_{T}\left(X,Y_2\right) G^{(0)}_{T}\left(X,Y_3\right)G^{(0)}_{T}\left(X,Y_4\right)\\
   - g_{12}g_{34} G^{(0)}_{T}\left(Y_1,Y_2\right) \int {\rm d}^{d+2}Y \,  G^{(m)}_{T}\left(Y_1,Y\right) G^{(0)}_{T}\left(Y,Y_3\right)G^{(0)}_{T}\left(Y,Y_4\right)\\
   - g_{12}g_{34} G^{(0)}_{T}\left(Y_1,Y_2\right) \int {\rm d}^{d+2}Y \,  G^{(m)}_{T}\left(Y_2,Y\right) G^{(0)}_{T}\left(Y,Y_3\right)G^{(0)}_{T}\left(Y,Y_4\right).
\end{multline}
We see that the first line on the rhs is indeed proportional to the exchange \eqref{exchmasslessdefA} generated by non-derivative cubic vertices. The second line is a four-point contact diagram (e.g. \eqref{contnpt}) generated by a non-derivative vertex \eqref{nptcontnond} of the external massless scalars $\phi_i$. The third and fourth lines are three-point contact diagrams of $\phi$ with $\phi_{3,4}$ multiplied by a massless propagator $G^{(0)}_{T}\left(Y_1,Y_2\right)$. In this paper we work modulo such terms, focusing on the contributions encoding the exchanged single-particle state---which in this example would be the first line of the exchange \eqref{derivexchA}.

\subsection{Massive spin-one exchange}\label{a::contact_massive_spin_1}

The expression \eqref{mass_spin_1_exch_dij} for the massive spin-1 exchange diagram \eqref{spin-1exchdef} is given modulo the contact terms. In this appendix we detail their contribution for completeness and to illustrate their form.

\vskip 4pt
The neglected contact terms take the form (see equation \ref{spin-1exchexpansion})
\begin{equation}\label{spin-1exchcontact}
    \frac{1}{m^2}\left(\partial^2_{Y_1}-\partial^2_{Y_2}\right)\left(\partial^2_{Y_3}-\partial^2_{Y_4}\right){\cal A}^{{\cal V}_{12 m}{\cal V}_{34 m}}\left(Y_1,Y_2,Y_3,Y_4\right),
\end{equation}
in terms of the massive scalar exchange \eqref{exchdef}. From the equation for the Feynman propagator it is straightforward to see that this reduces to a sum of products of Feynman propagators:
\begin{multline}
      \frac{1}{m^2}\left(\partial^2_{Y_1}-\partial^2_{Y_2}\right)\left(\partial^2_{Y_3}-\partial^2_{Y_4}\right){\cal A}^{{\cal V}_{12 m}{\cal V}_{34 m}}\left(Y_1,Y_2,Y_3,Y_4\right) \\ =   \frac{(-ig_{12})(-ig_{34})}{m^2}G^{(0)}_T(Y_1,Y_2)G^{(0)}_T(Y_3,Y_4) \left(G^{(m)}_T(Y_1,Y_3) - G^{(m)}_T(Y_1,Y_4) \right. \\
         \left. \phantom{x}- G^{(m)}_T(Y_2,Y_3) + G^{(m)}_T(Y_2,Y_4) \right).
\end{multline}

\subsection{Gauge boson exchange} 
\label{a::gauge_terms_spin_1}

The gauge dependent terms in the gauge boson propagator \eqref{gbprop} similarly contribute contact terms to the gauge boson exchange, which we show in this appendix.

\vskip 4pt
The gauge dependent part of the propagator reads
\begin{equation}
    G^\xi_{\mu \nu}(X,Y) = \partial^Y_\mu \partial^Y_\nu \Pi^\xi(X,Y), 
\end{equation}
with 
\begin{equation}
    \Pi_\xi(X,Y) = \frac{\Gamma\left(\frac{d}{2}\right)}{4 \pi^{\frac{d+2}{2}}} \frac{1 - \xi}{2(d-2)} \frac{1}{\left[(X-Y)^2 + i \epsilon \right]^{\frac{d}{2}-1}}. 
\end{equation}
In the exchange diagram, the corresponding gauge dependent contribution reads
\begin{multline}
        A^{\mathcal{V}_{12A}\mathcal{V}_{34 A}}_\xi(Y_1,Y_2,Y_3,Y_4) = -g_{12}g_{34} \int {\rm d}^{d+2}X_1 {\rm d}^{d+2}X_2 \left(G^{(0)}_T(Y_1,X_1)\overleftrightarrow{\partial_\mu}G^{(0)}_T(X_1,Y_2) \right)\\
         \times \partial_{X_1}^\mu \partial_{X_1}^\nu \Pi_\xi(X_1,X_2) \left(G^{(0)}_T(Y_3,X_2)\overleftrightarrow{\partial_\nu}G^{(0)}_T(X_2,Y_4) \right).
\end{multline}
Integrating by parts and using the Feynman propagator equations one obtains:
\begin{multline}
        A^{\mathcal{V}_{12A}\mathcal{V}_{34 A}}_\xi(Y_1,Y_2,Y_3,Y_4) = g_{12}g_{34}G^{(0)}_T(Y_1,Y_2)G^{(0)}_T(Y_3,Y_4) \\
        \times \left(\Pi_\xi(Y_1,Y_3) - \Pi_\xi(Y_1,Y_4)- \Pi_\xi(Y_2,Y_3) + \Pi_\xi(Y_2,Y_4) \right).
\end{multline}

\end{appendix}

\newpage

\bibliographystyle{JHEP}
\bibliography{refs}

\end{document}